\documentclass[11pt,oneside,final]{huthesis}
\usepackage{color,bbding}
\usepackage[textfont={sl}, labelfont={bf}, margin=10pt, labelsep=quad,    indent=1cm, justification=justified, singlelinecheck=false ]{caption}
\usepackage{epsfig,bm,epsf,float}
\usepackage{graphicx}
\usepackage{makeidx}
\usepackage{multicol}
\usepackage{geometry}
\usepackage{amsfonts}
\usepackage{amssymb}
\usepackage{amsmath}
\input{tcilatex}
 \def\lsim{\mathrel{\mathstrut\smash{\ooalign{\raise2.5pt\hbox{$<$}\cr\lower2.5pt\hbox{$\sim$}}}}}
\usepackage{amsmath,comment}
\usepackage{amsfonts}
\usepackage{amssymb,graphics,psfrag}
\usepackage{array,epsfig,stmaryrd,graphicx}
\usepackage{cite}
\usepackage{graphicx}
\usepackage{makeidx}
\usepackage{multicol}
\usepackage{geometry}
\usepackage{amsfonts}
\usepackage{mathrsfs}
\usepackage{amssymb}
\usepackage{amsmath}
\usepackage{slashed}
\usepackage{hyperref}

\definecolor{lightblue}{RGB}{0,255,255}
\definecolor{pink}{RGB}{255,213,213}
\definecolor{blue}{RGB}{0,0,255}
\definecolor{red}{RGB}{255,0,0}
\definecolor{green}{RGB}{0,128,0}
\definecolor{purple}{RGB}{128,0,128}
\definecolor{orange}{RGB}{255,102,0}
\definecolor{darkshade}{RGB}{153,153,153}
\definecolor{lightshade}{RGB}{204,204,204}
\begin{document}

\newcommand{\vev}[1]{\left|{#1}\right|}
\newcommand{\del}{\partial}
\newcommand{\Tr}{{\rm Tr}}

\newcommand{\eqnn}[1]{\begin{equation} {#1}\end{equation}}

\def\bPhi{{\bf \Phi}}

\def\C{{\bf C}}
\def\P{{\bf P}}
\def\R{{\bf R}}
\def\Z{{\bf Z}}

\def\a{\alpha}
\def\b{\beta}
\def\g{\gamma}
\def\e{\epsilon}
\def\h{\widehat}
\def\th{\theta}
\def\k{\kappa}
\def\l{\lambda}
\def\L{\Lambda}
\def\m{\mu}
\def\n{\nu}
\def\r{\rho}
\def\s{\sigma}
\def\t{\tau}
\def\f{\phi}
\def\F{\Phi}
\def\w{\omega}
\def\v{\varphi}
\def\z{\zeta}

\def\G{\Gamma}
\def\Th{\Theta}
\def\W{\Omega}

\def\d{\partial}
\def\dbar{{\overline\partial}}
\def\inv{^{-1}}
\def\Tr{{\rm Tr}}
\def\hf{{1\over 2}}

\def\half{\textstyle{1\over2}}
\def\third{\textstyle{1\over3}}
\def\quart{\textstyle{1\over4}}
\def\eighth{\textstyle{1\over8}}

\def\cA{{\cal A}}
\def\cB{{\cal B}}
\def\cC{{\cal C}}
\def\cD{{\cal D}}
\def\cE{{\cal E}}
\def\cF{{\cal F}}
\def\cG{{\cal G}}
\def\cH{{\cal H}}
\def\cI{{\cal I}}
\def\cJ{{\cal J}}
\def\cK{{\cal K}}
\def\cL{{\cal L}}
\def\cM{{\cal M}}
\def\cN{{\cal N}}
\def\cO{{\cal O}}
\def\cP{{\cal P}}
\def\cQ{{\cal Q}}
\def\cR{{\cal R}}
\def\cS{{\cal S}}
\def\cT{{\cal T}}
\def\cU{{\cal U}}
\def\cV{{\cal V}}
\def\cW{{\cal W}}
\def\cX{{\cal X}}
\def\cY{{\cal Y}}
\def\cZ{{\cal Z}}

\def\({\bigl(}
\def\){\bigr)}
\def\<{\langle\,}
\def\>{\,\rangle}
\def\cal#1{\mathcal{#1}}
 \def\be{\begin{equation}}
\def\ee{\end{equation}}
\def\bea{\begin{eqnarray}}
\def\eea{\end{eqnarray}}
\def\bean{\begin{mathletters}\begin{eqnarray}}
\def\eean{\end{eqnarray}\end{mathletters}}

\newcommand{\tbox}[1]{\mbox{\tiny #1}}
\newcommand{\pit}{\mbox{\small $\frac{\pi}{2}$}}
\newcommand{\sfrac}[1]{\mbox{\small $\frac{1}{#1}$}}
\newcommand{\mbf}[1]{{\mathbf #1}}
\def\text{\tbox}

\newcommand{\mV}{{\mathsf{V}}}
\newcommand{\mL}{{\mathsf{L}}}
\newcommand{\mA}{{\mathsf{A}}}
\newcommand{\lB}{\lambda_{\tbox{B}}}  
\newcommand{\ofr}{{(\mbf{r})}}       
\def\ofkr{(k;\mbf{r})}			
\def\ofks{(k;\mbf{s})}			
\newcommand{\ofs}{{(\mbf{s})}}       
\def\xt{\mbf{x}^{\tbox T}}		

\def\ce{\tilde{C}_{\tbox E}}		
\def\cew{\tilde{C}_{\tbox E}(\omega)}		
\def\ceqmw{\tilde{C}^{\tbox{qm}}_{\tbox E}(\omega)}	
\def\cewqm{\tilde{C}^{\tbox{qm}}_{\tbox E}}	
\def\ceqm{C^{\tbox{qm}}_{\tbox E}}	
\def\cw{\tilde{C}(\omega)}		
\def\cfw{\tilde{C}_{\cal F}(\omega)}		

\def\tcl{\tau_{\tbox{cl}}}		
\def\tcol{\tau_{\tbox{col}}}		
\def\terg{t_{\tbox{erg}}}		
\def\tbl{\tau_{\tbox{bl}}}		
\def\theis{t_{\tbox{H}}}		

\def\area{\mathsf{A}_D}			
\def\ve{\nu_{\tbox{E}}}			
\def\vewna{\nu_E^{\tbox{WNA}}}		

\def\dxcqm{\delta x^{\tbox{qm}}_{\tbox c}}	

\newcommand{\rop}{\hat{\mbf{r}}}	
\newcommand{\pop}{\hat{\mbf{p}}}

\newcommand{\sint}{\oint \! d\mbf{s} \,} 
\def\gint{\oint_\Gamma \!\! d\mbf{s} \,} 
\newcommand{\lint}{\oint \! ds \,}	
\def\infint{\int_{-\infty}^{\infty} \!\!}	
\def\dn{\partial_n}				
\def\aswapb{a^*\!{\leftrightarrow}b}		
\def\eps{\varepsilon}				

\def\dhdxt{\partial {\cal H} / \partial x}
\def\dhdx{\pd{\cal H}{x}}
\def\dhdxnm{\left( \pd{\cal H}{x} \right)_{\!nm}}
\def\dhdxnmsq{\left| \left( \pd{\cal H}{x} \right)_{\!nm} \right| ^2}

\def\bcs{\stackrel{\tbox{BCs}}{\longrightarrow}}	

%
\def\wx{\omega_x}
\def\wy{\omega_y}
\newcommand{\ofro}{({\bf r_0})}
\def\Eb{E_{\rm blue,rms}}
\def\Er{E_{\rm red,rms}}
\def\Es2{E_{0,{\rm sat}}^2}
\def\sb{s_{\rm blue}}
\def\sr{s_{\rm red}}

\def\ie{{\it i.e.\ }}
\def\eg{{\it e.g.\ }}
\newcommand{\etal}{{\it et al.\ }}
\newcommand{\ibid}{{\it ibid.\ }}

\def\gap{\hspace{0.2in}}

\newcounter{eqletter}
\def\mathletters{%
\setcounter{eqletter}{0}%
\addtocounter{equation}{1}
\edef\curreqno{\arabic{equation}}
\edef\@currentlabel{\theequation}
\def\theequation{%
\addtocounter{eqletter}{1}\arabic{chapter}.\curreqno\alph{eqletter}%
}%
}
\def\endmathletters{\setcounter{equation}{\curreqno}}

\def\kf{k_{\text F}}
\newcommand{\br}{{\bf r}}
\newcommand{\TLR}{{\text{L,R}}}
\newcommand{\VSD}{V_{\text{SD}}}
\newcommand{\GT}{\Gamma_{\text{T}}}
\newcommand{\DEL}{\mbox{\boldmath $\nabla$}}
\def\lf{\lambda_{\text F}}
\def\st{\sigma_{\text T}}
\def\stlr{\sigma_{\text T}^{\text{L$\rightarrow$R}}}
\def\strl{\sigma_{\text T}^{\text{R$\rightarrow$L}}}
\def\aeff{a_{\text{eff}}}
\def\aaeff{A_{\text{eff}}}
\def\gat{G_{\text{atom}}}
\newcommand{\LB}{Landauer-B\"{u}ttiker}

\dsp

\title{D-branes, Supersymmetry Breaking, and Neutrinos}
\author{Jihye Seo}
\degreemonth{May}
\degreeyear{2010}
\degree{Doctor of Philosophy}
\field{Physics}
\department{Physics}
\advisor{Cumrun Vafa}

\maketitle
\copyrightpage

\begin{abstract}
This thesis studies meta- and exactly stable supersymmetry breaking mechanisms in heterotic and type IIB string theories and constructs an F-theory Grand Unified Theory model for neutrino physics in which neutrino mass is determined by the supersymmetry breaking mechanism.

Focussing attention on heterotic string theory compactified on a 4-torus, stability of non-supersymmetric  states is studied. A non-supersymmetric state with robust stability is constructed, and its exact stability is proven in a large region of moduli space of $T^4$ against all the possible decay mechanisms allowed by charge conservation. Using string-string duality, the results are interpreted in terms of Dirichlet-branes in type IIA string theory compactified on an orbifold limit of a K3 surface.

In type IIB string theory, metastable and exactly stable non-supersymmetric systems are constructed using D-branes and Calabi-Yau geometry. Branes and anti-branes wrap rigid and separate
2-spheres inside a non-compact Calabi-Yau three-fold: supersymmetry is spontaneously broken. These metastable vacua are analyzed in a holographic dual picture on a complex-deformed Calabi-Yau three-fold where 2-spheres have
been replaced by 3-spheres with flux through them.
By computing bosonic masses, we identify location and mode of instability. The moduli space of this complex-deformed Calabi-Yau three-fold is studied, and methods for studying the global phase structure of supersymmetric and non-supersymmetric flux vacua are proposed.
By turning on a varying Neveu-Schwarz flux inside the Calabi-Yau three-fold, we build meta- and {\it exactly stable} non-supersymmetric configurations with D-branes but with no anti-D-branes.

Finally, a scenario for Dirac neutrinos in an F-theory $SU(5)$ GUT model is proposed. Supersymmetry breaking leads to an F-term for Higgs field $H_{d}^{\dag }$ of order ${%
F_{H_{d}}}\sim \mu H_{u}\sim M_{\mbox{\tiny weak}}^{2}$ which induces a
Dirac mass of $m_{\nu }\sim M_{\mbox{\tiny weak}}^{2}/\Lambda _{%
\mbox{\tiny
UV}}$.  A mild normal hierarchy with masses $%
(m_{3},m_{2},m_{1})\sim 50 \times (1,\alpha _{\rm GUT}^{1/2},\alpha _{\rm GUT})$ meV and large mixing angles $\theta _{23} \sim \theta _{12} > \theta _{13} \sim \theta _{C}\sim \alpha
_{\rm GUT}^{1/2}\sim 0.2$ are predicted.

\end{abstract}

\newpage
\addcontentsline{toc}{section}{Table of Contents}
\tableofcontents
\listoffigures

\begin{citations}

\vspace{0.8in}
\ssp

\noindent Large portions of Chapter \ref{Ch:metaIIB} have appeared in the following papers:
\begin{quote}
``Geometrically Induced Metastability and Holography'', M. Aganagic, C. Beem, {J. Seo}, C. Vafa,
Nucl. Phys. B 789: 382-412 (2008), arXiv:hep-th/0610249;
\end{quote}
\begin{quote}
``Phase Structure of a D5/Anti-D5 System at Large N'', J. Heckman, {J. Seo}, C. Vafa,
{JHEP {\bf 07}} 073 (2007), arXiv:hep-th/0702077.
\end{quote}
Most of Chapter \ref{Ch:exactIIB} has appeared in the following paper:
\begin{quote}
``Extended Supersymmetric Moduli Space and a SUSY/Non-SUSY Duality'', M. Aganagic, C. Beem, {J. Seo}, C. Vafa
Nucl. Phys. B 822: 135-171 (2009), arXiv:0804.2489 [hep-th].
 \end{quote}
The following paper forms the primary content of Chapter \ref{Ch:DiracF}:
  \begin{quote}
``F-theory and Neutrinos: Kaluza-Klein Dilution of Flavor Hierarchy'', V. Bouchard, J. Heckman, {J. Seo}, C. Vafa,
JHEP {\bf 01}, 061 (2010), arXiv:0904.1419 [hep-ph].
 \end{quote}

\noindent Electronic preprints (shown in {\tt typewriter font}) are available
on the Internet at the following URL:
\begin{quote}
	{\tt http://arXiv.org}
\end{quote}
\end{citations}

\begin{acknowledgments}

Five years ago, after attending a Physics Department Colloquium by Cumrun Vafa on crystal melting and topological string theory, I decided that I would do something similar: uncover surprising connections between various subjects in science. My passion still lies there, and I have never looked back. His contagious enthusiasm  has such a magical effect on me that discussing physics with him truly constitutes the most exciting and meaningful part of my life. His influence has significantly affected the way I picture physical situations and approach problems. I hope that some of his qualities have rubbed off on me and will remain with me.  At the same time, I thank him for allowing me space and encouraging me to grow in my own way.

I was very fortunate to have worked on various projects with  Mina Aganagic. Whenever I got stuck, she guided me like a compass. She pierced through apparent complications and found a way to compute things out. Our projects ran at full speed thanks to her. Collaborations with Christopher Beem, Vincent Bouchard, and Jonathan Heckman were thrilling adventures which taught me physics, math, and discipline.

The Harvard Theory Group is full of great minds and personalities. Andrew Strominger and Frederik Denef actively devoted much time for students.
I thank Andy for his friendship and all the wonderful times we had together. I have also had the great privilege of working on my first project with him and his students, which somewhat shaped me as a researcher.
I thank Chris Beasly,  Sergei Gukov, Daniel Jafferis, Subhaneil Lahiri, Joe Marsano, Andy Neitzke, Suvrat Raju, and  Xi Yin for giving me thoughtful answers to any questions I posed. I also value my comradeship with other students in the Theory Group: I will fondly miss Dionysis Anninos, Clay Cordova, Tom Hartman, Josh Lapan, Megha Padi, and David Simmons-Duffin, and I cherish the time spent with Wei Li, my soulmate and sister.

Emiliano Imeroni, Hanjun Kim, Ilya Nikokoshev, Kyriakos Papadodimas, and Ram Sriharsha introduced me to geometric transition, singularity resolution, a K3 surface, mirror symmetry, and fibration, respectively. I am deeply grateful to them for their patience and faith in me.
 Lubos Motl generously spent many hours teaching and encouraging me.  Melissa Franklin was there when I was lost and struggling. I am deeply grateful to Sheila Ferguson for her support and comfort.

I thank  Masahiro Morii for explaining particle experiments, and for having me in BaBar group meetings and New England Particle Physics Students Retreats during the early days of the graduate school. I derived great benefit from the advice and suggestions on my dissertation given by Cumrun, Frederik, and Masahiro. I am indebted to Mboyo Esole, Momin Malik, Chang-Soon Park, and Jon Tyson for their thorough reading of and helpful suggestions on my thesis.

Tudor Dimofte, Mboyo Esole, Rhiannon Gwyn, Ian-Woo Kim, Chang-Soon Park, Profesor Soo-Jong Rey, Sakura Schafer-Nameki, Minho Son, and Jaewon Song kindly shared with me their experience and knowledge and offered insight and courage as I prepare a transition from a student to an independent researcher.

I thank my family and friends for being there. I thank Baran Han and Yi-Chia Lin for loving me as who I am, Eunju Lee for warm caring, Suzanne Renna for sharing her wisdom, and Catherine Ulissey for teaching me that what matters is the journey itself rather than arriving at the destination.

My research was supported in part by NSF grants PHY-0244821 and DMS-0244464, and by the Korea Foundation for Advanced Studies.
\end{acknowledgments}

\dedication

\begin{quote}
\hsp
\em
\raggedleft

Dedicated to Sheila Ferguson, \\
 Lubos Motl, \\
and  Melissa Franklin. \\

\bigskip
\bigskip

I did not give up, because you were there for me.

\end{quote}

\newpage

\startarabicpagination

\chapter{Introduction}

This thesis examines supersymmetry breaking mechanisms and constructs a neutrino physics model in string theory using D-branes, with the intent of connecting string theory
to real world physics. In the heterotic and type IIB string theories, we study various non-supersymmetric configurations which are meta- or exactly stable. We also presents an F-theoretic minimal Grand Unified Theory model with Dirac neutrinos, whose mass scale is determined by a supersymmetry breaking mechanism.
The greatest strength of this thesis is that we use {\it minimal} ingredients and utilize
{\it geometry} maximally, making the whole process economical and natural.

Section \ref{sec:StringEarth} introduces the basic concepts\footnote{Interested readers are encouraged to consult the following books written for general audiences in the subjects of quantum mechanics \cite{Alice}, extra dimensions \cite{KakuHyperspace}, string theory \cite{ElegantUniverse}, supersymmetry \cite{KaneSUSY}, and physics beyond the Standard Model \cite{Randall}.} of string theory, supersymmetry breaking, and neutrinos. It aims to deliver the main idea of this thesis in non-technical language.
Section \ref{StringReview} supplies a springboard to follow the main arguments of the thesis: it introduces  string theories, string dualities, D-branes, and supersymmetry. (For a more complete introduction, the reader may consult  \cite{pol,GSW,BBS,KakuConformal,WessBagger,Nakahara,Vafa97Lecture,GreeneCY3,AspinwallK3,LykkenSUSYintro,TopReview}.)
 Section \ref{TowardStringPheno} is a concise introduction to various concepts necessary for constructing realistic particle theory models in string theory, focussing on supersymmetry breaking.

\section{String theory down to Earth \label{sec:StringEarth}}

The past century witnessed two great breakthroughs in our physical understanding of
Nature in two directions. The first is general relativity, which
explains gravity as an aspect of the curvature of spacetime. Its most popular application, the Global Positioning System (GPS) sits in our cars: the GPS would have been
impossible without such a precise understanding of gravity around the Earth.
The second breakthrough is the Standard Model of particle physics. It is a quantum field theory which explains all other fundamental forces, namely
electromagnetic, weak, and strong interactions.

Many real world applications fall in the domain of one but not both of these two
theories. It is either gravitational or quantum, not both. Roughly speaking,
quantum nature governs the small world, such as nuclei or carbon nanotubes,
while gravity dominates the big world, things at the scale of falling apples, orbiting
planets, and rotating galaxies.

\begin{figure}[!h]
\centerline{\includegraphics[width=.90\textwidth]{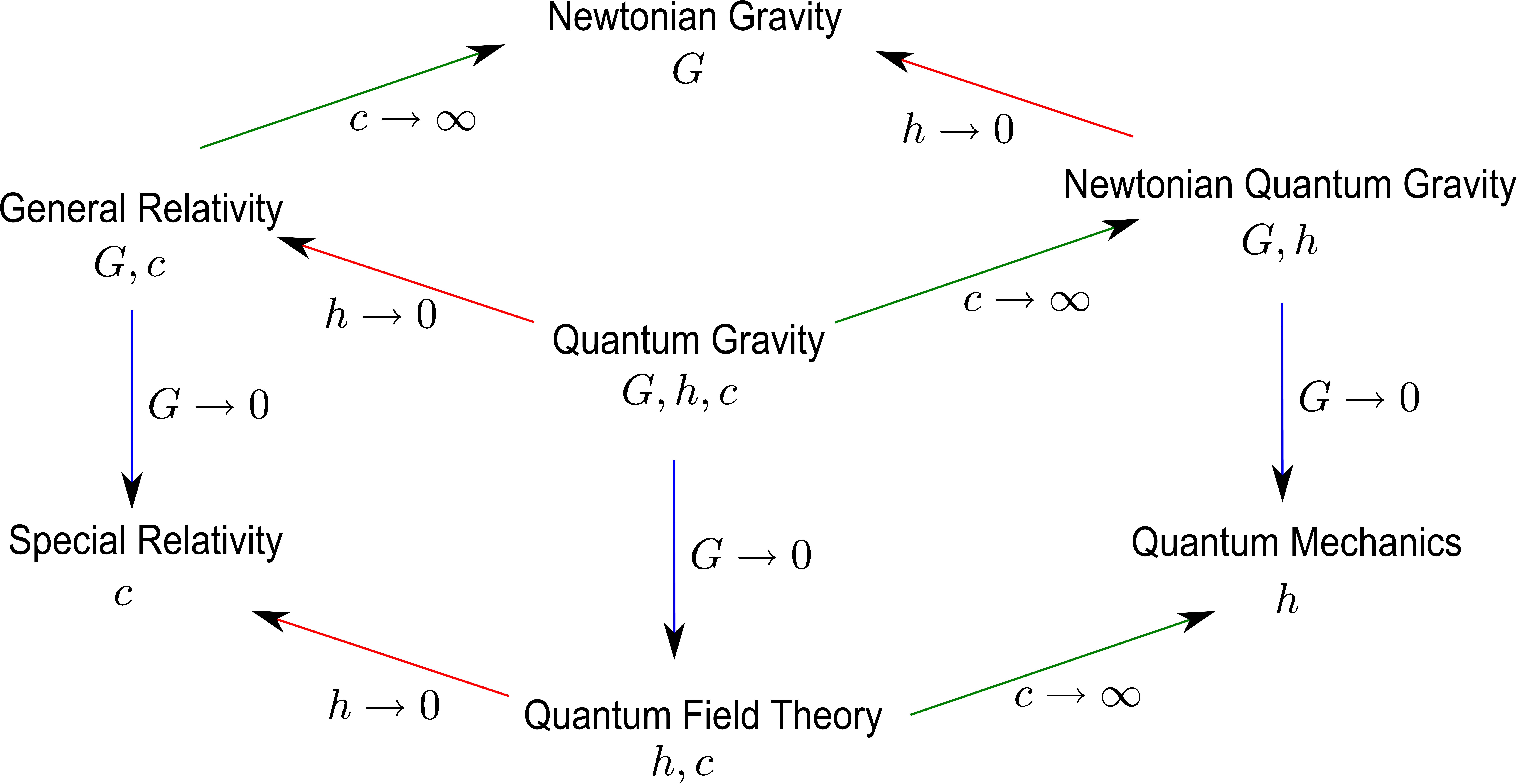}}
\caption[Quantum gravity merges quantum field theory and general relativity. As we take a limit of $G\rightarrow 0$, $h\rightarrow 0$, or $c\rightarrow \infty$, we arrive at less complete theories.]{Quantum gravity merges quantum field theory and general relativity. As we take limits of \textcolor{blue}{$G\rightarrow 0$}, \textcolor{red}{$h\rightarrow 0$}, or \textcolor{green}{$c\rightarrow \infty$}, we arrive at theories that are less complete with limited validity. A similar figure appeared in \cite{Nam}.}
\label{fig:QuantumGravity}
\end{figure}

However, there are instances where both
gravitational and quantum properties are important. For example, at the
horizon (boundary) of black holes, particles deal with their own quantum
business, such as the creation of particle-antiparticle pairs. At the same time,
black holes are infamous for having strong gravity. This is an example of where
more complete theory is needed to merge gravity and quantum physics together. It is
aptly called ``quantum gravity.'' As depicted in Figure \ref{fig:QuantumGravity}, quantum gravity properly deals with quantum uncertainty $h$, the strength of gravity $G$, and the finiteness and constancy of the speed of light $c$. Various established theories arise as limits in which some combination of these constants are taken to be negligible, as shown in Figure \ref{fig:QuantumGravity}. Quantum field theory assumes no gravity $G\rightarrow 0$, while general relativity assumes no quantum effect $h\rightarrow 0$.

String theory is a candidate for the ultimate theory of quantum gravity.
String theory can accommodate all the physical interactions, and having
gravity is an inevitable consequence of string theory\footnote{Whether string
theory accommodates the actual quantum field theory of our world is another
question, which we will discuss soon.}. Although the Universe appears to have
three spatial dimensions and one temporal
dimension, string theory predicts the existence of ten or more dimensions.
Although this first appears paradoxical, these extra dimensions can be seen as tools. Much as gravity is explained as the curvature of spacetime,
one hopes that the particle physics may be explained in terms of the geometry of these extra dimensions (as in Chapter \ref{Ch:DiracF}, for example).

Furthermore, going to higher dimensions may resolve singularities. For example,
the roller
coaster has a smooth track in 3D, but its shadow on the ground, a 2D projection of 3D object,
may show cusps, geometric singularities. Going to higher dimensions
can resolve these artificial singularities in mathematics \cite{HironakaInterview}, and the same
thing could happen in physics as well. Note that physics is not mathematics, but often
can be efficiently described in terms of geometry; historically, geometric thinking
actually accelerated the development of physics. For example, the way in which heavy objects curve the
spacetime fabric in general relativity is best phrased in terms of Riemann
geometry.

\begin{figure}[!h]
\centerline{\includegraphics[width=.70\textwidth]{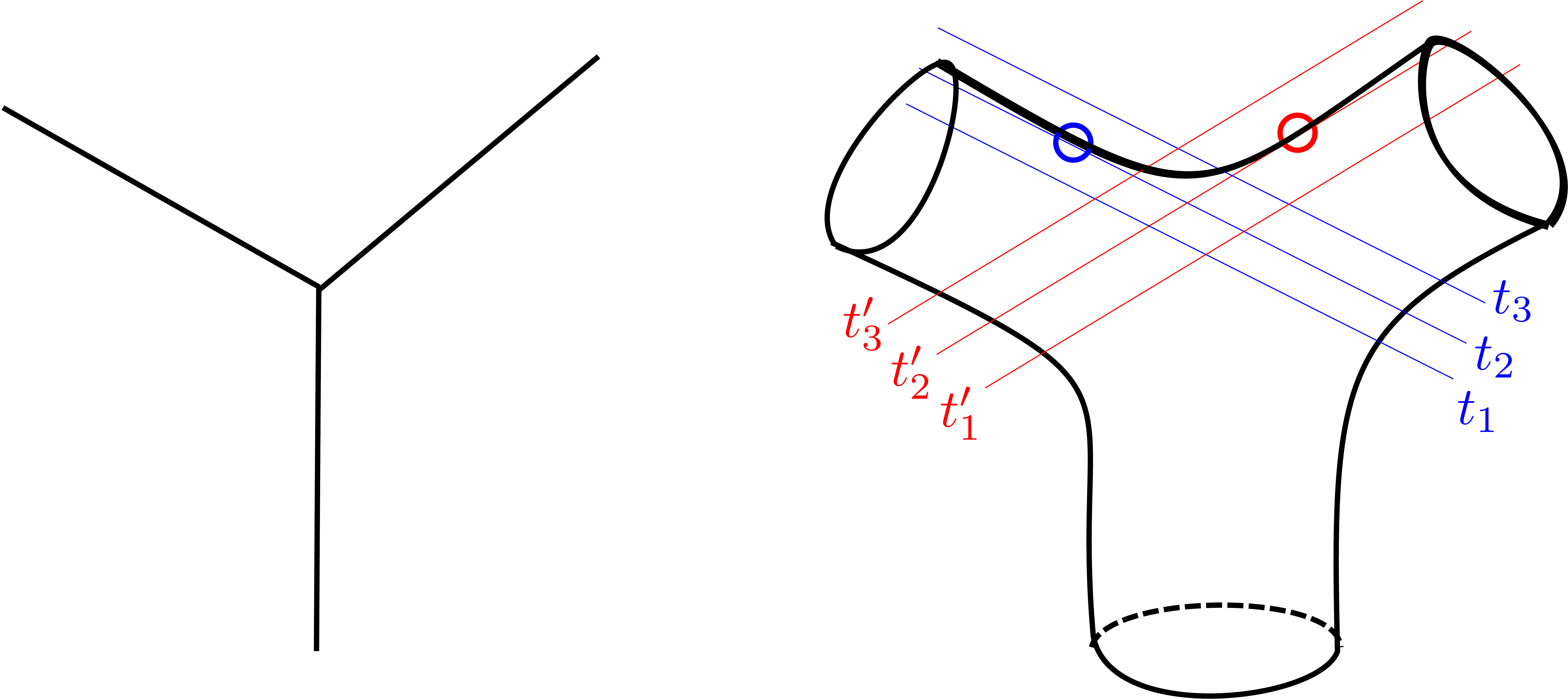}}
\caption[A Feynman diagram and a string pants diagram for a three-point interaction vertex: string theory may resolve
singularities of field theory]{A Feynman diagram of an interaction among three fields on the left, and a ``pants diagram'' on the right, which depicts a process where three closed string states interact. Different Lorentz observers with time coordinates \textcolor{blue}{$t$} and \textcolor{red}{$t^{\prime}$}, will have different interaction points (marked with circles).}
\label{fig:smooth}
\end{figure}

In quantum field theories, we come across singular behaviors. Scattering amplitudes for particle interactions are systematically organized in terms of
Feynman diagrams. At the points of interaction, sharp singular behavior can occur in the scattering amplitude. In string theory we replace Feynman
diagrams with pants diagrams, where each point particle is now replaced by a string, as in Figure \ref{fig:smooth}. In string theory, there is not one single point of interaction: instead, it is smeared out over the smooth surface. Depending on how we choose our time coordinate, or depending on how fast we travel relative to a given reference frame, the interaction will appear to happen at {\it different} points. Nothing singles out a point in a pants diagram of string interaction. String
theory fattens the {\it thin} Feynman diagrams into {\it thick} pants diagrams, removing the singularities at interaction
vertices. Notions of ``here'' and ``now'' are {\it spread out} in string theory.

There is a price we have to pay to accept string theory as an avenue toward the Theory
of Everything. For its own consistency, string theory, also called
superstring theory, demands supersymmetry, while our world is not
supersymmetric. However, there is no contradiction in describing the world by a theory whose general laws are symmetric, but whose solutions (vacua) are not. Also note that having supersymmetry in theory is very attractive. Supersymmetry unifies the coupling constants of gauge interactions - strong and weak nuclear forces and electromagnetic force. Supersymmetry is also helpful in solving the Higgs mass hierarchy problem, and it may provide dark matter candidates. Therefore, let us take the stance of starting from a supersymmetric theory and breaking supersymmetry to find a realistic non-supersymmetric vacuum in which we live. When symmetry is broken in a solution (vacuum) of a symmetric theory, we call that the symmetry is broken {\it spontaneously} in that vacuum.

 Therefore, we still deal with supersymmetric string
theory, but we will look for non-supersymmetric vacua where supersymmetry is spontaneously broken. More specifically, we want our vacua to have lifetimes long enough to allow our history of the Universe to fit
in. Therefore, we want to find non-supersymmetric vacua that are exactly
stable (with infinite lifetimes) or metastable (with finite lifetimes).

Breaking supersymmetry is only the tip of the iceberg of the string theory
to-do list. We need to find a string theory vacuum that explicitly displays the particle properties of our Universe. Allowing certain interactions is not enough: we also
want to be able to explain why each particle has certain properties, such as
mass. This is a very active field of research in
string theory, called string phenomenology, because it aims to explain
particle or astrophysical phenomenology in string theory framework.
Supersymmetry breaking in string theory and string phenomenology are
intertwined problems, because the way we break supersymmetry affects the
kinds of particle physics we get.

Among all the particles in the Standard Model of particle physics, neutrinos
are particularly interesting. In the Standard Cosmology, neutrinos are also believed to be the most abundant particles in the Universe after photons. Neutrinos are much lighter
than the other massive particles in the Standard Model, and they behave in very strange ways,
such as changing their {flavor}\footnote{There are three charged leptons: electron, lepton, and tauon. Neutrino flavor refers to the corresponding charged lepton with which neutrino interacts. For example, neutrinos with electronic flavor interact with an electron only.} with time, as is quantitatively captured in a neutrino mixing matrix. Neutrinos were thought to be massless until the 1970s, when flavor oscillations were observed in neutrinos arriving from the Sun. The only way to explain neutrino oscillation is to ascribe different
masses to neutrinos, one for each flavor composition.

Constructing a string phenomenology model for neutrinos is a challenging but
rewarding task. The principle difficulty is that relatively little experimental data is available: their masses have a large range of uncertainty, and it is unknown whether neutrinos are anti-particles of themselves (Majorana particles). However, they provide a window into
physics beyond the Standard Model of particle physics, allowing us opportunities to make predictions for
coming experiments.

\subsection{Supersymmetry}

Supersymmetry is the symmetry between fermions and bosons. Each elementary particle has a quantum number called spin, which has the same unit as angular momentum. Bosons have integer spin, and they like to clump together into the same quantum state. Fermions have half-integer spin, and they are mutually exclusive. For example, photons are bosons with spin one, and they stay together in the same state to form a strong coherent light in a laser. On the other hand, electrons are fermions with spin of a half, and (as explained by Pauli's exclusion principle) they cannot stay in the same quantum state: so they inhabit successive orbital shells in an atom, rather than all occupying the inner-most shell.

Supersymmetry pairs bosons and fermions into super-multiplets. They are superpartners of each other, and except for spin they share all properties, including mass and charge. If the superpartners have different masses, this difference (mass splitting) denotes the amount of supersymmetry breaking in a vacuum. If we lived in a supersymmetric vacuum, we would see a massless photino, a superpartner to massless photon, and a light selectron, a bosonic superpartner to a light electron. Since none of these have been observed, clearly we live in a non-supersymmetric vacuum and superpartners are too heavy to be observed at the current energy scale of experiments.

\begin{figure}[!h]
\centerline{\includegraphics[width=.50\textwidth]{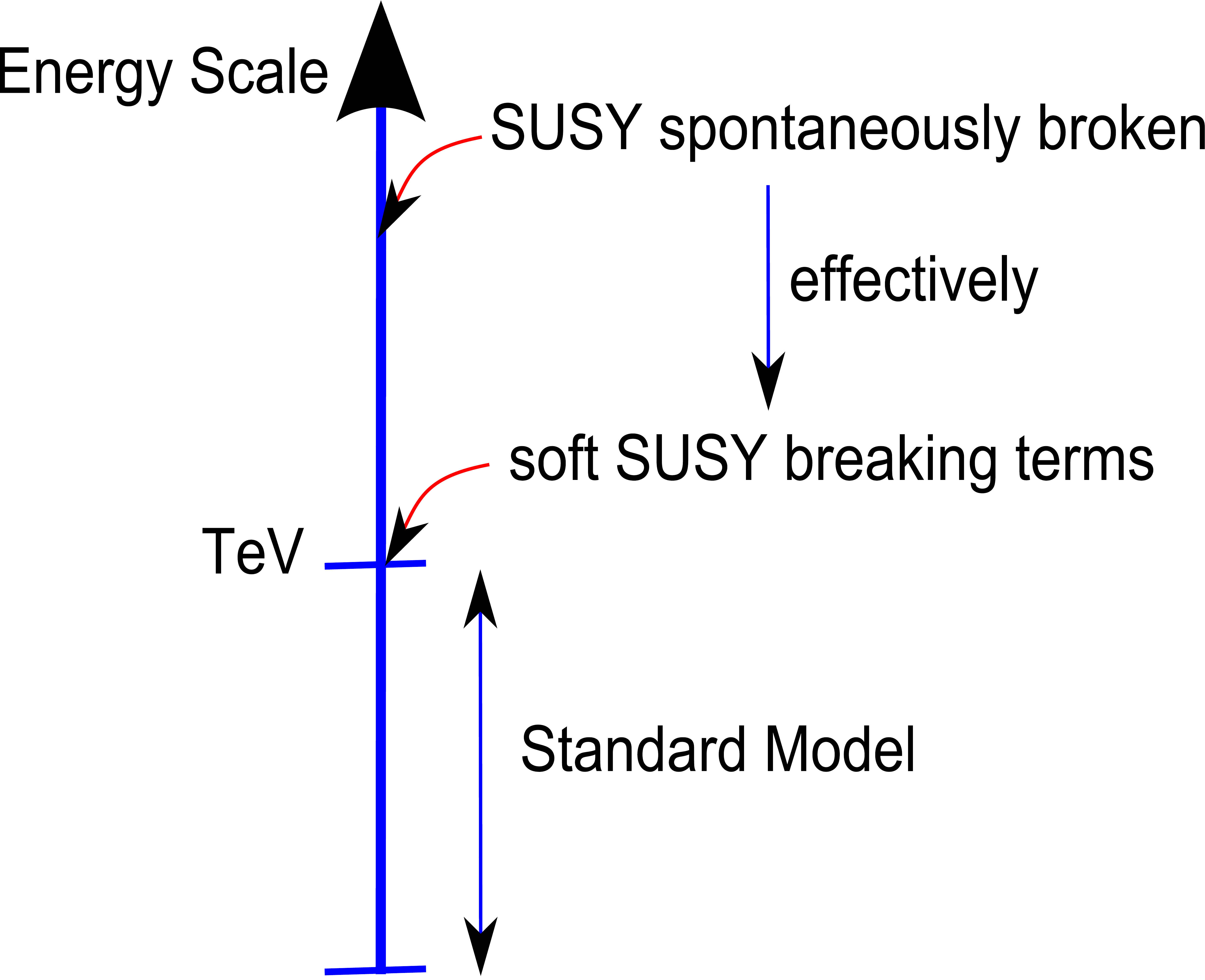}}
\caption{Spontaneous breaking of supersymmetry in a UV-complete theory gives soft SUSY breaking terms at the TeV scale for a Lagrangian of a low energy effective theory.}
\label{fig:CS}
\end{figure}

Currently, the most popular supersymmetry breaking scenario posits that we live in a supersymmetric, visible sector, and supersymmetry breaking happens spontaneously in hidden sectors at high energy, for example in string theory. There are messenger fields that mediate between visible and hidden sectors. The messenger's supersymmetry is broken by interactions with fields in hidden sectors. The interaction terms between visible and messenger sectors will provide explicit soft supersymmetry breaking terms in the effective theory in the visible sector. As drawn in Figure \ref{fig:CS}, spontaneous breaking of supersymmetry at high energy provides an explicit supersymmetry breaking terms in the Lagrangian of the corresponding low energy effective theory. This makes the low energy effective theory appear as a non-supersymmetric theory with added explicit supersymmetry breaking terms with soft UV behavior. Chapters \ref{Ch:nonBPShet}, \ref{Ch:metaIIB}, and \ref{Ch:exactIIB} explore various ways of breaking supersymmetry in string theory.

\subsection{Closed and open strings, and D-branes}

In string theory, we have open and closed strings: open strings are
like a path with 2 ends, and closed strings are like loops with no free ends. They have a topology of a
line segment and a circle, respectively. Open strings have two endpoints, which must reside on multi-dimensional objects called
D-branes, whereas closed strings can float around anywhere they like.
To explain these concepts by analogy: imagine we fly airplanes.
We could circle around in the sky, making a closed loop just like a
closed string. If we fly from one airport to the other, then our flight path
is an open string. The airport corresponds to D-branes where the flight path
has to end or start: we cannot land anywhere we like, we can only land where this is an airport.

 Dirichlet (D)-brane is a set of points where open strings can
end. D-branes can come in many dimensions.  D-branes and strings have tension energy proportional to their {\it volume}\footnote{Here, volume is an umbrella term for length (1D), area (2D), and volume (3D) and similar concepts in higher dimensions.}.
Due to the brane tension energy, a D-brane tries to minimize its volume, like a rubber band wrapped on the stem of a wine glass.
D-branes with opposite charges are called
anti-D-branes. D-branes and anti-D-branes are attracted to each other, and they annihilate each other when brought together, like matter and anti-matter.

\subsection{Intersection and wrapping of D-brane stacks}

We can stack multiple D-branes at a same location.
If we have two stacks of D-branes, an open string can stretch between them with one endpoint living on each stack. The mass (tension energy) of the open string is proportional to its length, i.e. or the separation between the stacks.
As we bring the brane stacks closer, the string get shorter and the string mode become lighter. When we intersect the D-branes, near the intersection locus, we have {\it matter fields} which are light, easily excitable, string modes.
When many stacks intersect at a point, we will have many fields, which are so close to one another that they can interact with one another, forming Yukawa interaction terms.

For our neutrino model in Chapter \ref{Ch:DiracF}, matter fields arise along the curve where two
stacks of D-branes intersect. Interaction terms between matter fields (Yukawa
couplings) arise where these curves or stacks of D-branes intersect at a point.

D-branes and anti-D-branes break different halves of supersymmetry, together they break all supersymmetry. If put together in a flat geometry, they also attract and mutually annihilate. On open strings stretched between these two stacks of branes, dangerous tachyonic modes will be excited, which correspond to fluctuations which breaks down the vacuum.  However, if two stacks are wrapped on different cycles that are far apart, with a barrier between them, then the system will become stable. A topological barrier gives exact stability we explore in Chapters \ref{Ch:nonBPShet} and \ref{Ch:exactIIB}, while a geometric barrier gives metastability we study in Chapters \ref{Ch:metaIIB} and \ref{Ch:exactIIB}.

\section{String dualities and D-branes \label{StringReview}}
\begin{figure}[!h]
\centerline{\includegraphics[width=.65\textwidth]{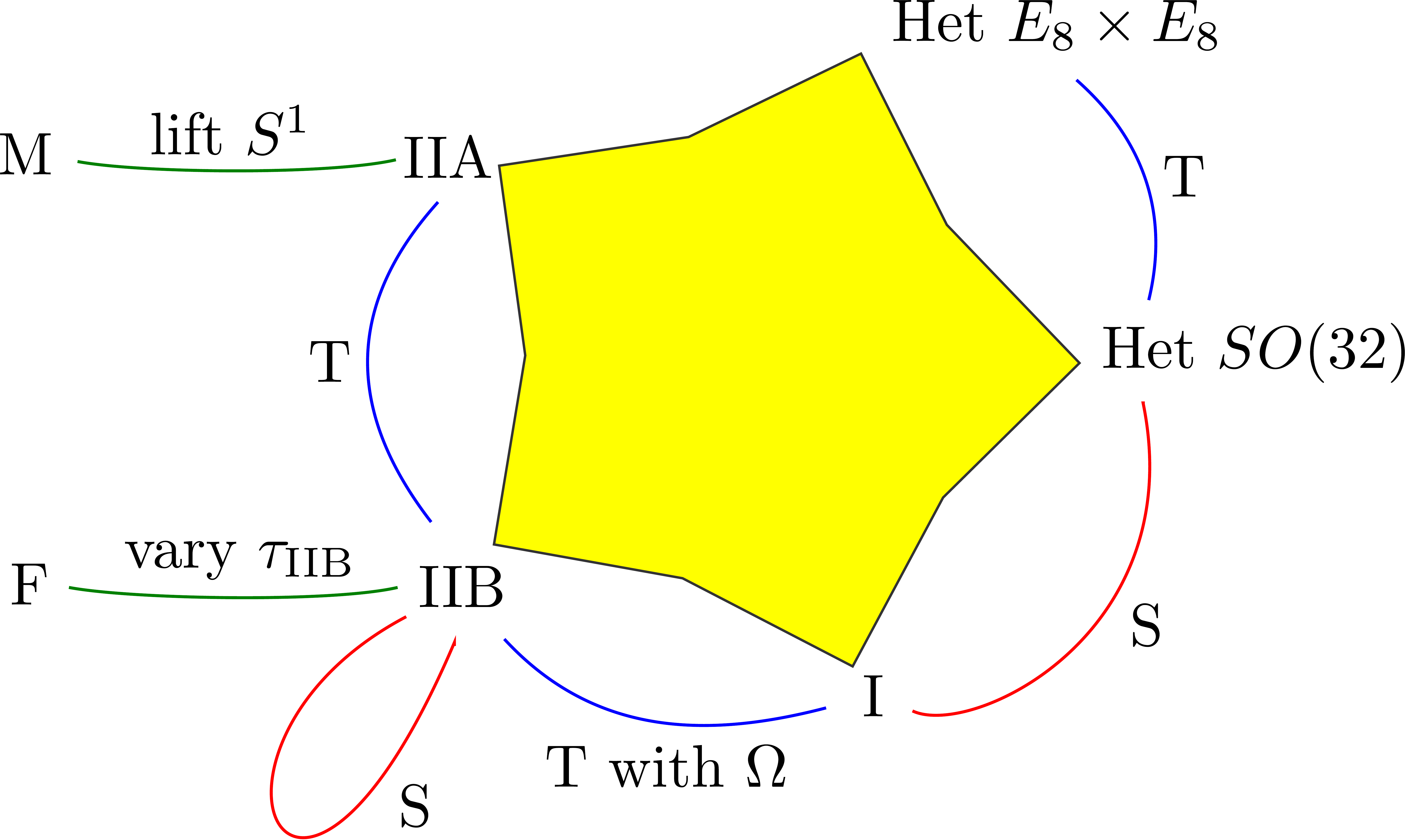}}
\caption[The map of dualities between string theories]{The map of dualities between string theories. S and T in the figure stand for \textcolor{red}{strong-weak} and \textcolor{blue}{toroidal} dualities.}
\label{fig:duality}
\end{figure}

There are many string theories, which are related to each other by dualities. See the Figure \ref{fig:duality} of string theory duality map. Start from type IIA, IIB, and I string theories, and then we discuss dualities, through which we arrive at other known string theories.

Type IIA, IIB, and I string theories are 10 dimensional The numbers I and II refer to the amount of supercharges seen in 10D\footnote{In lower dimensions, each supercharge spinor is smaller, and a 4D observer will see four times more supercharges than a 10D observer.}. They have open strings which end on D-branes. Strings and branes can carry Neveu-Schwarz (NS) and Ramond (RR) charges. These charges are conserved: when geometry removes D-branes, RR-flux has to appear to replace their effect. More details are discussed in Chapters \ref{Ch:metaIIB} and  \ref{Ch:exactIIB}. A varying NS-field appears in Chapter \ref{Ch:exactIIB} and provides exactly stable supersymmetry breaking vacua.

Type IIA and IIB string theories have left-and right-movers with opposite and equal chiralities.
Type I string theory is obtained by identifying the left- and right-moving modes of type IIB string theory, and therefore it has less supersymmetry and no orientation. Type IIA and IIB string theories are related by toroidal (T) duality. T-duality interchanges the big and small radii on a torus, and changes the dimensions of D-branes by 1. Type IIB string theory is T-dual to type I string theory with the action of a worldsheet  parity orientifold operator $\Omega$, which removes orientations of type IIB string theory which are already absent in type I string theory.

Strong-weak (S) duality interchanges strong and weak coupling constants. Inverting the coupling constant of type I string theory, one obtains $SO(32)$ heterotic string theory \cite{PolchinskiWittenHetIduality,WittenSDualityDimensions}. Heterotic string theory has no open strings or D-branes. Instead, it has closed strings with left- and right-movers, which are bosonic (26d) and supersymmetric (10d) respectively. The extra 16 dimensions of a bosonic left-mover should have the structure of $E_8 \times E_8$ or $SO(32)$ for the consistency. They correspond to $E_8 \times E_8$ and $SO(32)$ heterotic string theories, which are T-dual to each other \cite{GinspargHetTorus}. Heterotic string theory will appear in Chapter \ref{Ch:nonBPShet}, which studies stability of its non-supersymmetric states.

Type IIB string theory is self-dual under S-duality. This strange behavior of the coupling constant of type IIB string theory leads to F-theory, a non-perturbative version of type IIB string theory in which the coupling constant $\tau_{\rm IIB}$ is allowed to vary, taking a value in a 2-torus. F-theory is discussed in Chapter \ref{Ch:DiracF}, which constructs a model for neutrino physics.

M-theory is an 11-dimensional supergravity with no strings or branes. If compactified on a tiny circle, it becomes type IIA string theory.

There are other dualities not shown here. For example, heterotic string on $T^4$ is S-dual to type IIA string theory on K3 surface \cite{WittenSDualityDimensions,AspinwallK3}, a fact which will be used in Chapter \ref{Ch:nonBPShet}. Note that the amount of supersymmetry between these two theories matches because K3 surface kills half of the supersymmetry present, due to its holonomy.

\section{Toward a realistic string phenomenology \label{TowardStringPheno}}

This section is a concise introduction to various concepts necessary for constructing realistic particle theory models in string theory, focussing on supersymmetry breaking. More details discussions can be found in  \cite{Dine,MohapatraGUT,MohapatraNeutrino,LutyTASI}.

If supersymmetry is broken spontaneously in the low energy theory at the leading order (at the tree level with no quantum corrections), the predicted masses of superpartners \cite{FGPmass} are in contradiction with observations. Therefore the supersymmetry breaking needs to be explicit at low energy, with explicit supersymmetry breaking terms in the Lagrangian whose origin is quantum. However, in order to provide the Higgs mass with a soft UV behavior, supersymmetry breaking has to be spontaneous in a UV-complete theory, such as string theory. (See Figure \ref{fig:CS}.) In order to have a realistic Grand Unified Theory with chiral fermions, there is a further constraint for the theory to have only ${\cal N}=1$ supersymmetry at low energy
\cite{WittenN1dynamicbreaking}.%

The Minimally Supersymmetric Standard Model (MSSM) is a ${\cal N}=1$ version of the Standard Model. The MSSM lives in a visible sector, and supersymmetry breaks spontaneously in a hidden sector, usually at a UV complete theory such as string theory. Supersymmetry breaking at a hidden sector is mediated through messengers to MSSM at the visible sector. The interaction terms between the MSSM and the messengers provide the explicit supersymmetry breaking terms in the low energy Lagrangian. See \cite{LutyTASI} for a review of  various mediation mechanisms.

The Higgsino attains a mass in the MSSM through an operator of the form
\begin{equation}
{\cal L} = \int d^2 \theta \mu H_u H_d + h.c. . \label{muterm}
\end{equation}
A mass $\mu$ on the order of $100$ GeV roughly explains the weak scale. Explaining why $\mu$ takes this small value, which is near the soft SUSY breaking parameter, is the so-called ``$\mu$ problem.''

The Giudice-Masiero mechanism \cite{GiudiceMasiero} solves this problem by imposing a $U(1)$ Peccei-Quinn (PQ) symmetry and introducing messenger fields which mediate supersymmetry breaking. One starts by assigning PQ charges on Higgs fields $H_u$ and $H_d$ so that there will be no term of the form \eqref{muterm} in the leading order. The messenger field $X$ has the F-term vacuum expectation value  $\langle X \rangle=x + \theta^2 F_X $ with $ F_X  \ne 0$, providing a term like \eqref{muterm} in the subleading order. Therefore, PQ symmetry and messenger field $X$ together suppress the Higgsino mass. They  suppress neutrino masses in Chapter \ref{Ch:DiracF}, in agreement with the experimental results.

\section{Organization of the thesis}

Chapter \ref{Ch:nonBPShet} studies non-BPS objects in heterotic string theory and their stability region. In Chapters \ref{Ch:metaIIB} and \ref{Ch:exactIIB} discuss supersymmetry breaking mechanisms which maintain meta- or exact stability in type IIB string theory on a non-compact Calabi-Yau three-fold. Chapter \ref{Ch:DiracF} discusses a Grand Unified Theory model in F-theory, where the supersymmetry breaking mechanism provides the neutrino mass scale.

\chapter{Stability of non-BPS states in heterotic string theory \label{Ch:nonBPShet}}

Exactly stable non-BPS states have been studied in various string theories, and they may help us to describe non-supersymmetric field theories and to construct non-supersymmetric string compactification \cite{SenNonBPSReview,MukhiDDbarRepulsion}.
A non-BPS D0-brane is stable in type I string theory \cite{SenD0I,SenD0Iinteractions,WittenDK}, which is realized as a stable non-BPS pair of a D1-brane and an anti-D1-brane in type IIA string theory \cite{SenIspinorIIAD}. This stability holds in a particular region of moduli space. At the boundary of the stability region, tachyons become massless, the force between non-BPS objects vanishes, and there is exact degeneracy in the Bose-Fermi spectrum \cite{SenProofSpacetimeSusyTachyon,BoseFermiNoForce}. Non-BPS states and their stability against certain decay channels have been discussed for type IIA string theory compactified over a K3 surface \cite{SenK3orbifoldD} and a Calabi-Yau three-fold \cite{StefanskiCY3}.
 There are also stable non-BPS brane-antibrane constructions in type IIB string theory using D4-branes and anti-D4-branes hung between NS5-branes \cite{MukhiTong,MukhiDDbarRepulsion}.
 See these review papers \cite{SenNonBPSReview,SchwarzTASInonBPS,GaberdielLecture} on stable non-BPS string states.

We study stability of non-BPS states in heterotic string theory compactified on $T^4$ and map them to the dual type IIB theory on K3 surface as in \cite{BG,GaberdielLecture}. We systematically exhaust all the possible decay channels allowed by charge conservation, and then we find that a certain spinor representation has a large stability region, which contains those of other less stable non-BPS states as well.
We interpret these non-BPS and BPS heterotic string states in terms of D-branes wrapped over orbifold limit of K3 surface in type IIA string theory.

We introduce a set of transformation matrices in heterotic string side, which is equivalent to taking even number of T-dualities on $T^4$.  These $16 \times 16$ matrices form a subgroup of isometry group of compactified 16 dimensional momentum vector of the left-mover of a heterotic string state. The momentum is a conserved charge and limits possible decay modes. With
these new tools, one can study possible decay channels of given non-BPS states in a
systematic way.

Stability region of non-BPS states in heterotic string theory turns out to be large. Every other corner of moduli space allows stable non-BPS
states, and these corners are connected into one huge region of the moduli
space where a non-BPS is stable against all the possible decays allowed by charge conservation. This may be a fertile path for studying
non-supersymmetric field theory and supersymmetry-breaking hidden sectors for
realistic model building in string theory.

The organization of the rest of the chapter is as follows. Section \ref{HetIIAduality} reviews heterotic string theory and the duality chain
between heterotic string theory and type IIA string theory. Section \ref{sec:HeteroticOrganizer} introduces a new tool to keep track of conserved charges in heterotic
string and analyze various non-BPS and BPS states in
heterotic string using this tool. The stability
region of non-BPS states is discussed in section \ref{HetRegion}

\section{Heterotic string theory on $T^4$ and string-string duality\label{HetIIAduality}}

The section reviews heterotic string theory on $T^4$ and type IIA string theory on an
orbifold limit of a K3 surface and discusses the duality chain between them.

\subsection{Heterotic string theory on $T^4$}

Heterotic string theory has a fermionic 10-dimensional right-mover and a bosonic 26-dimensional left-mover. The left-mover has extra 16 dimensions, whose momentum is quantized as a 16-dimensional vector
${P}_{L} = V_K \in \Gamma^{16} $. A 16-dimensional even self-dual lattice $\Gamma^{16}$ is given as
\begin{equation}
\Gamma ^{16}=\left \{  (n_1, \cdots , n_{16}), \left(n_1+\frac{1}{2} , \cdots , n_{16} +\frac{1}{2} \right) \bigg| \sum_i n_i \in 2 \mathbb{Z} \right \} .
\end{equation}

Heterotic string compactified on a 4-torus \cite{HetTorus,NarainMassLevel} has Kaluza-Klein and winding excitations $n^i , w_i \in \mathbb{Z}$ in each direction $x_i$ ($i=1,2,3,4$) of a 4-torus $T^4$. We also choose four Wilson lines $A^i$ (superscripts on the right-hand-side denoting repetition of components):
\begin{eqnarray}
A^{1} &=&\left( \left( \frac{1}{2}\right) ^{8},0^{8}\right) , \qquad
A^{2}=\left( \left( \left( \frac{1}{2}\right) ^{4},0^{4} \right)^2   \right) \\
A^{3} &=&\left( \left( \left( \frac{1}{2}\right) ^{2},0^{2} \right)^4   \right), \qquad
A^{4} =\left( \left( \frac{1}{2},0\right) ^{8}\right),
\end{eqnarray}
so that this heterotic string theory is dual to type IIA string theory compactified on an orbifold limit of a K3 surface.
Now the left- and right-moving momenta in internal directions are given by
\begin{equation}
\mathbf{P}_{L}=(P_{L},p_{L})    , \qquad \mathbf{P}_{R} = p_{R} .
\end{equation}%
The momentum of the left-mover on 16-dimensional lattice \begin{equation} P_L =  V_{K}+ A_{K}^{i}w_{i}   \end{equation} is shifted by Wilson lines and winding, where summation over $i=1,2,3,4$ is implied.
Components for left- and right-moving momenta in $T^4$ are given as:
\begin{equation}
p_L^i =   \frac{p^{i}}{%
2R_{i}}+w^{i}R_{i}  , \qquad
p_R^i  =  \frac{p^{i}}{2R_{i}}%
-w^{i}R_{i} ,
\end{equation} where the index $i=1,2,3,4$ is not contracted.
  The physical momentum $p^{i}$ in the $T^{4}$ is also shifted by winding excitations and choice of Wilson lines as
\begin{equation}
p^i = n^i + B^{ij} w_j - V^K A^i_K -\frac{1}{2} A^i_K A^j_K w_j , \label{pVAw}
\end{equation}  with $w_i, n_i \in {\mathbb{Z} }$ and contraction over the index $j=1,2,3,4$.
For simplicity, we will assume $B^{ij}=0$ here.

The level matching condition has to be satisfied by any heterotic string state:

\begin{equation}
{\frac{1}{2}}P_{L}^{2}+N_{L}-1={\frac{1}{2}}P_{R}^{2}+N_{R}-C_{R}, \label{HetMatch}
\end{equation}%
with
\begin{equation}
C_{R}=\frac{1}{2}, \ \ {\rm (NS)};  \qquad C_{R}=  0,  \ \  {\rm(R)}
\end{equation}
for Neveu-Schwarz and Ramond sectors, respectively. Non-negative integers $N_{L}$ and $N_R-C_R$  denote oscillation numbers on the
bosonic left-mover and the fermionic right-mover, respectively.

The BPS states with
\begin{equation}
N_{R}=C_{R}  \label{BPScondition}
\end{equation}
saturate BPS\footnote{It is named after Bogomolny, Prasad, and Sommerfield.} bound,
and half BPS (resp. quarter BPS) states form a short (resp. ultrashort) multiplet and satisfy $N_{L}=1$ (resp. $N_{L}=0$).
The heterotic string state has mass given by:%
\begin{equation}
\frac{1}{8}m_{h}^{2}={\frac{1}{2}}P_{L}^{2}+N_{L}-1={\frac{1}{2}}%
P_{R}^{2}+N_{R}-C_{R} .  \label{HetMass}
\end{equation}

\subsection{A duality chain between heterotic
theory and type IIA string theory}

Type IIA string theory on an orbifold limit of a K3 surface is dual to heterotic string theory on $T^4$ \cite{WittenSDualityDimensions,AspinwallK3}, through the following chain of dualities \cite{BG}
\begin{equation}
\frac{\mbox{het}}{T^{4}}\;\;\overset{S}{\longrightarrow }\frac{\mbox{I}\;}{%
T^{4}}\;\overset{T^{4}}{\longrightarrow }\frac{\mbox{IIB}\;}{T^{4}/{\mathsf{%
Z\!\!Z}}_{2}^{\prime }}\;\overset{S}{\longrightarrow }\frac{\mbox{IIB}\;}{%
T^{4}/{\mathsf{Z\!\!Z}}_{2}^{\prime \prime }}\;\overset{T}{\longrightarrow }%
\frac{\mbox{IIA}\;}{T^{4}/{\mathsf{Z\!\!Z}}_{2}}\;\, . \label{chain}
\end{equation}
Here the $\mathbb{Z}_2$ actions are given as
\begin{equation}
{\mathsf{Z\!\!Z}}_{2}^{\prime }  =  (1,\Omega \mathcal{I}_{4}) ,\qquad
{\mathsf{Z\!\!Z}}_{2}^{\prime \prime}  =  (1,(-1)^{F_{L}}\mathcal{I}_{4}) ,\qquad
{\mathsf{Z\!\!Z}}_{2}  =  (1,\mathcal{I}_{4}),
\end{equation}
where the operator $\Omega $
reverses world-sheet parity and $F_{L}$ is the left-moving part of the
spacetime fermion number.  The operator
$\mathcal{I}_{4}$ implements reflection in all 4 compact directions $x_i$'s of a 4-torus \begin{equation}
\mathcal{I}_{4} :(x_1,x_2,x_3,x_4) \rightarrow (-x_1,-x_2,-x_3,-x_4). \end{equation} Since orbifolding in each direction gives 2 fixed points, the action of $\mathcal{I}_{4}$ on a 4-torus gives $2^4=16$ fixed points on an orbifold limit of a K3 surface.

This chain employs S-duality between type I
and heterotic string theories, self-S-duality of type IIB string theory, T-duality between
type I and IIB string theories along all four $x_i$ directions of $T^4$, and T-duality between type IIB and IIA string theories along $x_4$ direction.

Assuming a diagonal metric
tensor for $T^4$,
the coupling constant $g_h$ of heterotic string theory and the radii $R_{hi}$'s of $T^4$ are
written in terms of the coupling constant $g_A$ of type IIA string theory and the moduli $R_{Ai}$'s of an orbifold limit of a K3 surface as \cite{BG}
\begin{equation}
g_{h}=\frac{V_{A}}{8g_{A}R_{A4}},
\qquad
R_{hj}={\frac{1}{2}}\frac{\sqrt{V_{A}}}{R_{Aj}R_{A4}},
\qquad
R_{h4}=\frac{\sqrt{V_{A}}}{2}, \label{radii}
\end{equation}
with \begin{equation}
V_{A}\equiv R_{A1}R_{A2}R_{A3}R_{A4}, \qquad V_{h}\equiv
R_{h1}R_{h2}R_{h3}R_{h4}.\end{equation}
Radii along $x_j$ ($j=1,2,3$) directions and $x_4$ direction have different formula in \eqref{radii} due to an extra T-duality along $x_4$ direction between type IIA and IIB string theories in the duality chain of \eqref{chain}.
The masses of BPS states in type IIA and heterotic string theories are related to each other by \cite{BG}
\begin{equation}
m_{h}=\frac{ \sqrt{ V_{h}} }{g_{h}}m_{A}. \label{hetIIAmassBPS}
\end{equation}%

 \begin{table}[!h]
\caption{Mappings between BPS states in heterotic and type IIA string theories.
First two columns correspond to $p_L$ and $P_L$ of heterotic string states. Superscripts for $P_L$ denote repetition of components. A symbol $W_i$ denotes a set of these BPS objects with $w^i=1$ and $w^j=p=0$. Similarly, a set $M_i$ consists of the BPS excitation modes with minimal physical momentum $p^i =\frac{1}{2}$ in one of $T^4$ directions, with no other excitations $p^j = w =0$. The BPS heterotic string states are dual to BPS D-branes in type IIA string theory. The last column denotes the directions of cycles on which D-branes are wrapping.}
\begin{center}
\begin{tabular}{|c|c||c||c|}
\hline
$p_{L}$ & $P_{L}$ & symbol & K3 cycle  \\ \hline
$ \left( R_{h1},0,0,0 \right)  $ & $\left( (\pm {\frac{1}{2}})^{8},0^{8}\right)
 {\rm or}  \left( 0^{8},(\pm {\frac{1}{2}})^{8}\right) $ & $W_{1}$ &  $x_2, x_3$  \\
$\left(0,R_{h2},0,0 \right)$ & $\left(  \left( (\pm \frac{1}{2})^{4},0^{4} \right)^2 \right) {\rm or}  \left( \left( 0^{4},(\pm {\frac{1}{2}})^{4}\right)^2  \right) $ & $W_{2}$& $x_1,x_3$  \\
$\left(0,0,R_{h3},0 \right)$ & $\left( \left( (\pm \frac{1}{2})^{2},0^{2}\right)
^{4}\right)  {\rm or}  \left( \left( 0^{2},(\pm {\frac{1}{2}})^{2}\right)
^{4}\right) $ & $W_{3}$& $x_1, x_2$  \\
$\left(0,0,0,R_{h4} \right)$ & $\left( (\pm {\frac{1}{2}},0)^{8}\right)  {\rm or} \left( (0,\pm {%
\frac{1}{2}})^{8}\right) $& $W_{4}$ & $x_1,x_2,x_3,x_4$  \\
$\left(\frac{1}{4R_{h1}},0,0,0 \right)$ & $\left( 0^{a},\pm 1,0^{7},\pm 1,0^{7-a}\right) $& $M_{1}$ &
$x_1,x_4$  \\
$\left(0,\frac{1}{4R_{h2}},0,0 \right)$ & $(0^{a},\pm 1,0^{3},\pm 1,0^{11-a})$ &  $
M_{2} $  & $x_2, x_4$\\
$\left(0,0,\frac{1}{4R_{h3}},0 \right)$ & $(0^{a},\pm 1,0,\pm 1,0^{13-a})$  & $M_{3}$ & $x_3, x_4$ \\
$\left(0,0,0,\frac{1}{4R_{h4}} \right)$ & $\left( 0^{2a},\pm 1,\pm 1,0^{14-2a}\right) $ & $M_{4}$ &
fixed point  \\
$\left(0,0,0,\frac{1}{2R_{h4}} \right)$ & $0^{16}$  & $b$ & bulk$\ $  \\ \hline
\end{tabular}%
\label{table:matchIIAhetBPS}
\end{center}
\end{table}

\subsection{Type IIA string theory compactified on an orbifold limit of a K3 surface \label{IIAonK3}}
Consider compactification of type IIA string theory on an orbifold limit of a K3 surface, $T^4/{{\mathsf{Z\!\!Z} }_2}$. See \cite{Vafa97Lecture} for a review. D-even-branes are BPS states in type IIA string theory, and their masses may be expressed as the tension $\frac{1}{g_A}$ times the volume of D-brane
\begin{equation}
m_{{\rm BPS}, A} = \frac{ {\rm volume}}{g_A}.
\end{equation}

A bulk D0-brane has a unit volume in 0-dimension and has the mass of $\frac{1}{g_A}$. Fractional D0-branes sit
at the 16 fixed points of an orbifold limit of a K3 surface. After blowing up each fixed point into a 2-sphere, we can wrap D2-branes over these 2-cycles. The fractional D0-brane can be
thought of as a D2-brane wrapping a vanishing 2-cycle, which comes from a resolution of an orbifold singularity.
 Fractional D0-branes have one-half unit volume in 0-dimension because of the ${\mathsf{Z\!\!Z}}_{2}$ orbifolding in the K3 surface. Their mass
is  $\frac{1}{2g_A}$, which is half of that of a bulk D0-brane.
 A D4-brane wrapping the whole K3 surface has mass $\frac{V_A}{2g_A}$. D2-branes wrapping the torus $T^2$ in $x_j$ and $x_k$ directions have mass $\frac{%
R_{Aj} R_{Ak}}{2g_A}$.

By matching the charges and the masses, Table \ref{table:matchIIAhetBPS} lists mappings between BPS states in heterotic and type IIA string theories.
The first two columns correspond to $p_L$ and $P_L$ of heterotic string states, and symbols $W_i$ and $M_i$ in the third column denote BPS modes with minimum winding and momentum in $x_i$ direction, respectively.  Reading off from $p_L$, left-moving momentum in $T_4$, the first 4 rows correspond to BPS excitation modes with unit wrapping $w^i =1$ in one of $T^4$ directions, with no other excitations $w^j = p =0$. From \eqref{pVAw}, 16-dimensional momentum $P_L$ has 8 half-integer entries and 8 integer entries. Level matching \eqref{HetMatch} and BPS conditions \eqref{BPScondition} further restrict $P_L$ to have eight zeros and eight $\pm \frac{1}{2}$'s.  In order to satisfy \eqref{pVAw},
the signs before each $\pm \frac{1}{2}$ are chosen such that $P_L A^i \in {\mathbb{Z} }$ and $P_L A^j \in {\mathbb{Z} } +\frac{1}{2}$ for $j\ne i$. A symbol $W_i$ denotes a set of these BPS objects with $w^i=1$ and $w^j=p=0$.

Similarly, the next 4 rows of table \ref{table:matchIIAhetBPS} correspond to BPS excitation modes with minimal physical momentum $p^i =\frac{1}{2}$ in one of $T^4$ directions, with no other excitations $p^j = w =0$, which belong to a set $M_i$. Their $P_L \in {\mathbb{Z} }^{16}$ has 14 zero entries and two $\pm 1$ entries. The location of two $\pm 1$ entries are chosen to satisfy \eqref{pVAw}: there are $2^{4-i}-1$ zeroes between two $\pm 1$ entries, in order to satisfy $P_L A^i \in {\mathbb{Z} } +\frac{1}{2}$ and $P_L A^j \in \mathbb{Z}$ for $j\ne i$.
There are also BPS bound states of these objects having more than one of $M_i$ and $W_i$ excitations.

The BPS heterotic string states are dual to BPS D-branes in type IIA string theory. The last column of table \ref{table:matchIIAhetBPS} denotes the directions of cycles on which D-branes are wrapping. Heterotic states in $W_4$ and $M_4$ are dual to D4-branes and fractional D0-branes respectively, and heterotic states in $W_j$ and $M_j$ ($j=1,2,3$) are dual to D2-branes over 2-cycles over $x_k, x_l$ ($\{j,k,l\} =\{1,2,3\}$) and $x_j, x_4$ directions, respectively.

A non-BPS state may decay into a collection of $n$ BPS states
\begin{equation}
{\rm ( state)}_{\rm non-BPS} \rightarrow \sum_{i=1}^n {\rm ( state)}_{{\rm BPS},i}
\end{equation}
subject to charge conservation and non-creation of mass
\begin{eqnarray}
\mathbf{P}_{\rm non-BPS} & = & \sum_{i=1}^n \mathbf{P}_{{\rm BPS},i} \label{chargecons} \\
m_{\rm non-BPS} &\ge & \sum_{i=1}^n m_{{\rm BPS},i} \label{masscomp}.
\end{eqnarray}
Take the following strategy to test existence of an exactly stable non-BPS state:
\begin{enumerate}
 \item Start with the lightest possible non-BPS state whose mass does not depend on the moduli of $T^4$.
 \item Classify a collection of BPS states that holds \eqref{chargecons} and identify the lightest possible collection of BPS states in each class.
 \item Compare masses and find conditions on moduli which nullify \eqref{masscomp} for every class of BPS states.
 \end{enumerate}

\begin{figure}[!h]
\centerline{\includegraphics[width=1.5 in]{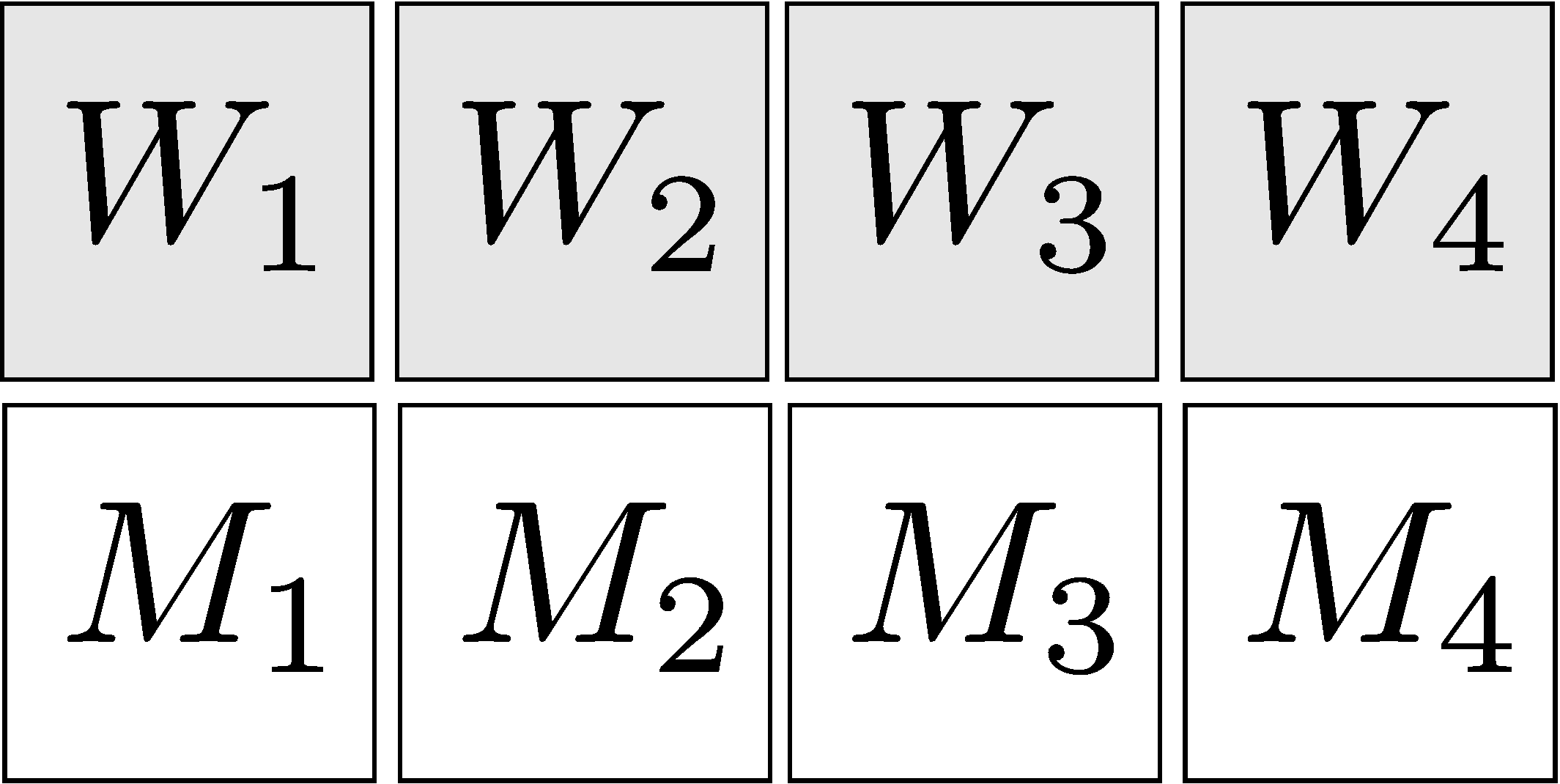}}
\caption[For a non-BPS state with $P_L \in \left( {\mathbb{Z} } +\frac{1}{2} \right)^{16}$, possible BPS-decay channels must contain one of the shaded objects, $W_i$'s here.]{For a non-BPS state with $p_R=p_L=0$ and $P_L \in \left( {\mathbb{Z} } +\frac{1}{2} \right)^{16}$, possible BPS-decay channels must contain one of the \colorbox{lightshade}{shaded} objects, $W_i$'s here.}
\label{fig:8boxW}
\end{figure}

 For example, start with a non-BPS state of $P_L \in \left( {\mathbb{Z} } +\frac{1}{2} \right)^{16}$, then the BPS decay products must contain some state which carries half-integers in some of these 16 entries of $P_L$. The BPS-states in $M_i$'s have only integer entries in $P_L$. Therefore, the decay channel must contain some of $W_i$'s as in Figure \ref{fig:8boxW}.
Consider a non-BPS state with $ p_L=p_R=0$, $w=p=0$,
\begin{equation}
P_{L}  = \left( \frac{1}{2},\frac{1}{2},\frac{1}{2},-\frac{1}{2%
}; \left(  \frac{1}{2},-\frac{1}{2},-\frac{1}{2},-\frac{1}{2}  \right)^2 ;-\frac{1}{2},-\frac{1}{2},-\frac{1}{2},\frac{1}{2}%
\right),
\end{equation}%
and mass $m_h = \sqrt{8({\frac{1}{2}}P_{L}^{2}-1)%
} =\sqrt{8}$.
Its BPS-decay products must contain states with $W_i$ excitations. The lightest possible collection of BPS-decay products are a pair of $W_i$ objects as discussed \cite{GaberdielLecture}, with total mass $2\times \sqrt{8({\frac{1}{2}}P_{R}^{2})} =4|P_{R}| =4{R_{hi}}$. No other decays are allowed because of the form of the $P_L \in \left( {\mathbb{Z} } +\frac{1}{2} \right)^{16}$.
The stability region for this against all the possible decays allowed by charge conservation is therefore a corner of moduli space with
\begin{equation}
\sqrt{8} < 4{R_{hi}}, \qquad i=1,2,3,4. \label{alllarge}
 \end{equation}

As discussed in \cite{GaberdielLecture}, this non-BPS state in heterotic string theory corresponds to a non-BPS $\widehat{D3}$-brane in type IIA string theory stretched along $x_1, x_2, x_3$ directions. The symbol $\widehat{\phantom{D}}$ over $\widehat{D}$-brane denotes that a D-brane has wrong dimensions and is a non-BPS object. D-even-branes (D-odd-branes) are BPS (non-BPS) objects in type IIA string theory. The lightest possible BPS decay products are a pair of D4-brane and anti-D4-brane, or a pair of D2-brane and anti-D2-brane spanning $i$ and $j$ directions with $ i,j \in \{1,2,3 \} $.

 This demonstrates restriction for decay modes of non-BPS states with $p_R=p_L=0$, $P_L=\left( (\pm {\frac{1}{2}})^{16} \right)$ to contain some of $W_i$'s. Similarly, one may ask whether there are non-BPS objects whose decay products must contain some of $M_i$'s instead. This answer is yes, due to symmetry.   The next section introduces a set of eight $16\times 16$
unitary matrices acting on $P_L$ and shows how charge
conservation constrains possible decay modes into BPS states in a systematic way.
For example, we will see that a similar constraint exists for a non-BPS object with $p_R=p_L=0$ and $P_L=\left( 2,0^{15}\right)$, and its decay modes must contain some of $M_i$.

\section{A systematic test of non-BPS stability \label{sec:HeteroticOrganizer}}
The transformations for 16-dimensional momentum vector $P_L$ of the left-mover of heterotic string states are given by the $16\times 16$ matrices ${\mathsf{1\!\!1}}_{16}, T_{1234}$, and $ T_{ij}$ with $i < j \in \{1,2,3,4\}$, given by the following formulas:
\begin{equation}
T_{12}   \equiv   \left(
\begin{array}{cccc}
L & 0 & 0 & 0 \\
0 & 0 & R & 0 \\
0 & R & 0 & 0 \\
0 & 0 & 0 & L%
\end{array} \right),
\end{equation}
\begin{eqnarray}
T_{13} &\equiv & U_{23} \cdot T_{12} \cdot U_{23}, \qquad
T_{14}  \equiv   U_{34} \cdot T_{13} \cdot U_{34}\\
T_{23} &\equiv & U_{12} \cdot T_{13} \cdot U_{12}, \qquad
T_{24}  \equiv   U_{34} \cdot T_{23} \cdot U_{34}, \qquad
T_{34}  \equiv   U_{23} \cdot T_{24} \cdot U_{23}\\
T_{1234} &\equiv  & T_{12} \cdot T_{34} = T_{13} \cdot T_{24} =T_{14} \cdot T_{23} ,
\end{eqnarray}
where $L$ and $R$ are $4 \times 4$ matrices given below:%
\begin{equation}
L  \equiv  \left(
\begin{array}{cccc}
+{\frac{1}{2}} & +{\frac{1}{2}} & +{\frac{1}{2}} & -{\frac{1}{2}} \\
+{\frac{1}{2}} & +{\frac{1}{2}} & -{\frac{1}{2}} & +{\frac{1}{2}} \\
+{\frac{1}{2}} & -{\frac{1}{2}} & +{\frac{1}{2}} & +{\frac{1}{2}} \\
-{\frac{1}{2}} & +{\frac{1}{2}} & +{\frac{1}{2}} & +{\frac{1}{2}}%
\end{array}%
\right), \qquad R \equiv  \left(
\begin{array}{cccc}
+{\frac{1}{2}} & -{\frac{1}{2}} & -{\frac{1}{2}} & -{\frac{1}{2}} \\
-{\frac{1}{2}} & +{\frac{1}{2}} & -{\frac{1}{2}} & -{\frac{1}{2}} \\
-{\frac{1}{2}} & -{\frac{1}{2}} & +{\frac{1}{2}} & -{\frac{1}{2}} \\
-{\frac{1}{2}} & -{\frac{1}{2}} & -{\frac{1}{2}} & +{\frac{1}{2}}%
\end{array}%
\right).
\end{equation}
 Each $16 \times 16$ unitary matrix $U_{ij}$ exchanges $x_i$ and $x_j$ directions in $T^4$ and written as:
\begin{eqnarray}
U_{12}  & \equiv &  \left(
\begin{array}{cccc}
{{\mathsf{1\!\!1}}}_4 & 0 & 0 & 0 \\
0 & 0 & {\mathsf{1\!\!1}}_4 & 0 \\
0 & {\mathsf{1\!\!1}}_4 & 0 & 0 \\
0 & 0 & 0 & {\mathsf{1\!\!1}}_4 %
\end{array} \right), \\
U_{23}  & \equiv &  \left(
\begin{array}{cc}
u_{23} & 0   \\
0 & u_{23}
\end{array} \right), \qquad U_{34}   \equiv   \left(
\begin{array}{cccc}
u_{34}  & 0 & 0 & 0 \\
0 & u_{34} & 0  & 0 \\
0 & 0 & u_{34} & 0 \\
0 & 0 & 0 & u_{34}  %
\end{array} \right),
\end{eqnarray}
with following $8\times 8$ and $4\times 4$ submatrices:
\begin{equation}
u_{23}   \equiv   \left(
\begin{array}{cccc}
{\mathsf{1\!\!1}}_2 & 0 & 0 & 0 \\
0 & 0 & {\mathsf{1\!\!1}}_2 & 0 \\
0 & {\mathsf{1\!\!1}}_2 & 0 & 0 \\
0 & 0 & 0 & {\mathsf{1\!\!1}}_2 %
\end{array} \right), \qquad u_{34}   \equiv   \left(
\begin{array}{cccc}
1  & 0 & 0 & 0 \\
0 & 0 & 1  & 0 \\
0 & 1  & 0 & 0 \\
0 & 0 & 0 & 1  %
\end{array} \right).
\end{equation}

All of $T$'s commute with one another, and have unit determinant and squares to an identity matrix,
\begin{equation}
T_{1234}^{2}  =  T_{ij}^{2}={\mathsf{1\!\!1}}_{16} , \qquad i < j \in \{1,2,3,4\}  .
\end{equation}%
A transformation matrix $T_{ij}$ corresponds to T-dualities on $x_i$ and $x_j$ directions in $T^4$ and exchanges excitations in $W_{i,j} \rightarrow M_{i,j}$ classes. Similarly, $T_{1234}$ corresponds to T-dualities on all the four directions in $T^4$ and exchanges excitations in $W_{1,2,3,4} \rightarrow M_{1,2,3,4}$ classes.

Starting from a BPS object in $M_4$ class with $P_L= \left( 1, 1, 0^{14} \right)$, one has
\begin{eqnarray}
P_{L} &=&P_{L}\cdot T_{12}=P_{L}\cdot T_{13} =P_{L}\cdot T_{23} =\left( 1,1,0^{14}\right)  \\
P_{L}\cdot T_{1234} &=&P_{L}\cdot T_{14}=P_{L}\cdot T_{24}=P_{L}\cdot T_{34} \\
&=&\left( \frac{1}{2},0,\frac{1}{2},0;
\frac{1}{2},0,-\frac{1}{2},0;\frac{1}{2},0,-\frac{1}{2},0;-\frac{1}{2},0,
-\frac{1}{2},0\right).
\end{eqnarray}%
The transformation by $T_{i4}$ or $T_{1234}$ turns $P_L$ of $M_4$ class into $P_L$ of $W_4$ class which has half-integer elements as seen in table \ref{table:matchIIAhetBPS}.

Similarly, starting from a BPS object in $M_1$ class with $P_L= \left( 1,0^{7},1,0^{7}\right)$, one has
\begin{eqnarray}
P_{L} &=&P_{L}\cdot T_{23}=P_{L}\cdot T_{24}=P_{L}\cdot T_{34}=\left( 1,0^{7}, 1,0^{7}\right)
\\
P_{L}\cdot T_{1234} &=&P_{L}\cdot T_{12}=P_{L}\cdot T_{13}=P_{L}\cdot T_{14} \\ &=& \left( \frac{1}{2}%
,\frac{1}{2},\frac{1}{2},-\frac{1}{2}, \frac{1}{2},-\frac{1}{2},-\frac{1}{2},%
-\frac{1}{2},0^{8}\right).
\end{eqnarray}%
The transformation by $T_{1j}$ or $T_{1234}$ turns $P_L$ of $M_1$ class into $P_L$ of $W_1$ class which has half-integer elements as seen in table \ref{table:matchIIAhetBPS}.

One may conclude that for a BPS object, $T_{ij}$ will reverse the form of $P_L$ of $W_a \leftrightarrow M_a$ for $a=i, j$ and
 $T_{1234}$ will reverse the form of $P_L$ of $W_a \leftrightarrow M_a$ for $a=1, 2, 3, 4$.
The transformations induced on a BPS object are partially shown here:
\begin{equation}
\begin{tabular}{|c||c|c|c|c||c|}
\hline
$P_{L}$ & $()_{\mathrm{given}}$ & $T_{1234}$ & $T_{23}$ & $T_{14}$ & $P_{L}$
\\ \hline
$\left( (\pm {\frac{1}{2}})^{8},0^{8}\right) ,\ \left( 0^{8},(\pm {\frac{1}{2%
}})^{8}\right) $ &  & $\leftrightarrow $ &  & $\leftrightarrow $ & $\left(
0^{a},\pm 1,0^{7},\pm 1,0^{7-a}\right) $ \\
$\left( \left( (\pm \frac{1}{2})^{4},0^{4}\right) ^{2}\right) ,\ \left(
\left( 0^{4},(\pm {\frac{1}{2}})^{4}\right) ^{2}\right) $ &  & $%
\leftrightarrow $ & $\leftrightarrow $ &  & $(0^{a},\pm 1,0^{3},\pm 1,0^{11-a})$
\\
$\left( \left( (\pm \frac{1}{2})^{2},0^{2}\right) ^{4}\right) ,\ \left(
\left( 0^{2},(\pm {\frac{1}{2}})^{2}\right) ^{4}\right) $ &  & $%
\leftrightarrow $ & $\leftrightarrow $ &  & $(0^{a},\pm 1,0,\pm 1,0^{13-a})$ \\
$\left( (\pm {\frac{1}{2}},0)^{8}\right) ,\ \left( (0,\pm {\frac{1}{2}}%
)^{8}\right) $ &  & $\leftrightarrow $ &  & $\leftrightarrow $ & $\left(
0^{2a},\pm 1,\pm 1,0^{14-2a}\right) $.\\ \hline
\end{tabular}
\end{equation}%
Here $\leftrightarrow$ on the $i$'th row indicates that $W_i$ and $M_i$ exchange the form of their $P_L$ charges.

\begin{figure} [!h]
\centerline{\includegraphics[width=1.5 in]{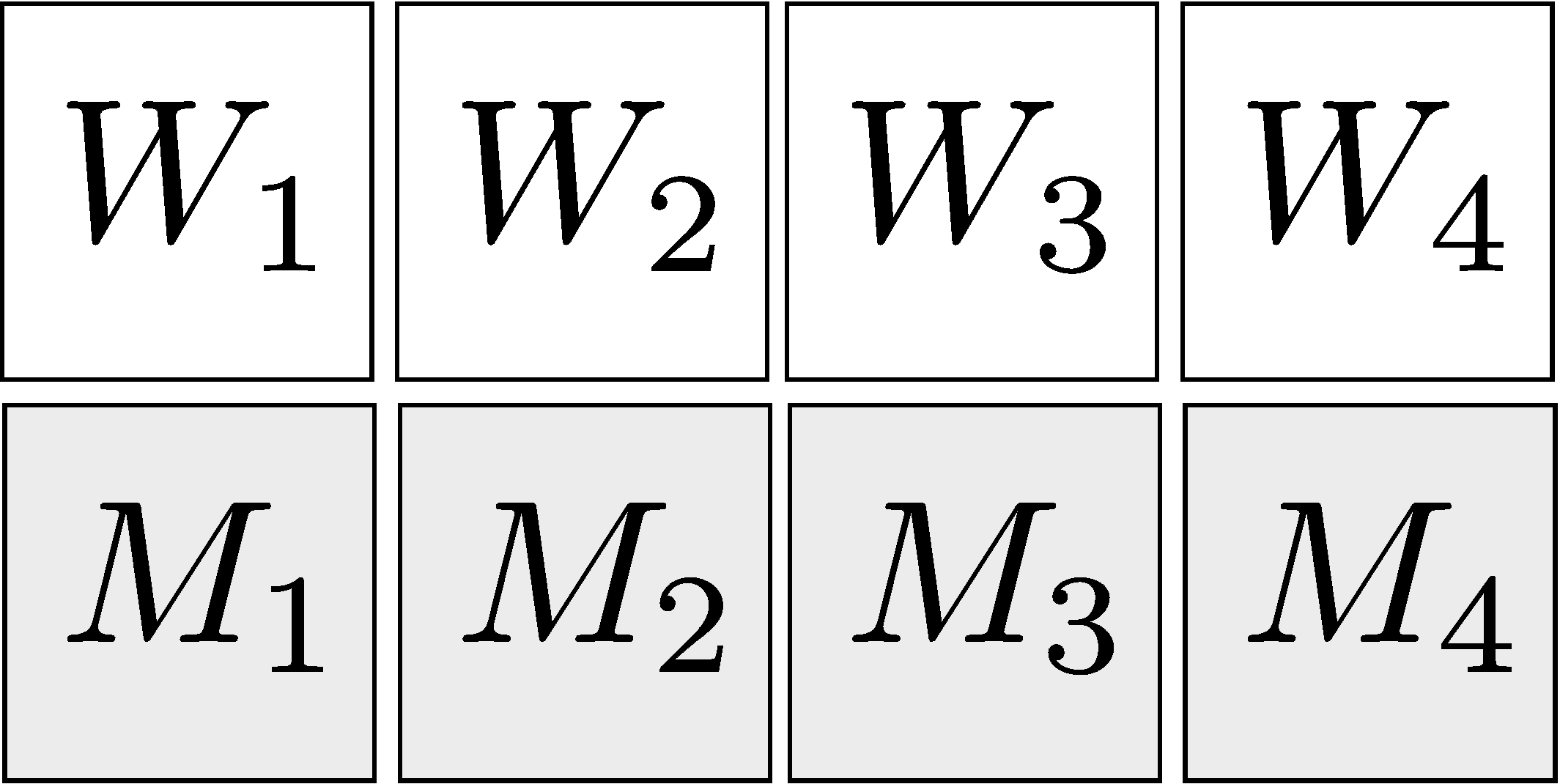}}
\caption[For a non-BPS state with $p_R=p_L=0$ and $P_L \cdot  T_{1234} \in \left( {\mathbb{Z} } +\frac{1}{2} \right)^{16}$, possible BPS-decay channels must contain one of the shaded objects, $M_i$'s here.]{For a non-BPS state with $p_R=p_L=0$ and $P_L \cdot  T_{1234} \in \left( {\mathbb{Z} } +\frac{1}{2} \right)^{16}$, possible BPS-decay channels must contain one of the \colorbox{lightshade}{shaded} objects, $M_i$'s here.}
\label{fig:8boxM}
\end{figure}

The transformation by
 $T_{1234}$ exchanges all $W_i \leftrightarrow M_i$. As promised earlier, we have now found a non-BPS state whose decay product now must contain some of $M_i$. A non-BPS state with $p_R=p_L=0$ and $P_L \cdot  T_{1234}  \in \left( {\mathbb{Z} } +\frac{1}{2} \right)^{16}$ can decay only into sets of BPS states that contain some of $M_i$'s, as shown in Figure \ref{fig:8boxM}.
 For example, a non-BPS state with $p_R=p_L=0$ and
 \begin{eqnarray}
P_{L} &=&\left( 2,0^{15}\right)  \\
P_{L}\cdot T_{1234} &=&\left( \frac{1}{2},\frac{1}{2},\frac{1}{2},-\frac{1}{2%
}; \left(  \frac{1}{2},-\frac{1}{2},-\frac{1}{2},-\frac{1}{2}  \right)^2 ;-\frac{1}{2},-\frac{1}{2},-\frac{1}{2},\frac{1}{2}%
\right)
\end{eqnarray}%
can decay into a brane-antibrane pair of any of $M_1, M_2, M_3, M_4$ as in \cite{BG,GaberdielLecture}, and we have shown that the lightest collection of BPS decay products {\it must contain} a pair of any of $M_i$'s. No other decays with less mass are possible.
This non-BPS object is interpreted as a non-BPS $\widehat{D1}$-brane stretched along $x_4$ direction \cite{BG,GaberdielLecture}. The possible BPS decay products allowed by charge conservation are into a pair of wrapped D0-brane and anti-D0-brane, or a
pair of D2-brane and anti-D2-brane spanning $x_i$ and $x_4$ directions with $i=1,2,3$.
The mass of non-BPS state, before decay, is $\sqrt{8 ({\frac{1}{2}} P_L^2 - 1)}  = \sqrt{8}$.
The mass on the BPS side, after decay, is $2 \times \sqrt{8({\frac{1}{2}} P_R^2)}  = 4 |P_R| =\frac{1}{R_{hi}}$
where $i = 1, 2, 3, 4$.
The stability region for this object against every possible BPS decay is
\begin{equation}
\sqrt{8} < \frac{1}{R_{hi}}, \qquad  i=1,2,3,4 \label{allsmall}
\end{equation}
which holds in another corner of moduli space.
\begin{figure}[!h]
\centerline{\includegraphics[width=1.5 in]{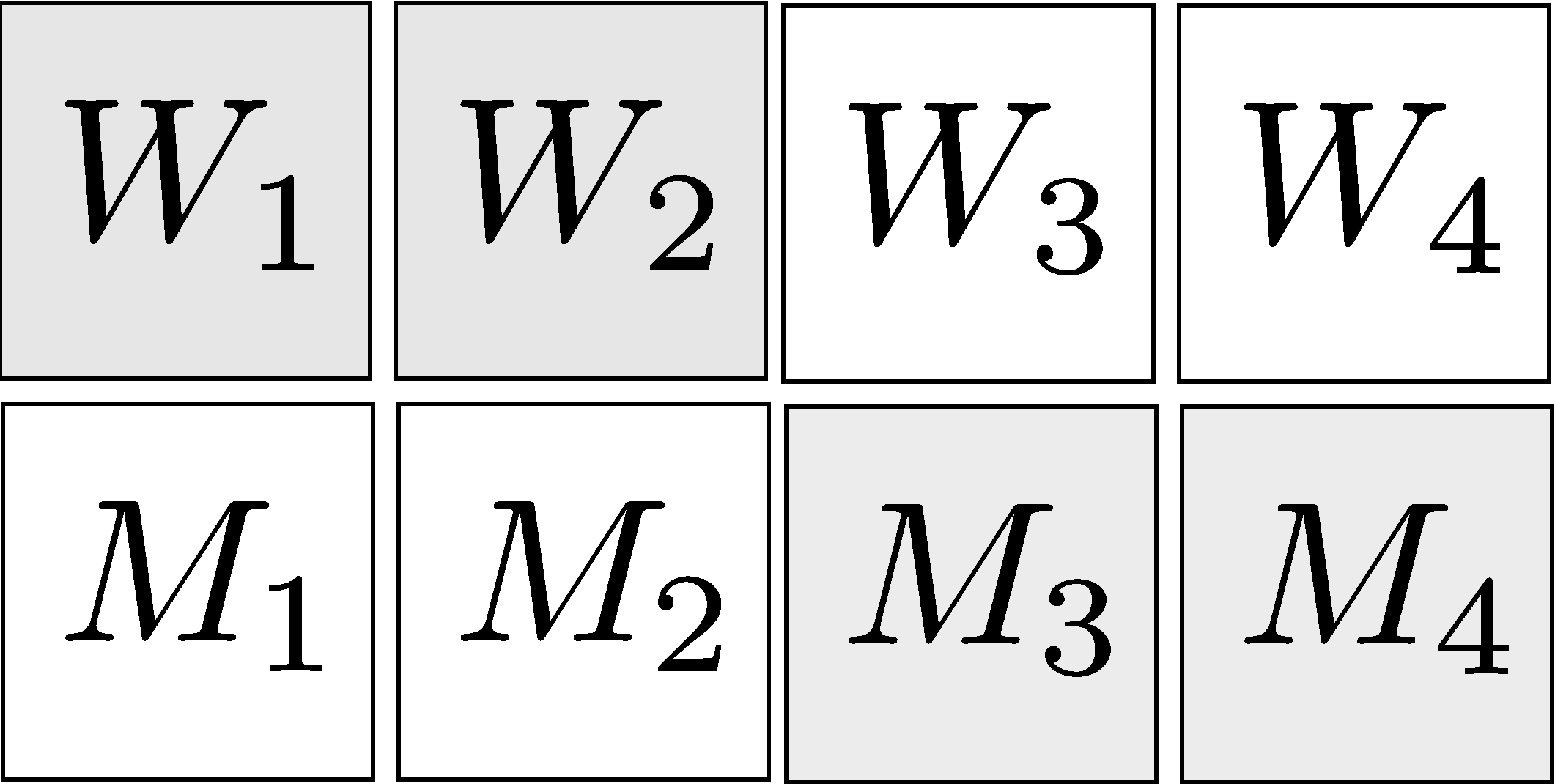}}
\caption[For a non-BPS state with $p_R=p_L=0$ and $P_L  \cdot T_{34} \in \left( {\mathbb{Z} } +\frac{1}{2} \right)^{16}$, possible BPS-decay channels must contain one of the shaded objects, $W_1, W_2, M_3$, and $M_4$ here.]{For a non-BPS state with $p_R=p_L=0$ and $P_L  \cdot T_{34} \in \left( {\mathbb{Z} } +\frac{1}{2} \right)^{16}$, possible BPS-decay channels must contain one of the \colorbox{lightshade}{shaded} objects, $W_1, W_2, M_3$, and $M_4$ here.}
\label{fig:8box34}
\end{figure}

Decay products of a non-BPS object with $p_R=p_L=0$ and
\begin{eqnarray}
P_{L} &=&\left( 1,1,1,-1,0^{12}\right)  , \qquad  P_{L}\cdot T_{12} =\left( 2,0^{15}\right)  \\
P_{L}\cdot T_{34} &=&\left( \frac{1}{2},\frac{1}{2},\frac{1}{2},-\frac{1}{2%
}; \left(  \frac{1}{2},-\frac{1}{2},-\frac{1}{2},-\frac{1}{2}  \right)^2 ;-\frac{1}{2},-\frac{1}{2},-\frac{1}{2},\frac{1}{2}%
\right)\end{eqnarray}
must contain some of $W_1, W_2, M_3, M_4$ as in Figure \ref{fig:8box34}.
Similarly, BPS decay products of a non-BPS state with $p_R=p_L=0$ and $P_L \cdot T_{ij}=\left( (\pm {\frac{1}{2}})^{16} \right)$ must contain some of $M_i, M_j, W_k, W_l$ where $\{i,j,k,l\}=\{1,2,3,4\}$.

A non-BPS object with $p_R =p_L=0$ and $P_L \cdot T_{i4}
= ( 0^a, \pm 2 , 0^b)$ has $P_L \cdot T_{jk}=(  (\pm {\frac{1}{2}
})^{16}) $ and the lightest possible BPS decay products must contain some of $M_i, M_4, W_j, W_k$, which are
D2-branes spanning $x_i$ and $x_4$ directions (where $l =1,2,3$) and D0-branes at fixed points of an orbifold limit of a K3 surface. Therefore, this corresponds to a non-BPS
 $\widehat{D1}$-brane, stretched along stretched along $x_i$ direction, with the non-BPS stability condition
 \begin{equation}
 \sqrt{8} < \frac{1}{R_{hi}} , 4{R_{hj}},4{R_{hk}}, \frac{1}{R_{h4}} , \qquad \{i,j,k\}=\{1,2,3\}. \label{slls}
 \end{equation}

Similarly, a non-BPS object with $p_R =p_L=0$ and $P_L \cdot T_{jk}
= (  0^a, \pm 2 , 0^b) $ has $P_L \cdot T_{i4} =(  (\pm {\frac{1}{2}
})^{16}) $ and the lightest possible BPS decay products must contain some of $W_i, W_4, M_j, M_k$, which are
D2-branes spanning $x_j$ and $x_k$ direction and D4-branes spanning all 4 directions. Therefore this corresponds to a non-BPS
 $\widehat{D3}$-brane, stretched along stretched along $x_j, x_k, x_4$ directions, with the non-BPS stability condition
 \begin{equation}
 \sqrt{8} < 4{R_{hi}}, \frac{1}{R_{hj}} , \frac{1}{R_{hk}}, 4{R_{h4}} , \qquad \{i,j,k\}=\{1,2,3\}. \label{lssl}
 \end{equation}

\begin{figure}
\centerline{\includegraphics[width=.5\textwidth]{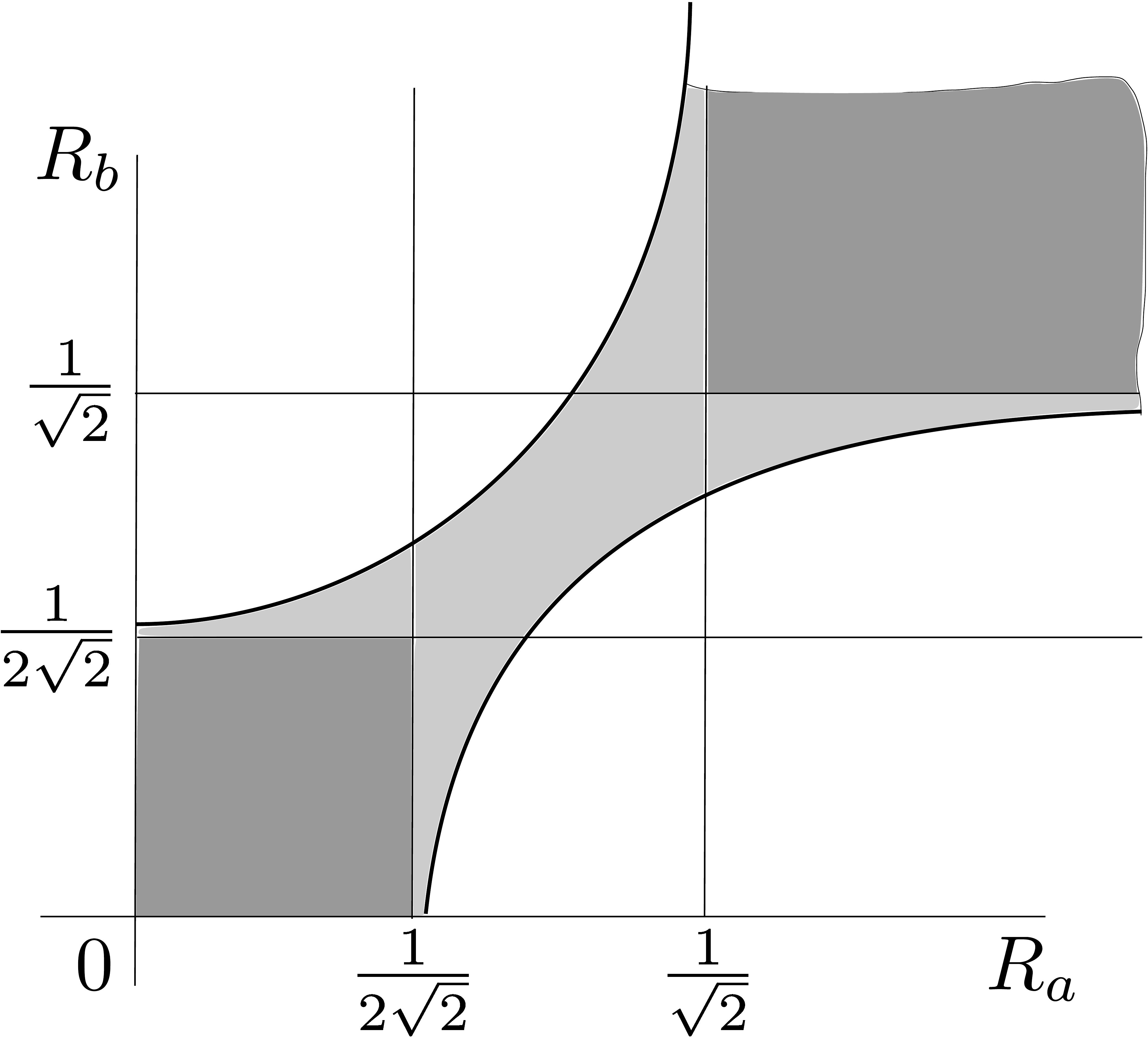}}
\caption[A 2D slice of stability regions of non-BPS states]{A non-BPS state of charge $P_L = \left( \left(   \frac{1}{2} \right)^{16} \right)$ is stable in all the shaded regions. Drawn is a 2d slice varying two radii $R_a$ and $R_b$ of the $T^4$, fixing both of the other two radii $R_c$ and $R_d$ very large or small - namely $R_c, R_d > \frac{1}{\sqrt{2}}$ or $R_c, R_d <  \frac{1}{2\sqrt{2}}$. If $R_c > \frac{1}{\sqrt{2}}$ and $R_d <  \frac{1}{2\sqrt{2}}$, then it will roughly appear as shown, with one of the axis now denoting $\frac{1}{4R}$ instead of $R$. As we choose less extreme values for $R_c$ and $R_d$, the \colorbox{lightshade}{light shade} will get {\it larger}.

\hskip 1.2cm An extra type of non-BPS states
 become exactly stable in the \colorbox{darkshade}{dark shades}.}
\label{fig:2Dphase}
\end{figure}

The results from \eqref{alllarge}, \eqref{allsmall}, \eqref{slls}, and \eqref{lssl} can be summarized as follows: A stable non-BPS object exists in every other corner of moduli space, where even number of radii are small and the rest even number of radii are large. In the dark shades in Figure \ref{fig:2Dphase}, one kind  of non-BPS states we considered become exactly stable. In type IIA string theory, they correspond to non-BPS $\widehat{D1}$-branes and non-BPS $\widehat{D3}$-branes.%
 \begin{figure}
\centerline{\includegraphics[width=4 in]{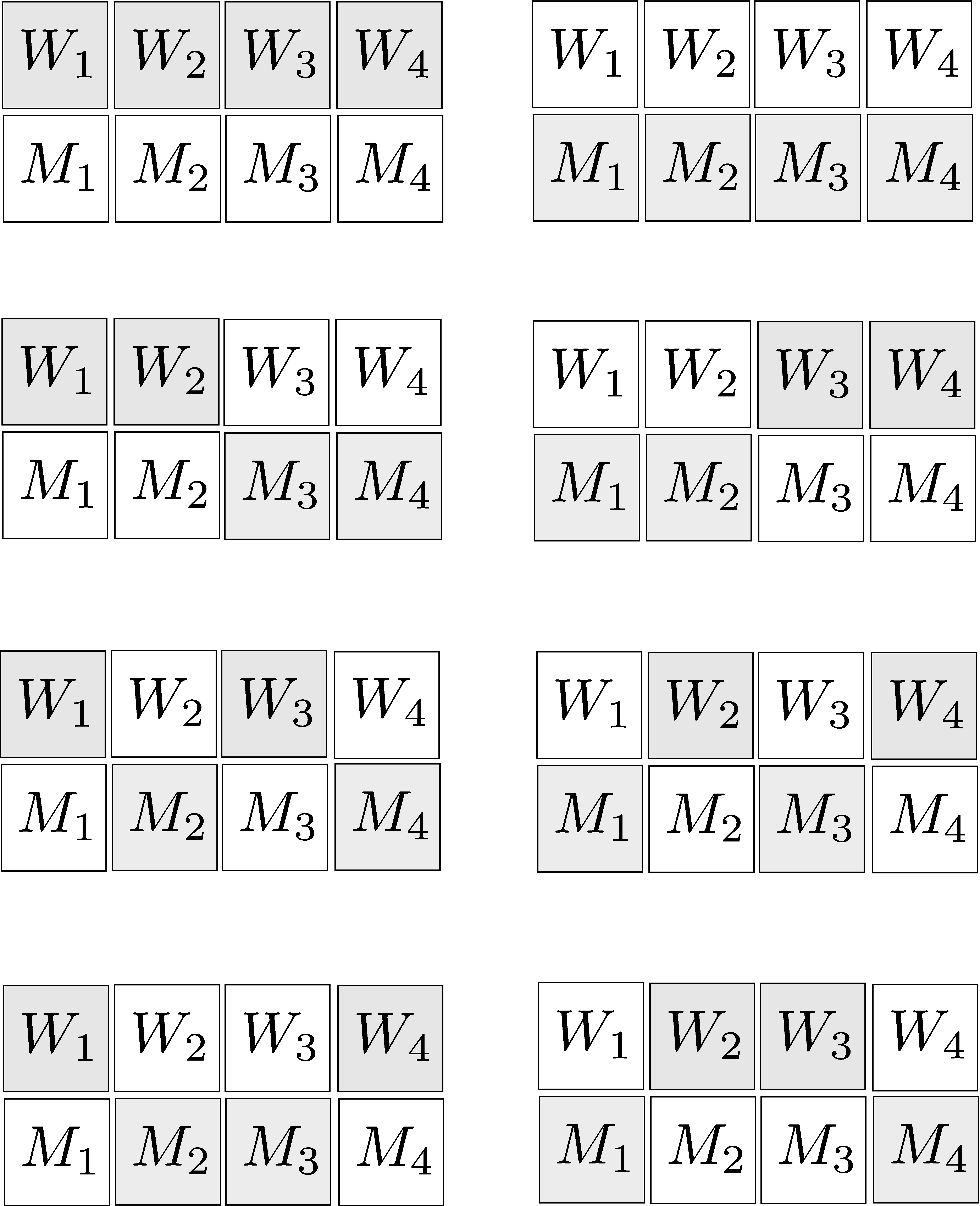}}
\caption[A non-BPS object with $p_R=p_L=0$ and $P_L  \cdot M_{ } \in \left( {\mathbb{Z} } +\frac{1}{2} \right)^{16}$ for all eight $T$'s can decay only into sets of BPS states that have overlap with all of these eight groups.]{A non-BPS object with $p_R=p_L=0$ and $P_L  \cdot M_{ } \in \left( {\mathbb{Z} } +\frac{1}{2} \right)^{16}$ for all eight $T$'s can decay only into sets of BPS states that have overlap with all of these eight groups.}
\label{fig:8box8necessary}
\end{figure}
 \begin{figure}
\centerline{\includegraphics[width=1.5 in]{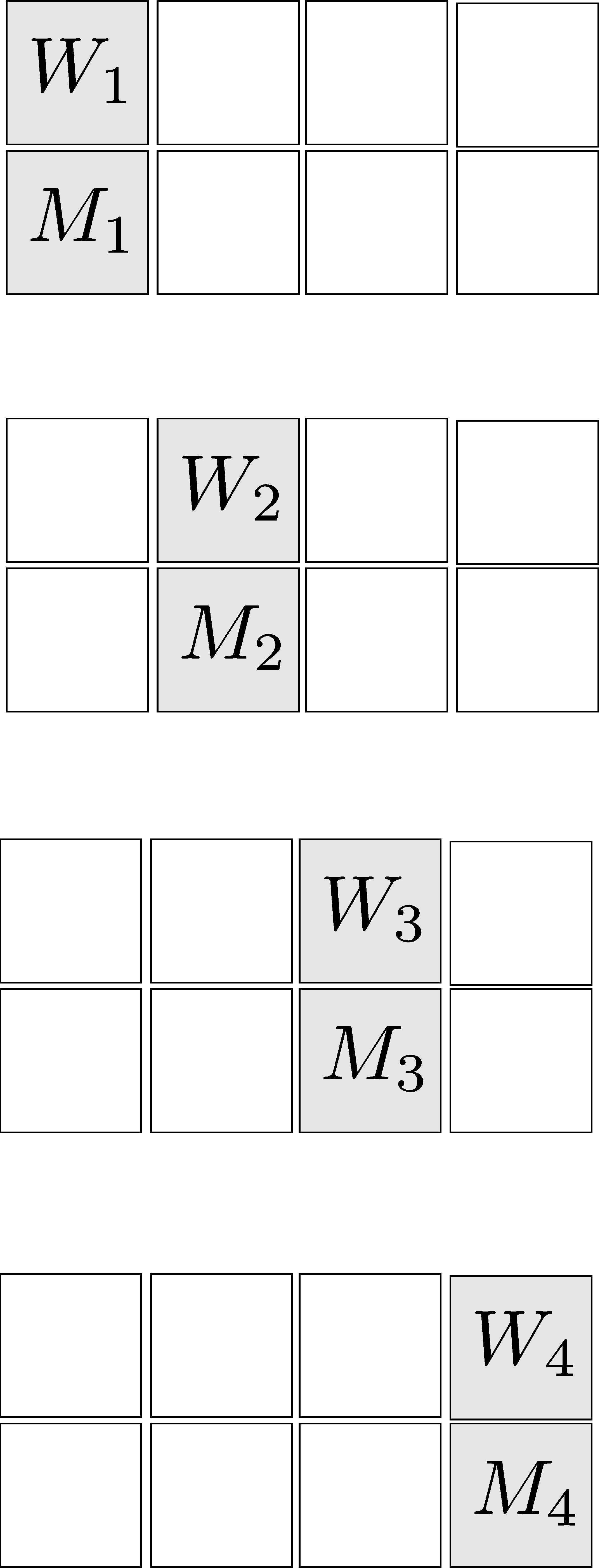}}
\caption[Charge conservation allows a non-BPS state with $p_R=p_L=0$ and $P_{L} =\left( \left( {\frac{1}{2}}\right) ^{16}\right)$ to decay into $M_a$ and $W_a$ pairs with $a=1, 2, 3, 4$, but energy prohibits those decays.]{Charge allows a non-BPS state with $p_R=p_L=0$ and $P_{L} =\left( \left( {\frac{1}{2}}\right) ^{16}\right)$ to decay into $M_a$ and $W_a$ pairs with $a=1, 2, 3, 4$, but energy prohibits those decays.}
\label{fig:8box4heavy}
\end{figure}
 \begin{figure}
\centerline{\includegraphics[width=4 in]{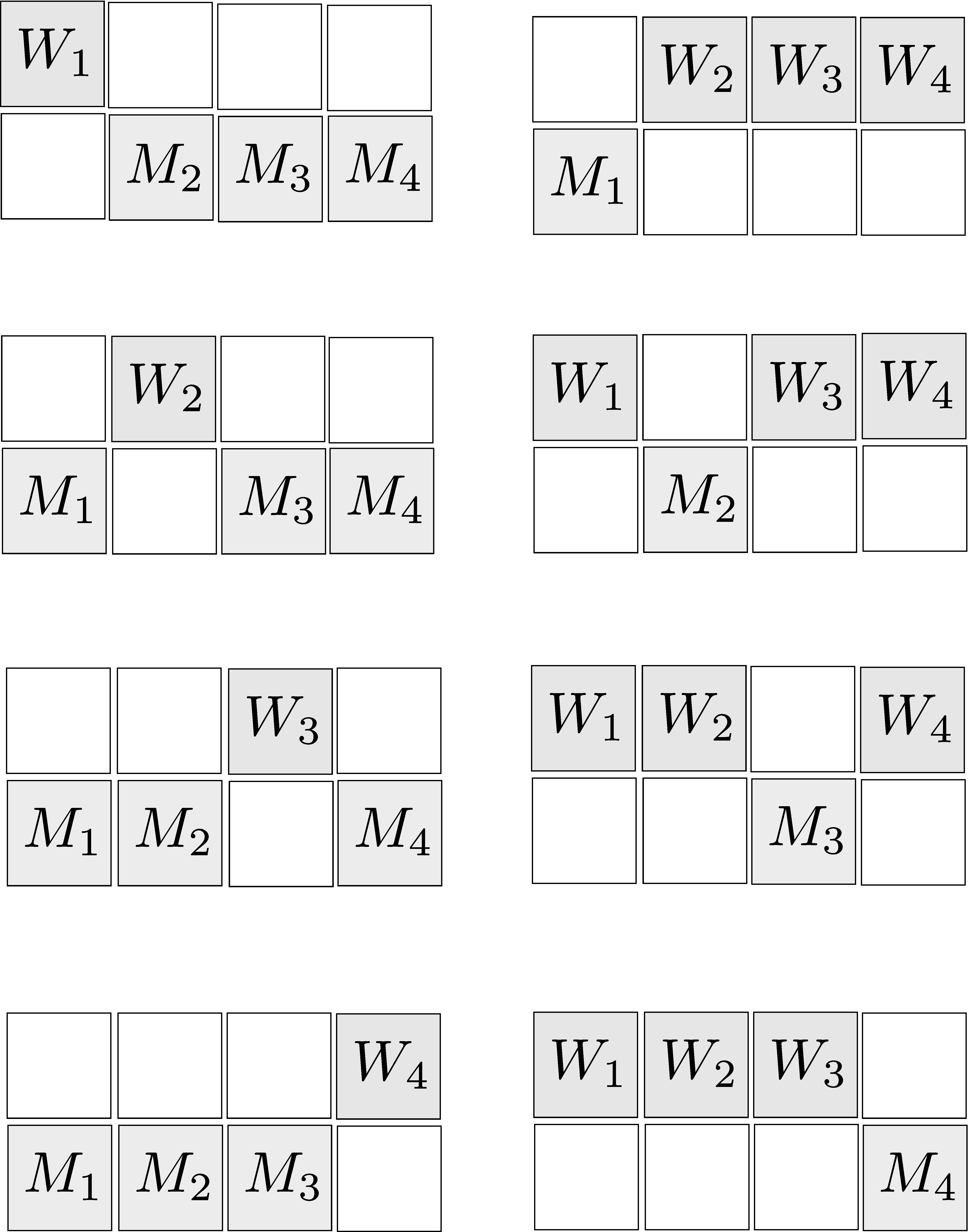}}
\caption[Charge conservation allows a non-BPS state with $p_R=p_L=0$ and $P_{L} =\left( \left( {\frac{1}{2}}\right) ^{16}\right)$ to decay into $2(W_{a}+M_{b}+M_{c}+M_{d})$ or $2(W_{a}+W_{b}+W_{c}+M_{d})$ with $\left\{ a, b, c, d \right\}  =\left\{ 1, 2, 3, 4 \right\} $.]{Charge allows a non-BPS state with $p_R=p_L=0$ and $P_{L} =\left( \left( {\frac{1}{2}}\right) ^{16}\right)$ to decay into $2(W_{a}+M_{b}+M_{c}+M_{d})$ or $2(W_{a}+W_{b}+W_{c}+M_{d})$ with $\left\{ a, b, c, d \right\}  =\left\{ 1, 2, 3, 4 \right\} $.}
\label{fig:8box8cheese}
\end{figure}

\section{Stability region of a non-BPS state in heterotic string theory \label{HetRegion}}

In the previous section, we showed that the form of $P_L$ can restrict possible BPS-decay modes. To maximize this effect, we now study a non-BPS object with $P_L  \cdot T_{ } \in \left( {\mathbb{Z} } +\frac{1}{2} \right)^{16}$ for all the eight $T$'s. For example, a non-BPS state with $p_R=p_L=0$  and $ P_{L}  = \left( \left( {\frac{1}{2}}\right) ^{16}\right) $ satisfies  $P_L  \cdot T_{ } \in \left(\pm \frac{1}{2} \right)^{16}$ for all the eight $T$'s, and its decay products must have overlap with all of the these eight groups $\{W_i\}, \{M_i\}, \{M_i, M_j, W_k, W_l\}$ with $\{i,j,k,l\} =\{1,2,3,4\}$ as depicted in Figure \ref{fig:8box8necessary}.

This severe restriction on decay channel comes from a ${{\mathbb{Z} }_2}^8$ parity, where each which ${{\mathbb{Z} }_2}$  determines whether $P_L  \cdot T_{ } \in   {\mathbb{Z} }^{16}$ or $P_L  \cdot T_{ } \in \left( {\mathbb{Z} } +\frac{1}{2} \right)^{16}$. This is reminiscent of ${{\mathbb{Z} }_2}$ symmetry of a SO(32) spinor representation of heterotic string theory in 10d, which does not decay due to conserved charge \cite{SpinorNotDecay}.

The lightest collections of BPS decay products are in following two kinds:
\begin{itemize}
\item $2(M_{a}+W_{a})$ as in Figure \ref{fig:8box4heavy}.
\item $2(W_{a}+M_{b}+M_{c}+M_{d})$ or $2(W_{a}+W_{b}+W_{c}+M_{d})$ with $\left\{ a, b, c, d \right\}  =\left\{ 1, 2, 3, 4 \right\} $ as in Figure
\ref{fig:8box8cheese}.
\end{itemize}
Mass on BPS side of the first kind
\begin{equation}
\frac{1}{R_{a}}+4R_{a} \geq 4>\sqrt{8}
\end{equation}
is always heavier than the original non-BPS state, so this decay is excluded by energy.
This set of heterotic BPS states are dual to two pairs of D-brane and anti-D-brane in type IIA string theory as following:
\begin{itemize}
\item a pair of D2-brane and anti-D2-brane over $x_i, x_j$ {\it and} another pair over $x_k, x_l$
\item a pair of D0-brane and anti-D0-brane {\it and} a pair of D4-brane and anti-D4-brane.
\end{itemize}
The non-BPS brane in type IIA string theory may correspond to a bound state of a non-BPS $\widehat{D1}$-brane and $\widehat{D3}$-brane, which was studied in \cite{GaberdielLecture}.

\begin{figure}
\centerline{\includegraphics[width=.50\textwidth]{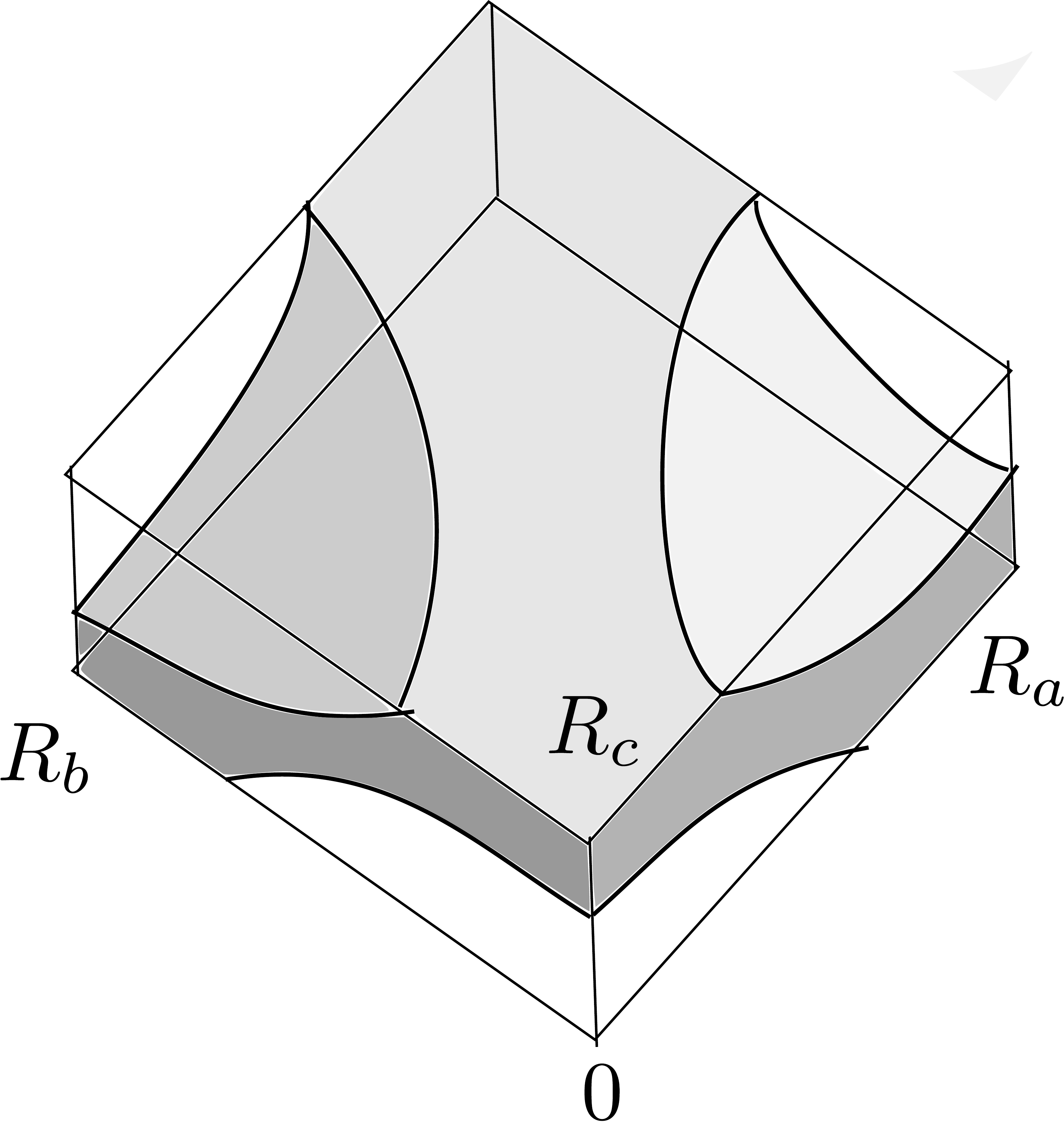}}
\caption[A cheese diagram: a 3d slice of stability region of a non-BPS state of charge $P_L = \left( \left( \pm \frac{1}{2} \right)^{16} \right)$]{A non-BPS state of charge $P_L = \left( \left( \pm \frac{1}{2} \right)^{16} \right)$ is stable inside the shaded object, which looks like a cheese whose every other corner is eaten. Here we drawn a 3d slice of phase structure varying three radii $R_a$, $R_b$, $R_c$ of the $T^4$, fixing   $R_d$ large, $R_c, R_d > \frac{1}{\sqrt{2}}$.

\hskip 1.2cm If they instead chose $R_d <  \frac{1}{2\sqrt{2}}$ to be small, then it will roughly look similar to this, with one of the axis denoting $\frac{1}{4R}$ instead of $R$. As we choose less extreme values for $R_d$, the stability region will get {\it thicker}.}
\label{fig:3Dphase}
\end{figure}

Masses of the second type of decay products sum up to
\begin{eqnarray}
\left(16R_{a}^2+\frac{1}{R_{b}^2}+\frac{1}{R_{c}^2}+\frac{1}{R_{d}^2} \right)^{\frac{1}{2}} \\
\left(16R_{a}^2+16R_{b}^2+16R_{c}^2+\frac{1}{R_{d}^2} \right)^{\frac{1}{2}}
\end{eqnarray}%
respectively\footnote{We thank Matthias Gaberdiel for helping us improve the mass relation by considering a BPS bound state instead adding masses of 4 BPS states separately}.
Therefore, this non-BPS object is exactly stable against decay into BPS states {if and only if} both of following hold:
\begin{eqnarray}
 16R_{a}^2+\frac{1}{R_{b}^2}+\frac{1}{R_{c}^2}+\frac{1}{R_{d}^2} & > & 8 \label{spinor1} \\
 16R_{a}^2+16R_{b}^2+16R_{c}^2+\frac{1}{R_{d}^2} & > & 8 . \label{spinor2}
\end{eqnarray}%
 If we consider the moduli space of $T^4$ as a 4d cube, then among 16 corners, this non-BPS state will be exactly stable in alternate corners and in the connecting region between them. The stability region looks like 4d cheese in a shape of a cube with every other corner (where odd number of radii are large and odd number of radii are small) eaten.  See Figures \ref{fig:2Dphase} and \ref{fig:3Dphase} for the 2d and 3d projection of stability region of this non-BPS state against decay into energetically competing BPS sides $2(W_{a}+M_{b}+M_{c}+M_{d})$ or $2(W_{a}+W_{b}+W_{c}+M_{d})$.  Each of uneaten 8 corners (where even number of radii are large and even number of radii are small) corresponds to where we have one kind of exactly stable non-BPS heterotic string states, which corresponds to a non-BPS $\widehat{D}$-brane (a non-BPS $\widehat{D1}$-brane or a non-BPS $\widehat{D3}$-brane) each along 4 possible choices of directions.
 The allowed modes $2(W_{4}+M_{i}+M_{j}+M_{k})$, $2(W_{i}+W_{j}+W_{k}+M_{4})$, $2(W_{i}+M_{j}+M_{k}+M_{4})$, and $2(W_{4}+W_{i}+W_{j}+M_{k})$ correspond to BPS bound states of D2-branes wrapped over $(x_i, x_l), (x_j, x_l)$ and $(x_k, x_l)$ directions and a D4-brane wrapped over $x_1, x_2, x_3, x_4$ directions, or BPS bound states of D2-branes over $(x_i, x_j), (x_j, x_k), (x_i, x_k)$ and a D0-brane.

A question remaining for future study is determination of
the non-BPS stability region in type IIA string theory. Duality may not be a sufficient test of stability of non-BPS objects. For example, a non-BPS D0-brane is unstable in type IIB string theory, but stable in type I string theory \cite{SenD0I,SenD0Iinteractions,WittenDK} because the orientifold action projects out tachyon modes in type I \cite{WittenDK}.

Non-BPS states made of D-branes in ${{\mathbb{Z} }_2} \times {{\mathbb{Z} }_2}$ orientifolds with torsion are studied in \cite{ShiuTorsionRegion}. The stability region is computed using boundary state formalism and demanding the tachyons to be massless \cite{ShiuTorsionRegion}. The non-BPS stability region delineated by \eqref{spinor1} and \eqref{spinor2} has a similar shape to those computed in \cite{ShiuTorsionRegion,QuirozStefanskiTorsion}.
In type IIA string theory, BPS D2-branes wrapped on 2-cycles of a Calabi-Yau 3-fold are studied, and a similar looking phase diagram appeared by considering decays between non-BPS $\widehat{D1}$-brane and non-BPS $\widehat{D3}$-brane \cite{MajumderSenIIAmoduliCY3}.

\chapter{Metastable vacua of D5-branes and anti-D5-branes \label{Ch:metaIIB}}
\begin{figure}[!h]
\centerline{\includegraphics[width=.70\textwidth]{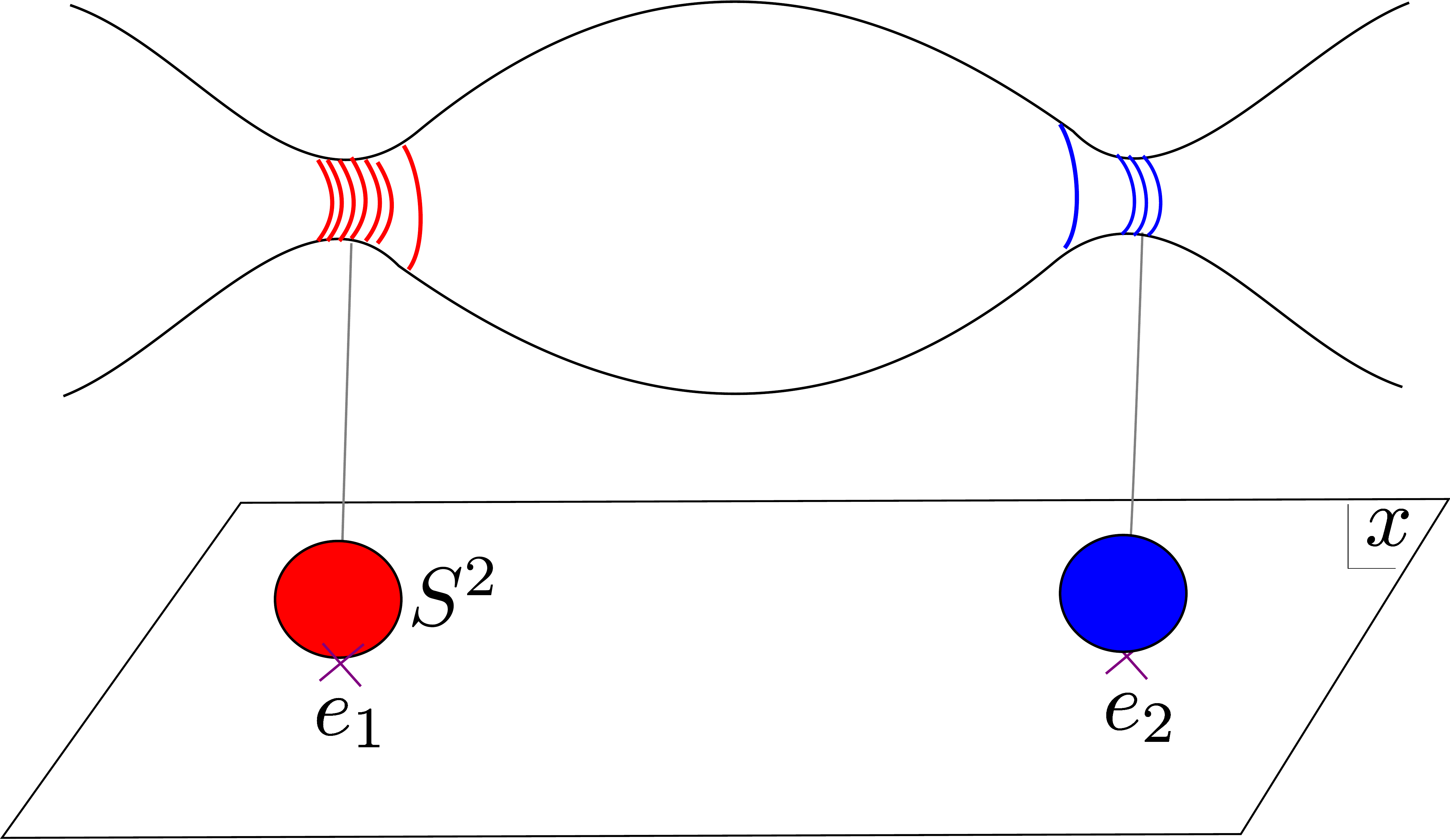}}
\caption[A metastable supersymmetry breaking system of D-branes and anti-D-branes.]{A metastable supersymmetry breaking system made of a stack of \textcolor{red}{D-branes} and another stack of \textcolor{blue}{anti-D-branes} wrapped on 2-spheres which are rigid and separated.}
\label{fig:Open}
\end{figure}

Encouragement by the existence of metastable vacua are generic in
supersymmetric gauge theories \cite{iss},
we now construct metastable supersymmetry breaking vacua in string theory, as suggested in \cite{VafaLargeN}. In this
scenario, we wrap branes and anti-branes on cycles of local Calabi-Yau three-folds, yielding metastability as a consequence of the geometry. The branes and the
anti-branes are wrapped over two separate rigid 2-cycles. The motion required for the branes to annihilate,  costs energy, since
the relevant minimal 2-spheres are rigid. This gives rise to a potential barrier, resulting in metastable configurations, as illustrated in Figure \ref{fig:Open}.

When the number of branes and anti-branes $N$ is large, the brane-antibrane metastable system has a dual description in a low energy effective theory. The dual description is obtained via a geometric transition
in which the 2-spheres shrink, and are subsequently replaced by 3-spheres with fluxes
through them.
This flux spontaneously breaks the
  $\mathcal{N}=2$ supersymmetry to an $\mathcal{N}=1$
subgroup in the case that only branes are present. When only anti-branes are present, we expect supersymmetry to be broken to a
\textit{different} $\mathcal{N}=1$ subgroup. With both branes \textit{and%
} anti-branes, the supersymmetry is completely broken $\mathcal{N}=0$. The vacuum
structure can be analyzed from an effective potential.  %
Unlike in the branes-only cases studied before, one expects to find a meta-stable vacuum which breaks
supersymmetry completely.

Next, restrict further to the cases $N_{1}=-N_{2}$ and $\left\vert
N_{1}\right\vert \gg \left\vert N_{2}\right\vert $ with the branch cuts
aligned along the real axis
of the complex $x$-plane. For sufficiently large 't Hooft
coupling, but far before the cuts touch, the theory undergoes a phase
transition and decay occurs.

The organization of the rest of the chapter is as follows. Section \ref{sec:paper1} presents metastable configurations of brane and anti-branes. Section \ref{sec:paper2} computes the masses to the higher order to identify the decay modes. Section \ref{sec:computeperiod} studies the moduli space of 2-cut geometry.

\begin{figure}[!h]
\centerline{\includegraphics[width=.90\textwidth]{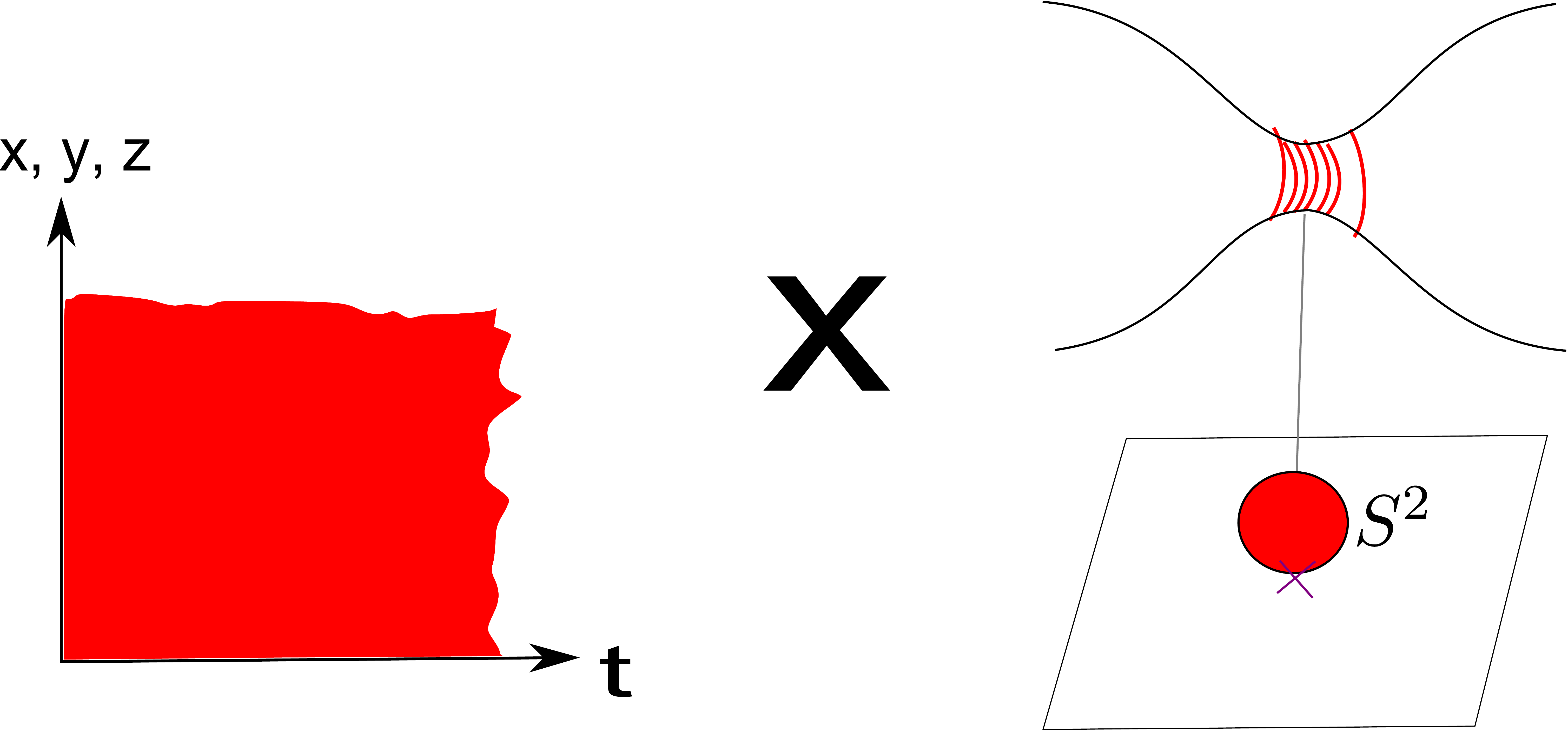}}
\caption[A stack of $N$ D5-branes fill spacetime and wrap a 2-cycle of internal Calabi-Yau three-fold.]{A stack of $N$ \textcolor{red}{D5-branes} fill spacetime and wrap a 2-cycle of internal Calabi-Yau three-fold.}
\label{fig:Mink2cycle}
\end{figure}
\section{Branes and anti-branes on the conifold \label{sec:paper1}}

Consider type IIB string theory with $N$ D5-branes wrapping the $S^{2}$
of a resolved conifold in a local Calabi-Yau three-fold, and the remaining 3+1
dimensions filling the Minkowski spacetime. See Figure \ref{fig:Mink2cycle}.

Type IIB string theory has $\mathcal{N}=2$ in 10-dimension. Compactifying string theory on a 6-dimensional manifold naively yields a theory in $d=4$ with $\mathcal{N}=8$. Choosing the 6-manifold to be a Calabi-Yau 3-fold breaks supersymmetry from $\mathcal{N}=8 $ into  $\mathcal{N}=2$, due to $SU(3)$ holonomy of a Calabi-Yau 3-fold.  D-branes breaks one-half of $\mathcal{N}=2$ and preserves $\mathcal{N}=1$, while adding anti-branes preserves
an orthogonal $\mathcal{N}=1$ subset. If we have multiple conifolds, then we
can put a stack of D-branes on a local conifold, and a stack of anti-D-branes on some other,
getting $\mathcal{N}=0$ system as in Figure \ref{fig:Open}.
\begin{figure}
\centerline{\includegraphics[width=.99\textwidth]{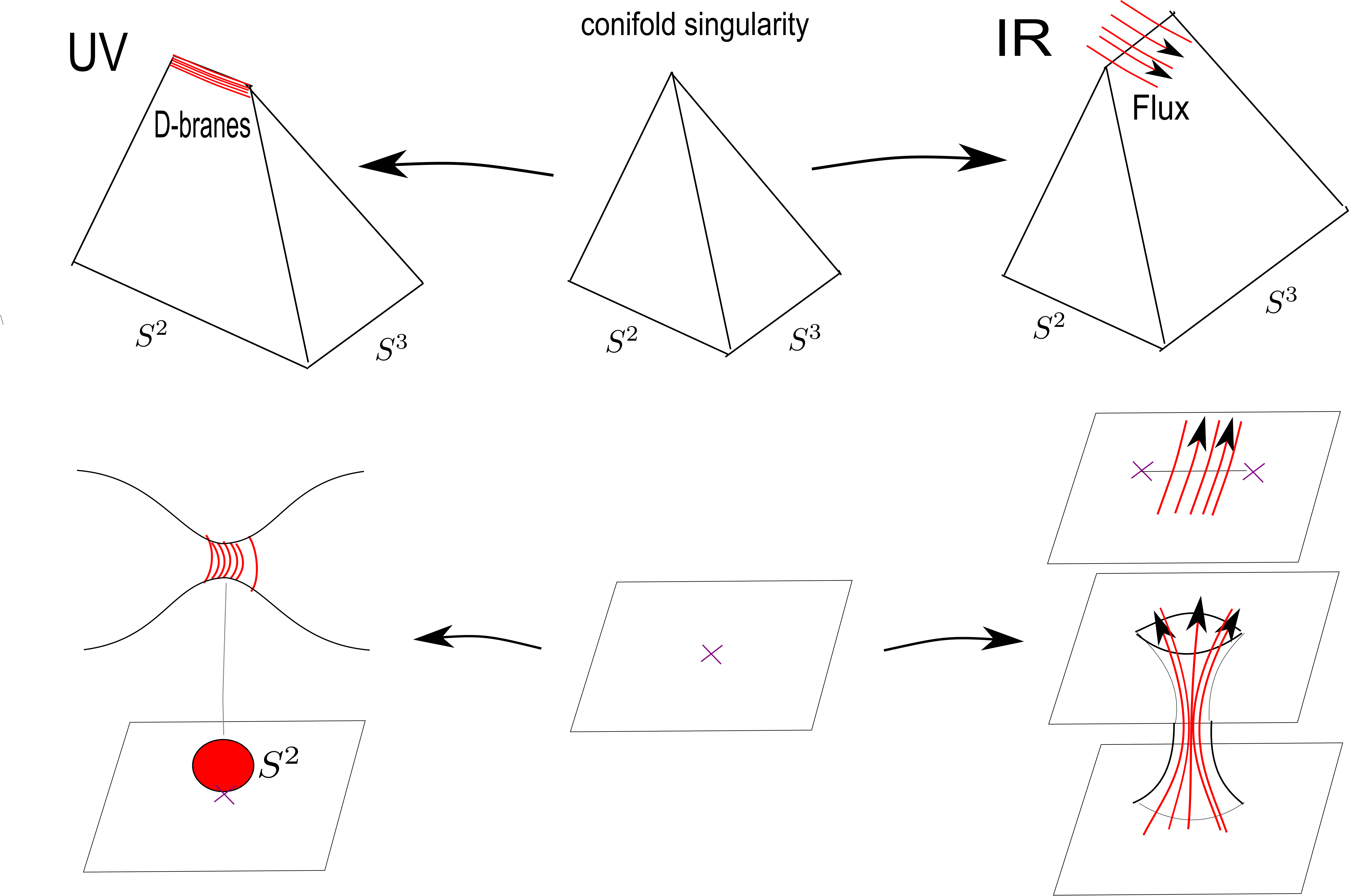}}
\caption[Blow-up (resolution) and complex deformation of a conifold singularity.]{Blow-up (resolution) and complex deformation of a \textcolor{purple}{conifold singularity}. In a UV theory, \textcolor{red}{D-branes} are wrapped over these non-vanishing $S^2$. One can also deform the singularity by complex deformation, which will now open up a new $S^3$ cycle through which \textcolor{red}{RR-flux} pierces in an IR theory.}
\label{fig:1cut}
\end{figure}
These stacks of D-branes and anti-D-branes attract,
as they try to annihilate each other. {In order to meet, they have to increase their volume in $S^2$ directions in the
Calabi-Yau geometry. This requires tension energy, which is proportional to
the volume.}

Guided by this qualitative understanding, we analyze this system in a low energy effective theory in a flux picture, where the branes are replaced by RR-flux. It is then straightforward to compute the effective potential, find its vacuum, and compute the masses and phase structure. Figure \ref{fig:1cut} shows how a conifold singularity is resolved in UV and IR pictures. One can resolve singularity by giving a size to a singular point of vanishing $S^2$. In a UV theory, D-branes are wrapped over these non-vanishing $S^2$. One can also deform the singularity by complex deformation, which will now open up a new $S^3$ cycle through which RR-flux pierces in an IR theory.

\subsection{Local multi-critical geometry}

\begin{figure} [!h]
\centerline{\includegraphics[width=.99\textwidth]{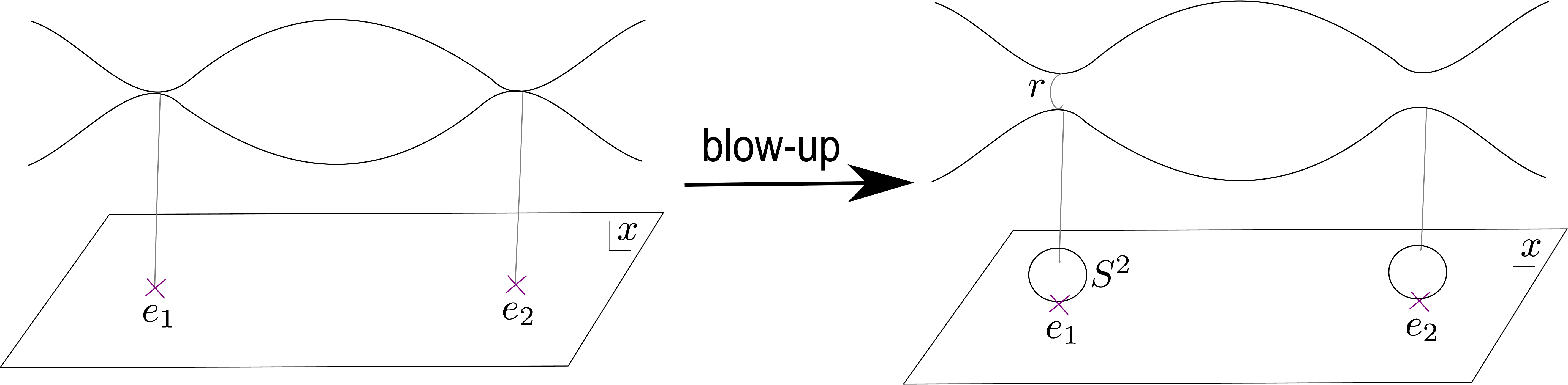}}
\caption[We blow up conifold singularities by giving size $r$ to $S^2$.]{We blow up \textcolor{purple}{conifold singularities} by giving size $r$ to $S^2$.}
\label{fig:BlowUp}
\end{figure}

Consider a Calabi-Yau three-fold given by
\begin{equation}
{\ uv=y^{2}+W^{\prime 2}}  \label{cyb}
\end{equation}%
where
\begin{equation}
W^{\prime }(x)=g\prod_{k=1}^{n}(x-e_{k}).
\end{equation}%
There are $n$ isolated ${S^2}$'s at $x=e_{k}$ whose area is given as
\begin{equation}
{A(x)=(|r|^{2}+\left|W^{\prime }\right|^2)^{1/2}.}  \label{area}
\end{equation}
The geometry in the $n=2$ case is drawn in Figure \ref{fig:BlowUp}

Consider wrapping some number of branes $N_{k}$, $k=1,\cdots n$
on each ${S^2}$. The case when all
the branes are D5-branes (with $N_{k}>0$ for all $k$) was studied in \cite{civ}, giving an $\mathcal{N}=1$ supersymmetric $U(N)$ gauge
theory on branes. The effective coupling constant $g_{\mathrm{YM}}$ of the four
dimensional gauge theory living on the brane is the area of the minimal $%
{S^2}$'s times $1/g_{s}$:
\begin{equation}
\frac{1}{g_{\mathrm{YM}}^{2}}=\frac{|r|}{g_{s}}.
\end{equation}

This chapter studies the case when some of the ${S^2}$'s are wrapped
with D5-branes and others with anti-D5-branes. When the ${S^2}$'s are (or $e_{i}$ are)
 widely separated, the branes and the anti-branes are expected to interact
weakly. However, the system should be
only meta-stable because supersymmetry is broken and there are lower energy
vacua available where some of the branes annihilate. For the branes to annihilate the anti-branes, they have to
climb the potential as in Figure \ref{fig:Open}. We have thus geometrically
engineered a metastable brane-antibrane configuration which breaks supersymmetry. The next subsection considers the large $N$ holographic dual for this system, which is a low energy effective theory at IR with RR-fluxes  replacing D-branes.

\begin{figure} [!h]
\centerline{\includegraphics[width=.60\textwidth]{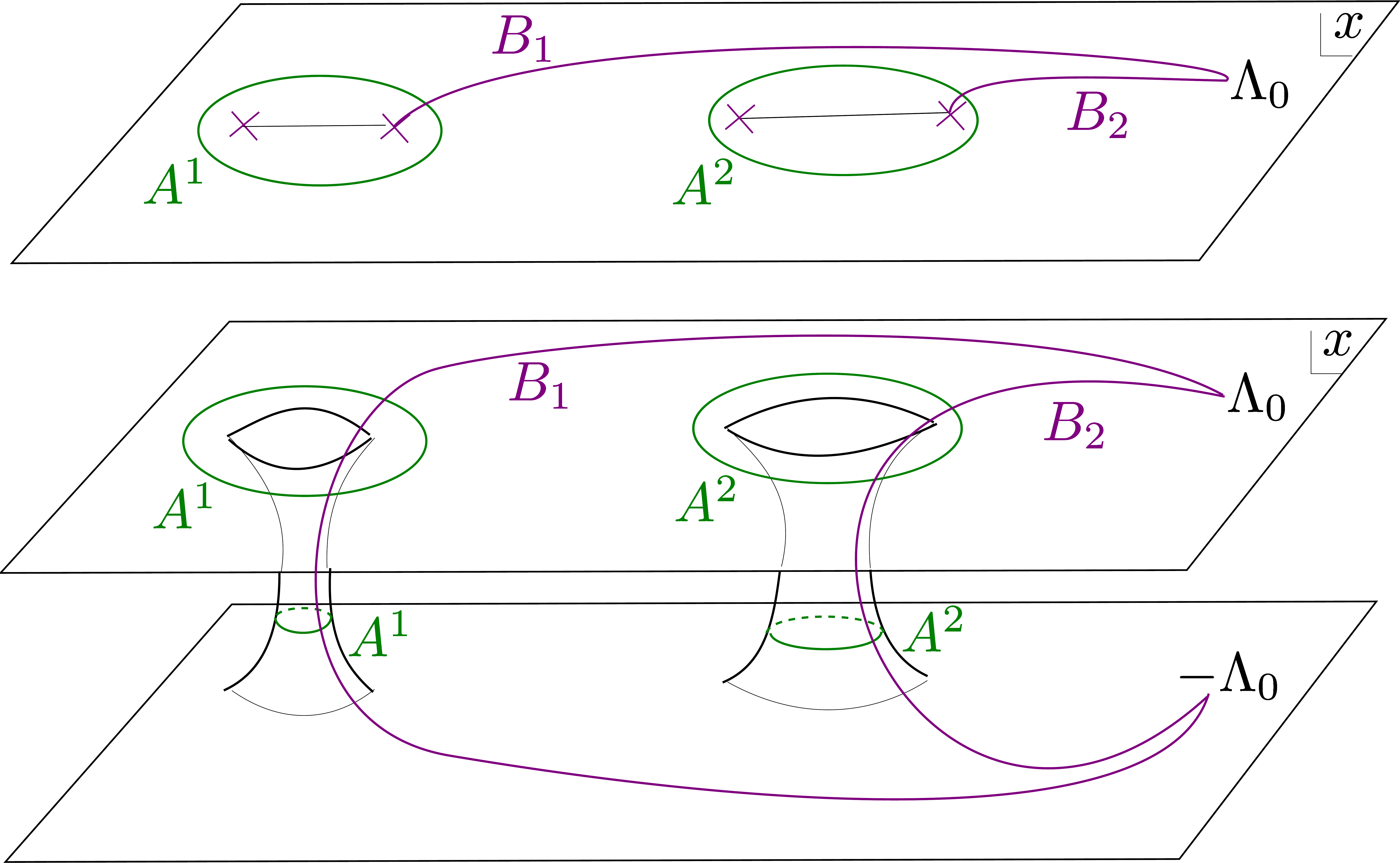}}
\caption[A Calabi-Yau three-fold with two local conifold singularities after complex deformation drawn on $x$-plane.]{A Calabi-Yau three-fold with two local conifold singularities after complex deformation drawn on $x$-plane. The top figure is drawn as a double sheet cover with 2 branch-cuts, while the bottom figure is drawn as a Riemann surface. Three-cycles of Calabi-Yau 3-fold project to 1-cycles on a Riemann surface. Here drawn are \textcolor{green}{compact $A^k$-cycles} and \textcolor{purple}{non-compact $B_k$-cycles}.}
\label{fig:FluxGeomOnly}
\end{figure}

\subsection{The large $N$ dual description}
This section considers the large $N$ limit of such brane/anti-brane
systems and find that the holographically dual closed string geometry is the
identical to the supersymmetric case with just branes, except that some of
the fluxes are negative. This leads, on the dual closed string side, to
a metastable vacuum with spontaneously broken supersymmetry.

The supersymmetric configuration of branes for this geometry was studied in
\cite{civ}, which proposes a large $N$ holographic duality. The relevant
Calabi-Yau geometry was obtained by a geometric transition of \eqref{cyb} whereby
the ${S^2}$'s are blown down and the $n$ resulting conifold
singularities at $x=e_{k}$ are resolved into $S^{3}$'s by deformations of
the complex structure. See Figure \ref{fig:1cut} for a depiction of singularity resolutions of a single conifold. The new complex-deformed geometry is given by
\begin{equation}
{\ uv=y^{2}+W^{\prime 2}+f_{n-1}(x),}  \label{cya}
\end{equation}%
where $f_{n-1}(x)$ is a degree $n-1$ polynomial in $x$. As explained in \cite{civ}, the geometry is effectively described by a Riemann surface which is a
double cover of the $x$-plane, where the two sheets come together along $n$
cuts near $x=e_{k}$ (where the ${S^2}$'s used to be), as in Figure \ref{fig:FluxGeomOnly}. The $A^k$ and $%
B_k $ 3-cycles of the Calabi-Yau three-fold project to 1-cycles on the Riemann surface. The geometry is characterized by the periods of the $(3,0)$ form $%
\Omega $,
\begin{equation}
S_{k}=\oint_{A^{k}}\Omega ,\qquad \partial _{S_{k}}\mathcal{F}%
_{0}=\int_{B_{k}}\Omega . \label{omegaint}
\end{equation}%

 \begin{figure} [!h]
\centerline{\includegraphics[width=.60\textwidth]{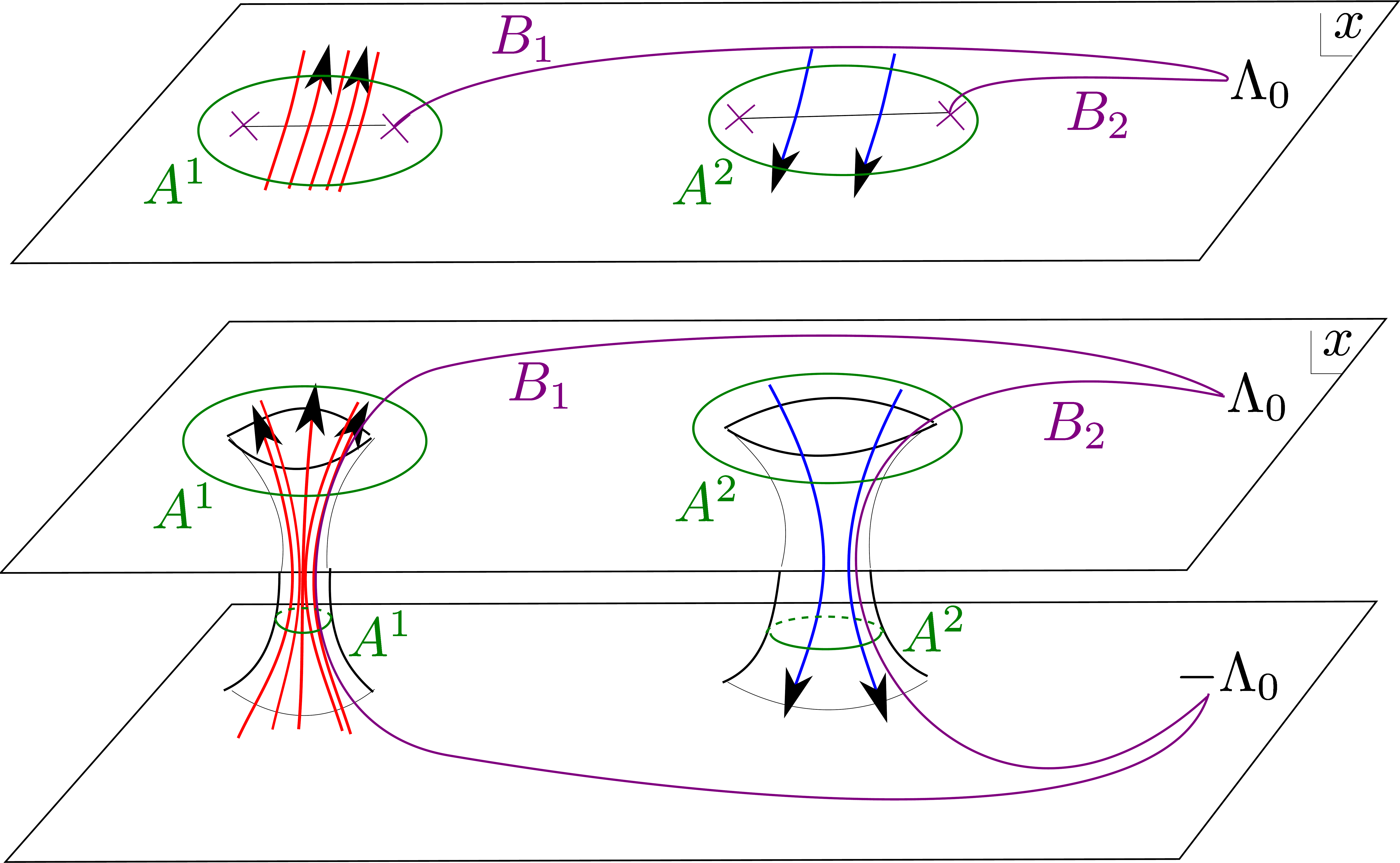}}
\caption[A Calabi-Yau three-fold with two local conifold singularities in a large $N$ holomorphic dual flux picture.]{Amount of RR fluxes \textcolor{red}{$N_1>0$} and \textcolor{blue}{$N_2<0$} through $A^1$ and $A^2$ 3-cycles have {\it mixed} signs.}
\label{fig:Flux}
\end{figure}
\begin{figure}[!h]
\centerline{\includegraphics[width=.90\textwidth]{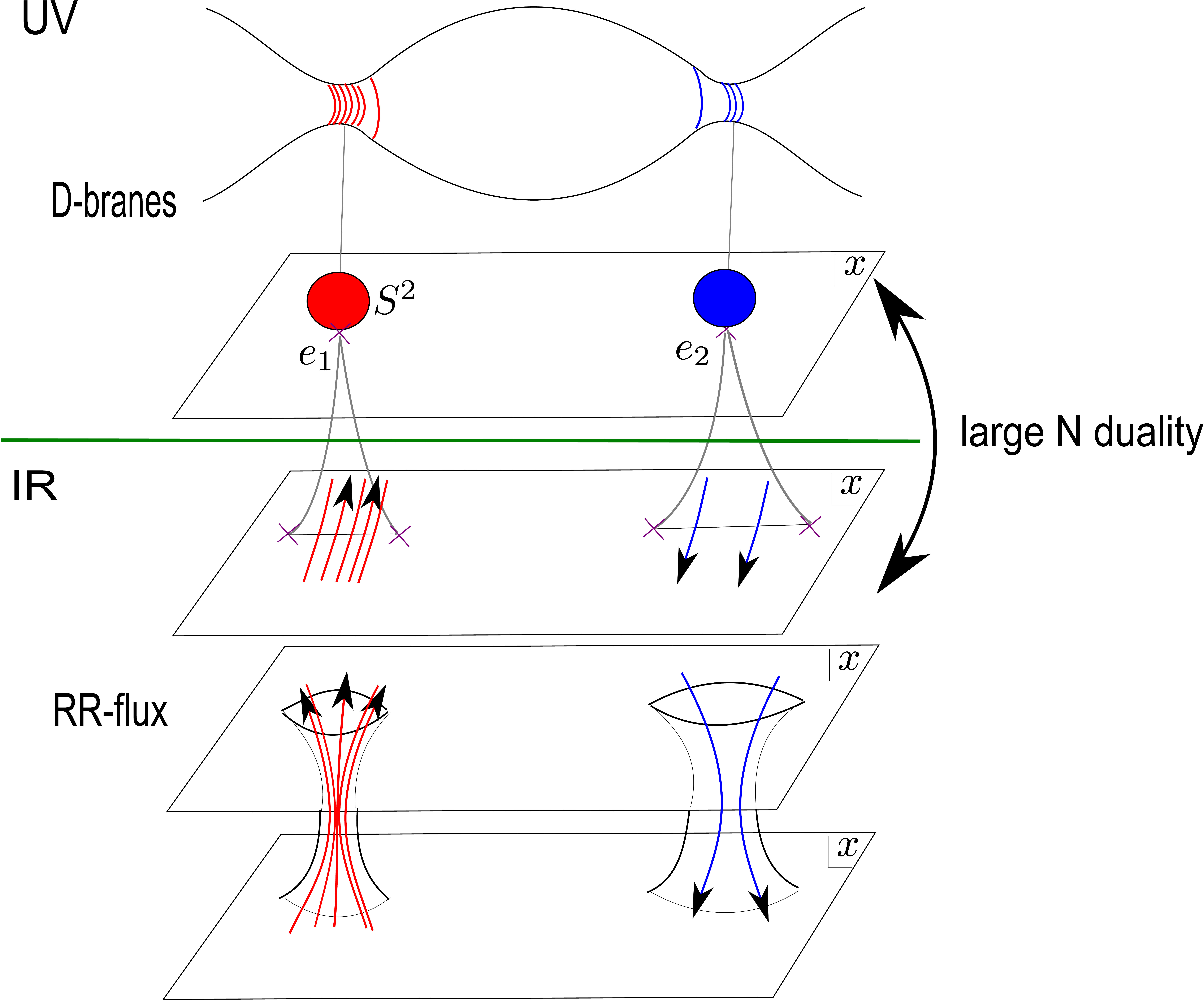}}
\caption[A metastable non-supersymmetric system of branes and anti-branes and its geometric transition]{We have two $S^2$ cycles (drawn as $S^1$ above) in same cohomology. where we wrap brane and anti-branes. Topology allows them to move, but geometry will make it hard, because the brane tension will increase. As the number of branes $N$ increases, there will be a {\it geometric transition}, now $S^2$s disappear and branch cuts will open up into $S^3$ cycles, through which we have RR-flux. The amount of \textcolor{red}{D-branes} (\textcolor{blue}{anti-D-branes}) wrapped along each $S^2$ is now translated into amount of \textcolor{red}{RR-fluxes} (\textcolor{blue}{negative RR-fluxes}) through each $S^3$ cycle.}
\label{fig:2cutmeta}
\end{figure}

If, before the transition, all of the $S^2$'s were wrapped with a
large number of branes, then the holographically dual type IIB string theory is
given by the geometry of \eqref{cya}, where the branes from before the transition
are replaced by fluxes
\begin{equation}
\oint_{A^{k}}H=N_{k},\qquad \int_{B_{k}}H=-\alpha .
\end{equation}%
We conjecture that this large $N$ duality holds even when $
N_{k} $'s have mixed signs as in Figure \ref{fig:Flux} .
The flux numbers $N_{k}$ are positive or
negative depending on whether D5-branes or anti-D5-branes wrap
the $k$'th $S^2$ before the transition, as in Figure \ref{fig:2cutmeta}. The flux through the $B_{k}$ cycles corresponds to the bare gauge coupling constant on the
D-branes wrapping the corresponding ${S^2}$. It is independent of $k$, since the ${S^2}$'s are all in the same homology class.
Turning on RR-fluxes generates a superpotential \cite{GVW}
\begin{equation}
\mathcal{W}=\int H\wedge \Omega ,
\end{equation}%
\begin{equation}
{\mathcal{W}(S)=\sum_{k}\alpha S_{k}+N_{k}{\partial _{S_{k}}}\mathcal{F}_{0}.%
}  \label{superpotential}
\end{equation}

In the supersymmetric case studied in \cite{civ}, the coefficients of the
polynomial $f(x)$ determine the dual geometry and the sizes of $S_{i}$ are
fixed by the requirement that
\begin{equation}
\partial _{S_{k}}W(S)=0,
\end{equation}%
giving a supersymmetric holographic dual. In the case of interest
for us, with mixed fluxes, we do not expect to preserve supersymmetry.
Instead we should consider the physical potential $V(S)$ and find the dual
geometry by extremizing
\begin{equation}
\partial _{S_{k}}V(S)=0,
\end{equation}%
which we expect to lead to a metastable vacuum. The effective potential $V$
is given in terms of the special geometry data and the flux quanta
\begin{equation}
V=g^{S_{i}{\bar{S}}_{j}}\,{\partial _{S_{i}}}\mathcal{W}\,\overline{{%
\partial _{S_{j}}\mathcal{W}}} \label{Veff}.
\end{equation}%
Here the K\"ahler metric is given by $g_{i{\bar{j}}}=\mathrm{Im}(\tau _{ij})$, where $\tau _{ij}$ is
the period matrix of the Calabi-Yau three-fold
\begin{equation}
\tau _{ij}={\partial _{S_{i}}\partial _{S_{j}}}\mathcal{F}_{0}.
\end{equation}%
As explained in \cite{absv}, the flux breaks $\mathcal{N}=2$ supersymmetry in a
rather exotic way. Namely, $which$ $\mathcal{N}=1$ is preserved off-shell
turns out to be a choice of a \textquotedblleft gauge\textquotedblright : one
can write the theory in such a way to manifest either the brane-type or the anti-brane-type of $%
\mathcal{N}=1$ supersymmetry. On shell, however, we have no such
freedom, and only one $\mathcal{N}=1$ supersymmetry can be preserved. Which
one this is depends only on whether the flux is positive or negative, and
not on the choice of the $\mathcal{N}=1$ supersymmetry made manifest by the
Lagrangian. The action $\mathcal{L}$ can be written in terms of $\mathcal{N}%
=2$ superfields, where turning on fluxes in the geometry corresponds to
giving a vacuum expectation value to some of the $\mathcal{N}=2$ F-terms \cite{APT}. Since $\mathcal{N}=2$ is softly broken by the flux terms, we conjecture
that the special K\"ahler metric is unaffected at the string tree level, but
it should be modified at higher string loops.

\subsection{The case of $2$ $S^3$'s \label{2cutmetasubsec}}

For simplicity, consider the case of just two $S^{3}$'s.
Before the transition, there are two shrinking ${S^2}$'s at $%
x=e_1, e_2$, as shown in Figure \ref{fig:BlowUp}. Let $\Delta $ denote the distance between them,
\begin{equation}
\Delta =e_{1}-e_{2}.
\end{equation}%
The theory has different vacua depending on the number of branes placed on
each ${S^2}$. The vacua with different brane/antibrane
distributions are separated by energy barriers due to brane tension. To overcome these barriers, the
branes must first become more massive.

The effective superpotential of the dual geometry, coming from the electric
and magnetic Fayet-Iliopoulos terms turned on by the fluxes, is
\begin{equation}
\mathcal{W}(S)=\alpha (S_{1}+S_{2})+N_{1}{\partial _{S_{1}}}\mathcal{F}%
_{0}+N_{2}{\partial _{S_{2}}}\mathcal{F}_{0} ,
\end{equation}
and the $B_k$-periods have been computed in terms of $A^k$-periods, $S_k$, explicitly in \cite{civ}.

To leading order, dropping the
quadratic terms in the $\frac{S_{k}}{g\Delta ^{3}}$'s, and higher, $\tau_{ij}$ matrix elements are given by:
\begin{eqnarray}
{\ 2\pi i\tau _{11}} &=&2\pi i\,{\partial _{S_{1}}^{2}}\mathcal{F}%
_{0}\approx \log (\frac{{{{S_{1}}}}}{{{{g\Delta ^{3}}}}})-\log (\frac{{{{%
\Lambda _{0}}}}}{{\Delta }})^{2} , \\
{2\pi i\tau _{12}} &=&2\pi i\,{\partial _{S_{1}}\partial _{S_{2}}}\mathcal{F}%
_{0}\approx -\log (\frac{{{{\Lambda _{0}}}}}{{\Delta }})^{2} , \\
{2\pi i\tau _{22}} &=&2\pi i\,{\partial _{S_{2}}^{2}}\mathcal{F}_{0}\approx
\log (\frac{{{{S_{2}}}}}{{{{g\Delta ^{3}}}}})-\log (\frac{{{{\Lambda _{0}}}}%
}{{\Delta }})^{2}.
\end{eqnarray}%
In particular, note that at the leading order ${\tau }_{12}$ is independent
of the $S_{i}$, so we can use $\tau _{ii}$ as variables. The physical high-energy cutoff
$\Lambda _{0}$ is used to compute periods of the non-compact $B_k$ cycles.
 It follows
that the minima of the potential occur when
\begin{eqnarray}
Re(\alpha )+Re(\tau )_{ij}N^{j} &=&0 , \\
\mathrm{Im}(\alpha )+\mathrm{Im}(\tau )_{ij}|N^{j}| &=&0.
\end{eqnarray}%
For example, with branes on the first ${S^2}$ and anti-branes on
the second, one has $N_{1}>0>N_{2}$, and the metastable vacuum solution is
\begin{equation}
{\ \left\vert {{S}_{1}}\right\vert =g\Delta ^{3}\;(\frac{{{{\Lambda _{0}}}}}{%
{\Delta }})^{2}}\overline{{(\frac{\Lambda _{0}}{\Delta })}}{^{2|\frac{N_{2}}{%
N_{1}}|}\;e^{-2\pi i\alpha /|N_{1}|},\qquad \left\vert {{S}_{2}}\right\vert =%
{\ g\Delta ^{3}}(\frac{\Lambda _{0}}{\Delta })^{2}}\overline{{(\frac{\Lambda
_{0}}{\Delta })}}{^{2|\frac{N_{1}}{N_{2}}|}e^{2\pi i{\ }\overline{{\alpha }}%
/|N_{2}|}.}  \label{nonsusy}
\end{equation}%
with its potential energy is given by
\begin{equation}
{\ V_{\ast }^{+-}=\frac{{8\pi }}{{g_{YM}^{2}}}\,(|N_{1}|+|N_{2}|)-}\frac{{2}%
}{{\pi }}{{|N_{1}||N_{2}|}\;\log |\frac{\Lambda _{0}}{\Delta }|^{2}}.
\label{pamin}
\end{equation}%
The first term, in the holographic dual, corresponds to the tensions of the
branes. The second term should correspond to the Coleman-Weinberg one loop
potential, which is generated by zero point energies of the fields. This
interpretation coincides nicely with the fact that this term is proportional
to $|N_{1}||N_{2}|$, and thus comes entirely from the $1-2$ sector of open
strings with one end on the branes and the other on the anti-branes. The
fields in the $1-1$ and $2-2$ sectors (with both open string endpoints on the
same type of branes) do not contribute terms proportional to $N_i^2$ to \eqref{pamin}, as those sectors are
supersymmetric and the boson and fermion contributions cancel. For comparison, in the case of where both $%
{S^2}$'s were wrapped by D5-branes, the potential at the critical
point $V_{\ast }^{++}$ equals
\begin{equation}
V_{\ast }^{++}=\frac{{8\pi }}{{g_{YM}^{2}}}\,(|N_{1}|+|N_{2}|)=V_{\ast }^{--}
\end{equation}%
and is the same as for all anti-branes. This comes as no surprise, since the
tensions are the same, and the interaction terms cancel since the theory is
now truly supersymmetric.

We now consider the masses of bosons and fermions in the brane/anti-brane
background. With supersymmetry broken, there is no reason to expect pairwise degeneracy of the four
real boson masses, which come from the fluctuations of $S_{1,2}$ around
the vacuum. The four bosonic masses are given by
\begin{equation}
\left( m_{\pm }(c)\right) ^{2}=\frac{{(a^{2}+b^{2}+2abcv)\pm \sqrt{%
(a^{2}+b^{2}+2abcv)^{2}-4a^{2}b^{2}(1-v)^{2}}}}{{2(1-v)^{2}}} \label{absvmass}
\end{equation}%
where $c$ takes values $c=\pm 1$, and
\begin{equation}
a\equiv \left\vert \frac{{N_{1}}}{{2\pi \Lambda _{1}^{3}\mathrm{Im}\tau _{11}%
}}\right\vert ,\qquad b\equiv \left\vert \frac{{N_{2}}}{{2\pi \Lambda
_{2}^{3}\mathrm{Im}\tau _{22}}}\right\vert , \qquad v\equiv \frac{{(\mathrm{Im}\tau _{12})^{2}}}{{r\mathrm{Im}\tau _{11}\mathrm{%
Im}\tau _{22}}}.
\end{equation}%
Indeed this vacuum is metastable, because all the masses squared are strictly positive. This
 follows from the above formula and the fact that $v<1$ in the regime
of interest $|S_{i}/g\Delta ^{3}|<1$. This is a nice check on our
holography conjecture, as the brane/anti-brane construction was clearly
metastable.
Moreover, we see that there are four real bosons, whose masses
are generically non-degenerate, as expected for the spectrum with broken
supersymmetry.

Since supersymmetry is completely broken from $\mathcal{N}=2$ to $\mathcal{N}%
=0$, we expect to find $2$ massless Weyl fermions, which are the Goldstinos.
Masses of the fermions are computed and we indeed find two massless fermions. Since supersymmetry is broken these are interpreted as
the Goldstinos. There are also two massive fermions, with masses
\begin{equation}
m_{\mathrm{f1}}=\frac{{a}}{{1-v}},\qquad m_{\mathrm{f2}}=\frac{{b}}{{1-v}}%
.\qquad
\end{equation}%
Note that $v$ controls the strength of supersymmetry breaking. In particular
when $v\rightarrow 0$ the 4 boson masses become pairwise degenerate and
agree with the two fermion masses $a$ and $b$, as expected for a pair of $%
\mathcal{N}=1$ chiral multiplets.

The mass splitting between bosons and fermions is a measure of the
supersymmetry breaking. In order for supersymmetry breaking to be weak,
these splittings have to be small. There are two natural ways to make
supersymmetry breaking small. One way is to take the number of
anti-branes to be much smaller than the number of branes, and the other way is
to make the branes and anti-branes be very far from each other.

\section{Breakdown of metastability\label{sec:paper2}}

This section considers a 2-cut geometry of subsection \ref{2cutmetasubsec}, and identifies location and mode of the eventual decay by computing masses of metastable vacua.  Two loop contributions to effective potential $V_{\mathrm{eff}}$
\ generate a preferred confining vacuum which aligns the phases of the
glueball fields.  In the closed string
dual this preferred vacuum corresponds to a\ configuration where the branch
cuts align along a common axis. We restrict to the case where branch cuts are along the real axis, small, and far apart.

Start with a heuristic
derivation of the value of the 't Hooft coupling for which we expect higher
order corrections to $V_{\mathrm{eff}}$ to lift the metastable vacua present
at weak coupling. \ Recall from equation (\ref{pamin}) that the
leading order energy density of the brane/anti-brane system is:%
\begin{equation}
E^{(0)}=\frac{8\pi}{g_{\mathrm{YM}}^{2}}\left( \left\vert N_{1}\right\vert
+\left\vert N_{2}\right\vert \right) -\frac{2}{\pi}\left\vert
N_{1}\right\vert \left\vert N_{2}\right\vert \log\left\vert \frac{\Lambda_{0}%
}{\Delta}\right\vert ^{2}\mbox{\tiny .}
\end{equation}
The first term corresponds to the bare tension of the branes and the second
term corresponds to the Coulomb attraction between the branes.

When $\left\vert N_{1}\right\vert \gtrsim\left\vert
N_{2}\right\vert $ and $N_{1}>0>N_{2}$, one has:
\begin{equation}
E^{(0)}\geq\frac{8\pi}{g_{\mathrm{YM}}^{2}}\left( N_{1}+N_{2}\right).
\label{ebounded}
\end{equation}
Loss of metastability is expected precisely when the
Coulomb attraction contribution to the energy density becomes comparable to
the bare tension of the branes. \ This is near the regime where $E^{(0)}$ is
close to saturating inequality (\ref{ebounded}). \ This yields the following
estimate for the breakdown of metastability:
\begin{equation}
\frac{1}{g_{\mathrm{YM}}^{2}\left\vert N_{1}\right\vert }\sim\log\left\vert
\frac{\Lambda_{0}}{\Delta}\right\vert ^{2}  \label{firstestimate} ,
\end{equation}
where all factors of order unity are omitted.

This breakdown in metastability is calculable
near the semi-classical expansion point, when restricted consideration to flux configurations which produce metastable vacua with
the branch cuts aligned along the real axis of the complex $x$-plane.

\subsection{Masses and the mode of instability: $N_{1}=-N_{2}$}

In the case that $N_{1}=-N_{2}\equiv N$ and $\theta _{\mathrm{YM}}=0$, the lowest-energy metastable
vacuum corresponds to two equal size branch cuts
aligned along the real axis of the complex $x$-plane.

  Here we compute the bosonic masses in order to
search for the mode of instability. \   We now show that the unstable mode of the system corresponds to the cuts
remaining equal in size and expanding towards each other. All the other modes are stable up to this
point. These facts are established by computation of
the bosonic mass spectrum:%
\begin{align}
m_{{\rm RA}}^{2} & =\frac{a^{2}}{1-v}+2a\left\vert N\right\vert \left( -\frac{10%
}{1+\sqrt{v}}+\frac{7}{(1-v)\pi\func{Im}\tau_{11}}\right)  \label{RA} \\
m_{{\rm RS}}^{2} & =\frac{a^{2}}{1-v}+2a\left\vert N\right\vert \left( -\frac{10%
}{1-\sqrt{v}}+\frac{7}{(1-v)\pi\func{Im}\tau_{11}}\right) \\
m_{{\rm IS}}^{2} & =\frac{a^{2}}{\left( 1+\sqrt{v}\right) ^{2}}+2a\left\vert
N\right\vert \left( \frac{10}{1+\sqrt{v}}+\frac{-3}{\left( 1+\sqrt {v}%
\right) ^{2}\pi\func{Im}\tau_{11}}\right) \\
m_{{\rm IA}}^{2} & =\frac{a^{2}}{\left( 1-\sqrt{v}\right) ^{2}}+2a\left\vert
N\right\vert \left( \frac{10}{1-\sqrt{v}}+\frac{17}{\left( 1-\sqrt {v}%
\right) ^{2}\pi\func{Im}\tau_{11}}\right)  \label{IA} .
\end{align}
Here in the above, ${\rm RA}$ denotes the real anti-symmetric mode corresponding
to both $S_{i}$'s real with one cut growing while the other shrinks, ${\rm RS}$
denotes the real symmetric mode corresponding to both $S_{i}$'s real with
both cuts growing in size together, and ${\rm IS}$ and ${\rm IA}$ are similarly defined
for the imaginary components of the $S_{i}$'s. \ Further, we have introduced
the parameters:%
\begin{equation}
a=\frac{|N|}{2\pi t\func{Im}{\tau}_{11}}\mbox{\tiny , }v=\frac {\func{Im}{{%
\tau}_{12}}^{2}}{\func{Im}{\tau}_{11}\func{Im}{\tau}_{22}}.
\end{equation}
 \ In equations
(\ref{RA}-\ref{IA}), the term proportional to $a^{2}$ corresponds to the
leading order contribution to the masses squared computed in \eqref{absvmass},
and the term proportional to $2a\left\vert N\right\vert $ corresponds to the
two loop correction to this value. \ As expected from symmetry, we find that as a function of $\left\vert N/\alpha\right\vert $, $%
m_{RS}^{2}$ approaches zero.

It is also of interest to consider the difference in masses between the
bosonic and fermionic fluctuations dictated by the underlying $\mathcal{N}=2$
structure of the theory. \ We find that the masses of the fermions naturally
group into two sets of values. \ At leading order in $1/N$, the $\mathcal{N}%
=2$ supersymmetry of the theory is spontaneously broken. \ This indicates
the presence of two massless goldstinos. \ Labeling the fermionic
counterparts of the gauge bosons and the $S_{i}$'s respectively by $%
\psi_{A}^{(i)}$ and $\psi_{S}^{(i)}$, we find that when $N_{1}=-N_{2}$, the
non-zero masses of the canonically normalized fermionic fields are all equal
and given by the value:%
\begin{equation}
|m_{\psi}|=\frac{a}{\left( 1-v\right) }+\left\vert N\right\vert \frac{7+10%
\sqrt{v}}{1-v}  \label{fermionmass}.
\end{equation}
As before, the first term corresponds to the leading order mass and
the second term is the two loop correction to this value.

We find more generally that for vacua which satisfy $S_{1}=-\overline{S_{2}}$%
, the system develops an instability at a similar value of $\left\vert
N/\alpha \right\vert $. \ In this case, the mode of instability causes the
cuts to expand in size and rotate towards the real axis of the complex $x$%
-plane. \ This is in agreement with the physical expectation that the flux
lines annihilate most efficiently when the branch cuts are aligned along the
real axis.

\subsection{Breakdown of metastability: $\left\vert N_{1}\right\vert
\gg\left\vert N_{2}\right\vert $\label{ONELARGE}}

We now study the behavior of $V_{\mathrm{eff}}$ for flux configurations with
$\left\vert N_{1}\right\vert \gg \left\vert N_{2}\right\vert $ and $\theta _{%
\mathrm{YM}}=0$, also requiring that $N_{1}$ is small enough for
the two loop approximation of $V_{\mathrm{eff}}$ to be valid. \ In this case,
the modulus $t_{1}\equiv S_{1}/g\Delta ^{3}$ fluctuates much less than $%
t_{2}\equiv -S_{2}/g\Delta ^{3}$. \

It is important to compute the masses squared of the bosonic fluctuations at
the metastable minimum in order to determine the mode of instability for
this flux configuration.   \ We find that the
unstable mode corresponds to the smaller branch cut increasing in size at a
much faster rate than its larger counterpart.

With the kinetic terms of the
Lagrangian density canonically normalized, the $4\times 4$ bosonic mass
squared matrix $m_{\mathrm{Bosonic}}^{2}$ takes the block diagonal form:%
\begin{equation}
m_{\mathrm{Bosonic}}^{2}=A^{(R)}\oplus A^{(I)}
\end{equation}%
where the $A^{(R,I)}$ are $2\times 2$ matrices of the form:%
\begin{equation}
A^{(R,I)}=\left(
\begin{array}{cc}
\frac{\left( \partial _{1}^{(R,I)}+\partial _{2}^{(R,I)}\right) ^{2}V_{%
\mathrm{eff}}}{1+v} & -\frac{\left( \partial _{1}^{(R,I)}-\partial
_{2}^{(R,I)}\right) \left( \partial _{1}^{(R,I)}+\partial
_{2}^{(R,I)}\right) V_{\mathrm{eff}}}{\sqrt{1-v^{2}}} \\
-\frac{\left( \partial _{1}^{(R,I)}-\partial _{2}^{(R,I)}\right) \left(
\partial _{1}^{(R,I)}+\partial _{2}^{(R,I)}\right) V_{\mathrm{eff}}}{\sqrt{%
1-v^{2}}} & \frac{\left( \partial _{1}^{(R,I)}-\partial _{2}^{(R,I)}\right)
^{2}V_{\mathrm{eff}}}{1-v}%
\end{array}%
\right)
\end{equation}%
and $A^{(R)}$ (resp. $A^{(I)}$) corresponds to the mass matrix for the real
(resp. imaginary) components of the $S_{i}$'s. \ In the above we have defined %
\begin{eqnarray}
\partial _{j}^{(R)}&=&\frac{1}{\sqrt{\func{Im}\tau _{jj}}}\frac{\partial }{%
\partial \func{Re}S_{j}}, \qquad v=\frac{\func{Im}{{\tau }_{12}}^{2}}{\mathrm{\func{Im}}{\tau }_{11}\func{Im}{%
\tau }_{22}}, \\
\partial _{j}^{(I)}&=&\frac{1}{\sqrt{%
\func{Im}\tau _{jj}}}\frac{\partial }{\partial \func{Im}S_{j}},\mbox{\tiny  }%
\end{eqnarray}%
and for future use we also introduce:%
\begin{equation}
a=\frac{|N_{1}|}{2\pi t_{1}\func{Im}{\tau }_{11}}\mbox{\tiny
, } \qquad b=\frac{|N_{2}|}{2\pi t_{2}\func{Im}{\tau }_{22}}.
\end{equation}%
In the above expressions the components of $\tau _{ij}$ correspond to their
values at the critical point of $V_{\mathrm{eff}}$ and hereafter will be
treated as constants. \ When $\left\vert N_{1}\right\vert >>|N_{2}|$, the
masses squared and eigenmodes of the block $A^{(R)}$ are %
\begin{align}
m_{\func{Re}S_{1}}^{2}& =\frac{b^{2}}{(1-v)^{2}} & \left( \sqrt{\frac{1-%
\sqrt{v}}{1+\sqrt{v}}},1\right) _{R}& \oplus \left( 0,0\right) _{I} \\
m_{\func{Re}S_{2}}^{2}& =a^{2}-2a\left( 10|N_{2}|-\frac{2|N_{1}|+5|N_{2}|}{%
\func{Im}{\tau }_{11}\pi }\right) & \left( -\sqrt{\frac{1+\sqrt{v}}{1-\sqrt{v%
}}},1\right) _{R}& \oplus \left( 0,0\right) _{I},
\end{align}%
and the masses squared and eigenmodes of the block $A^{(I)}$ are similarly given by
\begin{align}
m_{\func{Im}S_{1}}^{2}& =\frac{b^{2}}{(1-v)^{2}} & \left( \sqrt{\frac{1-%
\sqrt{v}}{1+\sqrt{v}}},1\right) _{I}\oplus & \left( 0,0\right) _{R} \\
m_{\func{Im}S_{2}}^{2}& =a^{2}+2a\left( 10|N_{2}|+\frac{2|N_{1}|+5|N_{2}|}{%
\func{Im}{\tau }_{11}\pi }\right) & \left( -\sqrt{\frac{1+\sqrt{v}}{1-\sqrt{v%
}}},1\right) _{I}\oplus & \left( 0,0\right) _{R}\mbox{\tiny .}
\end{align}%
Grouping the fermions according to the supermultiplet structure inherited
from the $\mathcal{N}=1$ supersymmetry of the branes, the non-zero fermion
masses are %
\begin{align}
m_{\psi _{S}}& =\frac{1}{1-v}\left( a+\frac{2|N_{1}|+5|N_{2}|+10|N_{1}|\frac{%
\func{Im}{\tau }_{12}}{\func{Im}{\tau }_{22}}}{\func{Im}{\tau }_{11}\pi }%
\right) \\
m_{\psi _{A}}& =\frac{1}{1-v}\left( b+\frac{2|N_{2}|+5|N_{1}|+10|N_{1}|\frac{%
\func{Im}{\tau }_{12}}{\func{Im}{\tau }_{11}}}{\func{Im}{\tau }_{22}\pi }%
\right),
\end{align}%
with similar notation to that given above equation (\ref{fermionmass}). \ By
inspection of the above formulae, we see that the two loop correction
increases the difference between the bosonic and fermionic masses already
present at leading order.

Keeping $g_{\mathrm{YM}}^{2}\left\vert N_{2}\right\vert $ fixed, we now
determine the mode which develops an instability as the 't Hooft coupling $%
g_{\mathrm{YM}}^{2}\left\vert N_{1}\right\vert $ approaches the critical
value where the original metastable vacua disappear. \ The determinant of
each block of the mass matrix is:%
\begin{align}
\det A^{(R)}& =\frac{4096\pi ^{8}\log t_{1}\log t_{2}}{%
g_{YM}^{8}t_{1}^{2}t_{2}^{2}\left( \log t_{1}\log t_{2}-\log \frac{|{\Lambda
}_{0}|^{2}}{{|\Delta |}^{2}}\left( \log t_{1}+\log t_{2}\right) \right) ^{6}}
\\
& \times \left(
\begin{array}{l}
\log t_{1}\log t_{2}-20t_{1}\left( \log t_{1}\right) ^{2}\left( \log \frac{|{%
\Lambda }_{0}|^{2}}{{|\Delta |}^{2}}-\log t_{1}\right) \\
-20t_{2}\left( \log t_{2}\right) ^{2}\left( \log \frac{|{\Lambda }_{0}|^{2}}{%
{|\Delta |}^{2}}-\log t_{2}\right) +\cdots%
\end{array}%
\right) \\
\det A^{(I)}& =\frac{4096\pi ^{8}\log t_{1}\log t_{2}}{%
g_{YM}^{8}t_{1}^{2}t_{2}^{2}\left( \log t_{1}\log t_{2}-\log \frac{|{\Lambda
}_{0}|^{2}}{{|\Delta |}^{2}}\left( \log t_{1}+\log t_{2}\right) \right) ^{6}}
\\
& \times \left(
\begin{array}{l}
\log t_{1}\log t_{2}+20t_{1}\left( \log t_{1}\right) ^{2}\left( \log \frac{|{%
\Lambda }_{0}|^{2}}{{|\Delta |}^{2}}-\log t_{1}\right) \\
+20t_{2}\left( \log t_{2}\right) ^{2}\left( \log \frac{|{\Lambda }_{0}|^{2}}{%
{|\Delta |}^{2}}-\log t_{2}\right) +\cdots%
\end{array}%
\right) .
\end{align}%
It follows from the last line of each expression that only $m_{\func{Re}%
S_{1}}^{2}$ or $m_{\func{Re}S_{2}}^{2}$ can vanish. \ Furthermore, because $%
m_{\func{Re}S_{1}}^{2}>>m_{\func{Re}S_{2}}^{2}$, the mode of instability
will cause the smaller cut to expand towards the larger cut. \ For $%
\left\vert N_{1}\right\vert \gg \left\vert N_{2}\right\vert ~$this occurs at
a value of $t_{1}$ given by:%
\begin{equation}
1\sim 20\left( -\log t_{1}+\log \left\vert \frac{{\Lambda }_{0}}{{\Delta }}%
\right\vert ^{2}\right) t_{1}.
\end{equation}%

\newpage

\section{Toward a global phase structure of a 2-cut metastable system \label{sec:computeperiod}}

 So far we have studied the case where two branch cuts are small and far apart
from each other. Already this small region of moduli space exhibits
a rich phase structure. It is interesting to ponder the
global phase structure of this system. For example, one may ask, what happens when two
cuts are near each other, or when their sizes grow? We do not yet know
whether this configuration supports any supersymmetric configuration, let
alone non-supersymmetric cases. In order to study the global phase
structure, one needs to understand the special geometry in the whole moduli
space, and compute the special geometry period there.

The organization of this section is as follows. The subsection \ref{modulispace} studies the structure of the moduli space, focussing on the
properties of the singular points and their six-fold duality. The subsection \ref{computeperiod} performs the integrations
for the period,
while restricting to the case of the real locus.

\subsection{Study of the structure of moduli space \label{modulispace}}

Consider a geometry with local deformed conifolds of \eqref{cya} with $n=2$ denoting number of conifolds.\begin{equation}
{\ uv=y^{2}+W^{\prime 2}+f_{1}(x),}  \label{cya2}
\end{equation}%
with
\begin{equation}
W^{\prime } =g(x-e_1)(x-e_2), \qquad f_{1}(x) = b_1 x + b_0 .
\end{equation}

 As explained in \cite{civ}, the geometry is effectively described by a Riemann surface which is a
double cover of the $x$ plane, where the two sheets come together along two branch-cuts as in Figure \ref{fig:FluxGeomOnly}. The geometry is characterized by the periods of the $(3,0)$ form $
\Omega $ over $A$ and $B$ 3-cycles as in \eqref{omegaint}. Equivalently, the period is computed from the integrating 1-form over corresponding 1-cycles of the Riemann surface as in \cite{civ},
\begin{equation}
dx\sqrt{W^{\prime 2}(x)+f_{1}(x)}=dx\ g\sqrt{%
(x-a_{1})(x-a_{2})(x-a_{3})(x-a_{4})} \label{ellipticintegrand}
\end{equation}%
in various segments over $x$-plane. The points at $x=a_i$ are the endpoints of the branch-cuts.
Widths of the branch-cuts are $2\Delta_{43}$ and $2\Delta_{21}$, and $I$ denotes the distance between centers of branch-cuts. They are related by
\begin{equation}
\left( a_{1},a_{2},a_{3},a_{4}\right) =\left( -\Delta _{21}-\frac{I}{%
2},\Delta _{21}-\frac{I}{2},-\Delta _{43}+\frac{I}{2},\Delta _{43}+\frac{I}{2%
}\right),
\end{equation}
and we have following relations
\begin{eqnarray}
 \frac{I^{2}}{2}-\frac{\Delta ^{2}}{2}+\Delta _{21}^{2}+\Delta _{43}^{2} =0
\label{ellipticVar} \\
b_{1}=I(\Delta _{21}^{2}-\Delta _{43}^{2})\sim \left( S_{1}+S_{2}\right) .
\end{eqnarray}

In previous sections we used the periods computed in \cite{civ} for the small region of moduli space where $\left\vert \Delta
_{21},\Delta _{43}\right\vert \ll \left\vert I,\Delta \right\vert $. These
periods are written in terms of two small expansion parameters $S_1 \sim
\Delta _{21}/I$ and $S_2\sim \Delta _{43}/I$.
To consider a global phase structure, one needs to compute the periods beyond the
region of $\left\vert \Delta _{21}, \Delta _{43}\right\vert \ll \left\vert
I\right\vert$, so that we can obtain expressions for
superpotential and effective physical potential. We would like to discover
 whether there still exist supersymmetric and
non-supersymmetric vacua, and what kind of stability they have.

\begin{figure} [!b]
\centerline{\includegraphics[width=.6\textwidth]{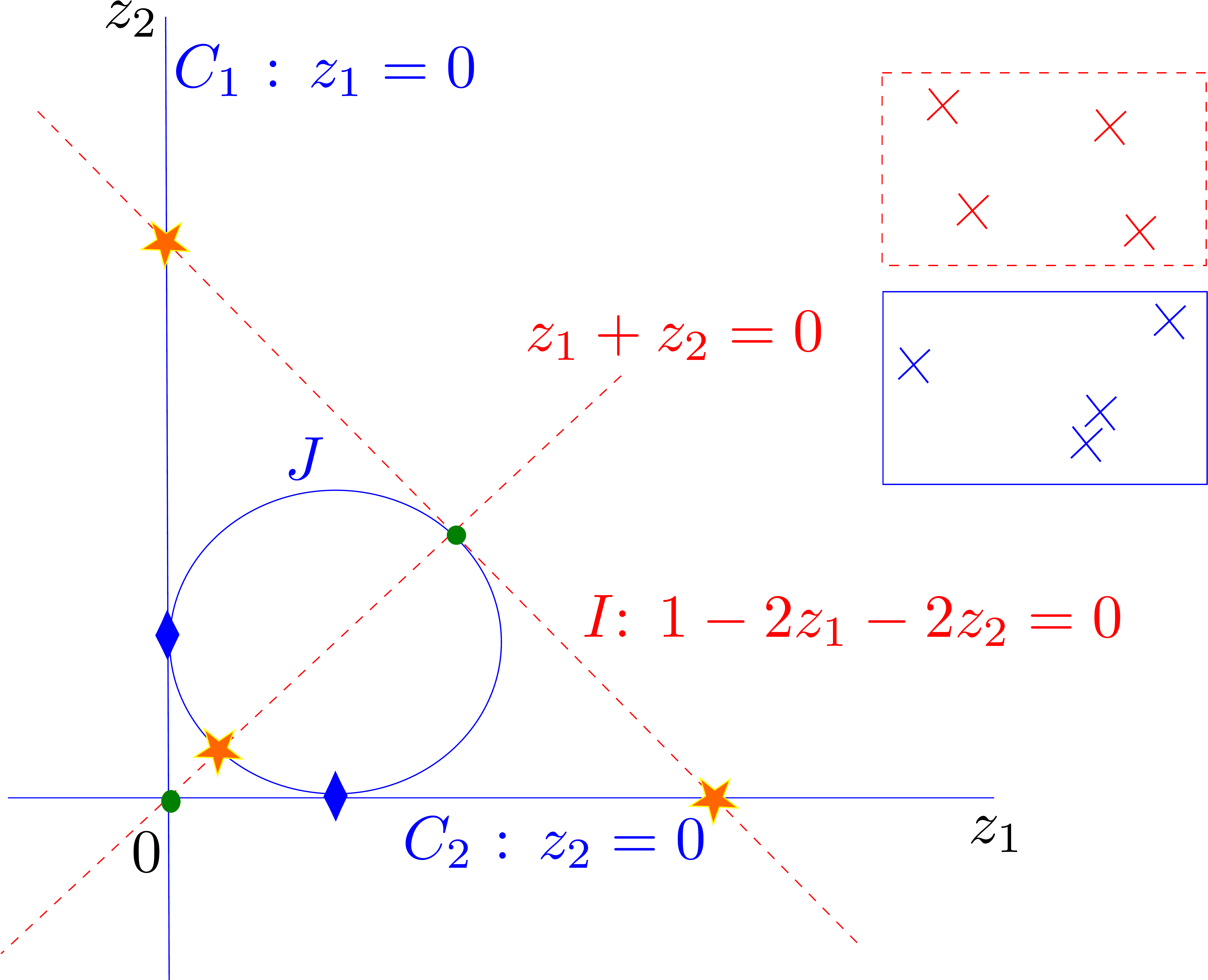}}
\caption[Moduli space and singular divisors of 2-cut geometry]{Moduli space and singular divisors of 2-cut geometry with $\Delta=1$. Under repairing of the endpoints of two branch cuts, some divisors will be identified with each other. Two sets of equivalent singular divisors of 2-cut geometry $\{ C_{i} , C_{j},  J \} $ (\textcolor{blue}{solid}) and $ \{ I ,z_{1}=z_{2} \}$ (\textcolor{red}{dashed}) are drawn.

\hskip 1.2cm The intersection points of divisors also get identified among themselves. They are grouped into following three sets of points:
\begin{description}
 \item[\textcolor{green}{\CircleSolid }] $ (z_1,z_2)=(0,0), \left(\frac{1}{4},\frac{1}{4}\right)$
  \item[\textcolor{orange}{\FiveStar }]  $(z_1,z_2)=\left(\frac{1}{8},\frac{1}{8}\right) , \left(0,\frac{1}{2}\right), \left(\frac{1}{2},0\right)$
 \item[\textcolor{blue}{\DiamondSolid }]   $(z_1,z_2)=\left(0,\frac{1}{3}\right),\left(\frac{1}{3},0\right)$
 \end{description}
 }
\label{fig:triangle}
\end{figure}
Before computing the periods, let us examine the moduli space.
With the
variables
\begin{equation}
z_{1}  = \frac{1}{4}\left( a_{2}-a_{1}\right) ^{2}=\Delta
_{21}^{2}, \qquad
 z_{2}  =  \frac{1}{4}\left( a_{4}-a_{3}\right) ^{2}=\Delta _{43}^{2},
  \end{equation}
  there are four divisors (drawn in Figure \ref{fig:triangle}) given by the following formulas
\begin{eqnarray}
C_{1} &:&z_{1}=\frac{1}{4}\left( a_{2}-a_{1}\right) ^{2}=0, \qquad C_{2} :z_{2}=\frac{1}{4}\left( a_{4}-a_{3}\right) ^{2}=0
\\
I &:&I^{2}=\Delta^2-2z_{1}-2z_{2}=0 \\
J &:&J=(a_{1}-a_{2})(a_{2}-a_{3})(a_{3}-a_{4})(a_{1}-a_{4}) =0.
\end{eqnarray}
They satisfy
\begin{eqnarray}
 \sum a_{i}^{2}&=&\Delta^2 ,\qquad I(C_{1}-C_{2})  = b_{1}={\rm constant} \\
\sum a_{i} &=&0, \qquad C_{1}C_{2}J^{2}  =  {\rm constant}.
\end{eqnarray}
The limit of $\Delta \rightarrow 0$ is another divisor $C_{\infty }$,
which corresponds to branch-cuts growing much larger than the separation between
cuts.

We propose following six-fold duality among intersection points of
singular divisors, coming from different ways to choose two branch-cuts by pairing up four possible endpoints of branch-cuts.
Divisor $I$ corresponds to the centers of cuts colliding. Additionally, one
can also consider the $z_{1}=z_{2}$ locus, where two cuts are equal in size and
direction (phase). The four points make a parallelogram, as in the upper box (dashed) of the Figure \ref{fig:triangle}.
Divisors $C_{i}$ correspond to the cuts shrinking to small sizes, bringing $%
a_{2i}, a_{2i-1}$ together. A divisor $J$ corresponds to the branch-cuts touching each
other. Two of the four endpoints will be close to each other, as in the lower box (solid) of the Figure \ref{fig:triangle}.
Under the permutation of $\left( a_{1}, a_{2}, a_{3}, a_{4}\right) $ it follows
easily that
\begin{equation}
C_{i}  \leftrightarrow  C_{j}\leftrightarrow J , \qquad
I  \leftrightarrow  z_{1}=z_{2}
\end{equation}

\subsection{Computation in real locus \label{computeperiod}}

\begin{figure}[!h]
\centerline{\includegraphics[width=.8\textwidth]{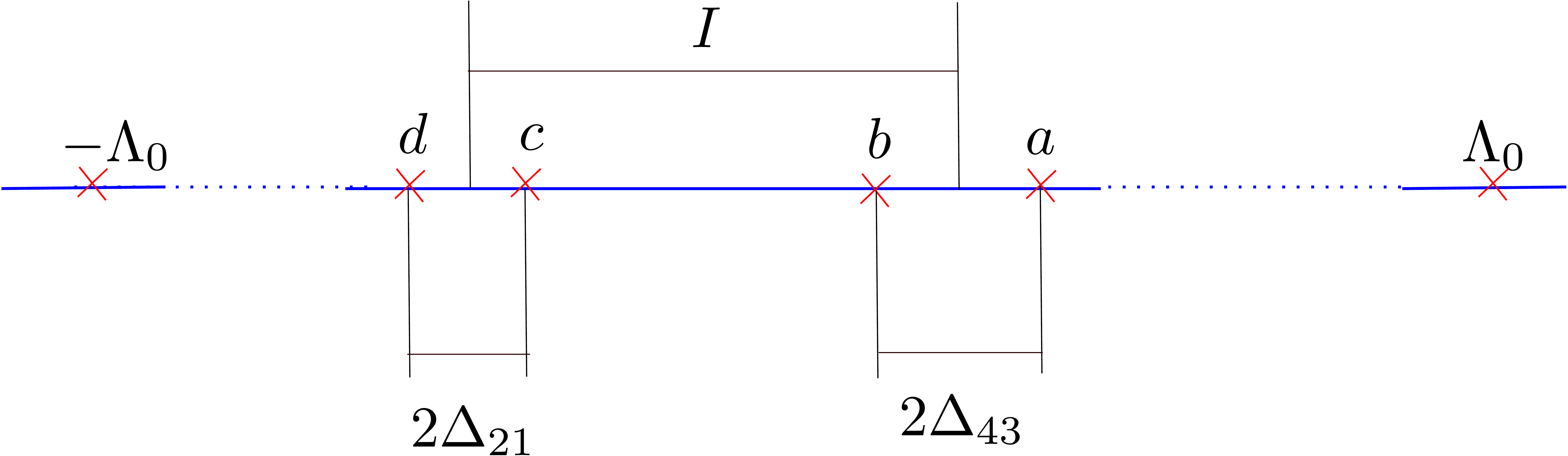}}
\caption{Two branch cuts are aligned along the real axis and separated.}\label{fig:sep}
\end{figure}

This section computes various integrals that are needed to obtain the periods in the
real locus $\left( a_{1},a_{2},a_{3},a_{4}\right) , \Lambda _{0}\in \mathbb{R}$.
First consider the case where two cuts are separated by a distance; with $S_2$ on the right and $S_1$ on the left, as in Figure \ref{fig:sep}. The cut-off scale $\Lambda
_{0}$ is on the far right of the endpoints of branch-cuts, $\Lambda_0 \gg a>b>c>d$, which are ordered as
\begin{equation}   \left( d, c, b, a\right) =\left( -\Delta
_{21}-\frac{I}{2}, \Delta _{21}-\frac{I}{2}, -\Delta _{43}+\frac{I}{2}, \Delta
_{43}+\frac{I}{2}\right).  \label{separate}
\end{equation}%
Without loss of generality one may assume that the first cut is smaller than the second cut: $\Delta _{21} \leq \Delta _{43}$.

\begin{figure}[!h]
\centerline{\includegraphics[width=.8\textwidth]{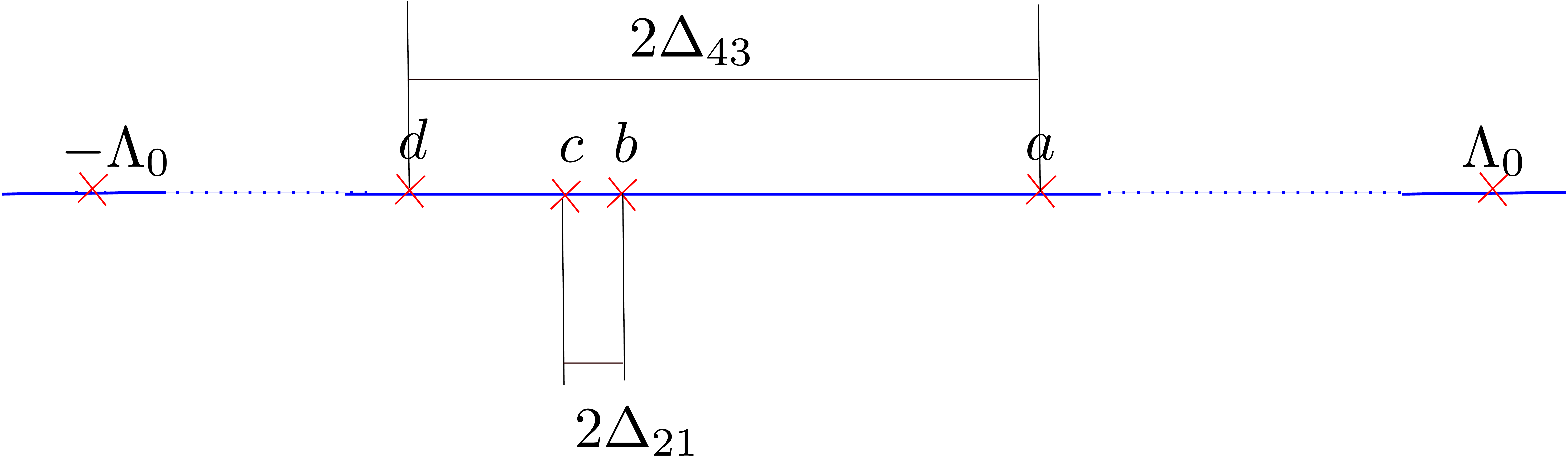}}
\caption{One small branch cut is living inside a larger branch cut.}\label{fig:ins}
\end{figure}

Now compute integrals over compact cycles and then over a non-compact cycle between the
cutoff $\Lambda _{0}$ and $S_2$ as in \eqref{ellipticintegrand}
\begin{eqnarray}
I_{R} &\equiv &\int_b^{a} dx \sqrt{|(x-a)(x-b)(x-c)(x-d)|} \\
I_{L} &\equiv &\int_d^{c} dx \sqrt{|(x-a)(x-b)(x-c)(x-d)|} \\
I_{M} &\equiv &\int_c^{b} dx \sqrt{|(x-a)(x-b)(x-c)(x-d)|}   \\
I_{NR} &\equiv &\int_a^{\Lambda_0} dx \sqrt{|(x-a)(x-b)(x-c)(x-d)|} .
\end{eqnarray}%
The integrals over compact cycles in the real locus is expressed in
closed form in terms of elliptic integrals as in \cite{HKperiod}. The non-compact cycle
is Taylor-expanded below in terms of a small variable.

The integrals $I_L, I_R, I_M, I_{NR}$ are given below to 4th order in $\Delta_{21}$ by
\begin{eqnarray} \nonumber
I_R &=& \frac{\pi }{2\sqrt{1+2s^{2}}}\left( \Delta_{43}^2
+(-1+c-s^{2})\Delta _{21}^{2}+\frac{\left( 1+2s^{2}\right)
(-1-7c^{2}+4c^{3}+4c^{5})}{8c^{3}}\Delta _{21}^{4}\right), \\ \nonumber
I_L &=&\frac{\pi }{2\sqrt{1+2s^{2}}}\left( c\Delta _{21}^{2}+\frac{\left(
1+2s^{2}\right) \left( -1-7c^{2}\right) }{8c^{3}}\Delta _{21}^{4}\right),
\end{eqnarray}
\begin{eqnarray}
\nonumber I_{NR} &=&\frac{\Lambda _{0}^{3}}{3}-\frac{\Lambda _{0}}{4}+\frac{1-6s^{2}\log \left[
\frac{2\Lambda _{0}}{\Delta_{43}}\right] }{12\left( 1+2s^{2}\right) ^{%
\frac{3}{2}}}+\frac{-2c^{2}\tanh ^{-1}c+(1+s^{2})\log \left[
\frac{2\Lambda _{0}}{\Delta_{43}}\right] }{\ 2\sqrt{1+2s^{2}}}\Delta _{21}^{2} \\ \nonumber
&+&\frac{\sqrt{1+2s^{2}}}{16c^{4}}\left( 2c\left( -1-7c^{2}\right) \tanh
^{-1}c-4c^{2}\left( 2-s^{2}\right) \log \left[
\frac{2\Lambda _{0}}{\Delta_{43}}\right] +\left( 5-6s^{2}\right) \right) \Delta _{21}^{4}, \end{eqnarray}
\begin{eqnarray}
\nonumber I_M &=&
\frac{1}{12\left( 1+2s^{2}\right) ^{\frac{3}{2}}}\left(
\begin{array}{c}
2c(1+s^{2})+3s^{3}(\pi -2\theta ) \\
+6cs^{2}\log \left[ \frac{\Delta _{21}^{{}}}{4c^{2}}s\sqrt{1+2s^{2}}\right]%
\end{array}%
\right) \\ \nonumber
&+&\frac{\Delta _{21}^{2}}{8c^{3}\sqrt{1+2s^{2}}}\left(
\begin{array}{c}
-2+9s^{2}-13s^{4}+5s^{6} \\
+c^{3}s(-2+s^{2})(\pi -2\theta ) \\
+(-4+s^{2}+2s^{4})s^{2}\log \left[ \frac{\Delta _{21}^{{}}}{4c^{2}}s\sqrt{%
1+2s^{2}}\right]%
\end{array}%
\right) \\ \nonumber
&+&\frac{\sqrt{1+2s^{2}}\Delta _{21}^{4}}{256c^{7}}\left(
\begin{array}{c}
-48+304s^{2}-1422s^{4}+2121s^{6}-1286s^{8}+284s^{10} \\
+8c^{7}s(-4+11s^{2})(\pi -2\theta ) \\
+2s^{2}(8-252s^{2}+481s^{4}-340s^{6}+88s^{8})\log \left[ \frac{\Delta
_{21}^{{}}}{4c^{2}}s\sqrt{1+2s^{2}}\right]%
\end{array}%
\right),%
\end{eqnarray}
where we denote $(c,s) =(\cos \theta,\sin \theta)$ where
 $\theta $ is given by %
\begin{equation}
\Delta _{43}=\frac{\sin \theta }{\sqrt{2\sin ^{2}\theta +1}}.
\end{equation}%

Now consider the case where a small cut $S_1$ is inside a larger cut $S_2$ as in Figure \ref{fig:ins}. The endpoints of branch-cuts are ordered as
\begin{equation}
\left( a_{1},a_{2},a_{3},a_{4}\right) =\left( d,c,b,a\right) =\left( -\Delta
_{43}+\frac{I}{2},-\Delta _{21}-\frac{I}{2},\Delta _{21}-\frac{I}{2},\Delta
_{43}+\frac{I}{2}\right). \label{inside}
\end{equation}
The integrals $I_R, I_L, I_M, I_{NR}$ are given to 4th order to $\Delta_{21}$ by
\begin{eqnarray}
 \nonumber I_R&=&\frac{I}{4cs}\left(
\begin{array}{c}
\left( \frac{c}{6}+\frac{s(\pi +2\theta^\prime )}{4(2+s^{2})}\right) +\frac{\Delta
_{21}^{2}}{4}\left( c(-1+2\log \left[\frac{\Delta _{21} s}{4c^{2}I}\right])-2s(\pi +2\theta^\prime )\right) \\
+\frac{\Delta _{21}^{4}(2+s^{2})}{64c^{3}}\left(
17-28s^{2}+8s^{4}+4(-5+4s^{2})\log \left[\frac{\Delta _{21} s}{4c^{2}I}\right]\right)%
\end{array}%
\right) ,\\
 \nonumber I_L &=& \frac{I}{4s}\left(
\begin{array}{c}
\left( \frac{c}{6}-\frac{s(\pi -2\theta^\prime )}{4(2+s^{2})}\right) +\frac{\Delta
_{21}^{2}}{4}\left( c(-1+2\log \left[\frac{\Delta _{21} s}{4c^{2}I}\right])+2s(\pi -2\theta^\prime )\right) \\
+\frac{\Delta _{21}^{4}(2+s^{2})}{64c^{3}}\left(
17-28s^{2}+8s^{4}+4(-5+4s^{2})\log \left[\frac{\Delta _{21} s}{4c^{2}I}\right] \right)%
\end{array}%
\right) , \\
\nonumber I_M &=&\frac{c\pi }{4}\left( \Delta _{21}^{{}}\frac{1-s}{2+s^{2}}+\frac{\Delta
_{21}^{2} I}{s}+\frac{\Delta _{21}^{3}\ }{8c^{4}}\left(
9+4s-8s^{2}-4s^{3}\right) +\frac{\Delta _{21}^{4}s}{8c^{4}I}\left(
-5+4s^{2}\right) \right), \\
 \nonumber I_{NR} &=&\frac{\Lambda _{0}^{3}}{3}-\frac{\Lambda _{0}}{4}+\frac{I^3}{12s^{2}}\left( s^{2}-6\log \left[ \frac{2\Lambda _{0}s}{I}\right]\right) +\frac{%
\Delta _{21}^{2} I}{4s}\left( 4s\log \left[ \frac{2\Lambda _{0}s}{I}\right]+c\left( \pi -2\theta^\prime \right) \right)  \\
 \nonumber&-&\frac{\Delta _{21}^{4}s}{32c^{3}I}\left( 2sc\left(
-3+2s^{2}\right) +(5-4s^{2}\right) (\pi -2\theta^\prime ))   \end{eqnarray}
 again $(c,s) =(\cos \theta^\prime,\sin \theta^\prime)$ where $\theta^{\prime}$ is given by
\begin{equation}
I=\frac{\sin \theta^{\prime} }{\sqrt{2+\sin ^{2}\theta^\prime }}.
\end{equation}

\chapter{Stable vacua with D5-branes and a varying Neveu-Schwarz flux \label{Ch:exactIIB}}

This chapter considers $\mathcal{N}=1$ supersymmetric theories with a
field-dependent gauge coupling. A novel mechanism for spontaneous supersymmetry breaking is observed to result from negative squared gauge couplings, and we obtain meta- and {it exactly stable} vacua.

 The set-up is following: Consider the case of D5-branes wrapped on vanishing cycles in local Calabi-Yau
3-folds, and then turn on a Neveu-Schwarz background $B$-field $\alpha ( x )$ which depends
holomorphically on one complex coordinate $x$ of the 3-fold.
Using large $N$ duality via a geometric transition as in the previous chapter, we
show how the strongly coupled IR dynamics can be
understood using string theoretic techniques.

If the background field $\alpha ( x )$ is chosen appropriately,
there are vacua exhibiting broken supersymmetry. A suitable
choice of higher-dimensional operators can lead to negative values of $g_{\mathrm{YM}}^{2}$ for certain factors of the gauge group, which leads to supersymmetry breaking. In the string theory framework, this
arises from the presence of anti-branes in a holomorphic $B$-field background.

 The organization of the rest of the chapter is as follows. Section \ref{sec:exactn1} provides the string theory construction in terms of $N$ D5-branes.
  The closed string dual at large $N$ is presented in section \ref{sec:closedNS}. Section \ref{sec:nonsusyNS} studies supersymmetry breaking mechanisms.

\section{The string theory construction \label{sec:exactn1}}
Consider type IIB string theory compactified on the Calabi-Yau 3-fold defined by
\begin{equation}
{uv=y^{2}-W^{\prime 2},}  \label{open}
\end{equation}
with the superpotential $W$
\begin{equation}
W(x)=\sum_{k=1}^{n+1}a_{k}x^{k} , \qquad W^{\prime }(x)=g\prod_{i=1}^{n}(x-e_{i}).
\end{equation}%
 At each of $n$ critical points $x=e_i$ of $W$, the geometry develops a conifold singularity, which
is resolved by a minimal ${S^2}$. (See Figures \ref{fig:1cut} and \ref{fig:BlowUp}.) We choose $N_{i}$
of the D5-branes to wrap the $i$'th ${S^2}$. This is similar to
the considerations of the section \ref{Ch:metaIIB}, except that we allow a varying holomorphic
Neveu-Schwarz field. In particular, the tree-level gauge coupling for the branes
wrapping the ${S^2}$ at $x=e_{i}$ is given by
\begin{equation}
{\int_{{S^2}_{i}}B_{0}=\left( \frac{\theta }{2\pi }+\frac{4\pi i}{%
g_{\mathrm{YM}}^{2}}\right) _{i}=\alpha (e_{i}).}  \label{ggcouple}
\end{equation}

\section{The closed string dual \label{sec:closedNS}}

The open-string theory on D5-branes at UV has a dual description in terms of
pure geometry with fluxes at IR.
In flowing to the IR, the D5-branes deform the geometry around them so that
the ${S^2}$'s they wrap vanish, while the $S^{3}$'s
surrounding the branes obtain finite sizes, as depicted in Figure \eqref{fig:1cut} for a single conifold. After the geometric transition, the geometry is complex-deformed from that given by \eqref{open} to the manifold
\begin{equation}
{uv=y^{2}-W^{\prime 2}+f_{n-1}(x) } . \label{closed}
\end{equation}%
Here $f_{n-1}(x)$ is a polynomial in $x$ of degree $n-1$, the $n$
coefficients of which govern the sizes of the $n$ resulting $S^{3}$'s.

The effective superpotential is classical in the dual geometry and is
generated by fluxes
\begin{equation}
\mathcal{W}_{\mathrm{eff}}=\int_{CY}H\wedge \Omega ,
\end{equation}%
where $\Omega $ is a holomorphic three-form on the Calabi-Yau three-fold
\begin{equation}
\Omega =\frac{{dx\wedge dy\wedge dz}}{z}.
\end{equation}%
The effective superpotential is computed in \cite{absv2}
\begin{equation}
{{\mathcal{W}}_{\mathrm{eff}}=\sum_{k=1}^{n}N_{k}}\frac{{\partial }}{{{%
\partial S_{k}}}}{{\ \mathcal{F}}_{0}+\oint_{A_{k}}\alpha (x)ydx.}
\label{delW}
\end{equation}

\section{Supersymmetry breaking \label{sec:nonsusyNS}}

This section studies the phase structure of the $\mathcal{N}=1$ models
introduced in the section \ref{sec:exactn1}. We find that there is a region in the parameter space
where supersymmetry is broken. This leads to novel and calculable
mechanisms for breaking supersymmetry.

The organization of the rest of the section is as follows.
Subsection \ref{allnega} studies the situation of all $\mathrm{Im}\,\alpha
(e_{i})$ negative. Subsection \ref{subsec:susyNSbreak} turns on NS-field which varies holomorphically, and the internal dynamics of the gauge theory softly
breaks supersymmetry. Subsection \ref{subsec:metaNS}
considers metastable supersymmetry breaking mechanism of the multi-sign case, and subsection \ref{subsec:metaNSdecay} studies its decay.

\subsection{Negative gauge couplings and flop of $S^2$ \label{allnega}}

Consider $N$ D5-branes on the resolved conifold geometry with a single $%
{S^2}$, and turn on a $constant$ $B$-field through the ${S^{2}}$,
\begin{equation*}
\alpha =\frac{{\theta }}{{2\pi }}+\frac{{4\pi i}}{{g_{\mathrm{YM}}^{2}}}%
=\int_{S_{x}^{2}}\bigl(B_{\mathrm{RR}}+\frac{{i}}{{g_{s}}}{\ B_{\mathrm{NS}}}%
\bigr).
\end{equation*}
By changing the $B$-field, an $S^{2}$ undergoes a flop\footnote{String theory makes sense even when going through a flop. See \cite{Mflop,GreeneCY3} for discussion.},
 into a \textit{new} ${S^2}$ with {\it negative} area. Moreover, the
charge of the wrapped D5-branes on this flopped ${S^2}$ is opposite
to what it was before the flop. Therefore, in order to conserve D5-brane
charge across the flop, anti-D5-branes appear on the new $%
{S^2}$ instead of D5-branes.

In the case of constant $B$-field, we again obtain a $U(N)$
gauge theory with $\mathcal{N}=1$ supersymmetry at low energies.
However,
the $\mathcal{N}=1$ supersymmetry that the theory preserves after the flop
has to be \textit{orthogonal} to the original one, since branes and
antibranes preserve different supersymmetries.

\begin{figure} [t]
\centerline{\includegraphics[width=.5\textwidth]{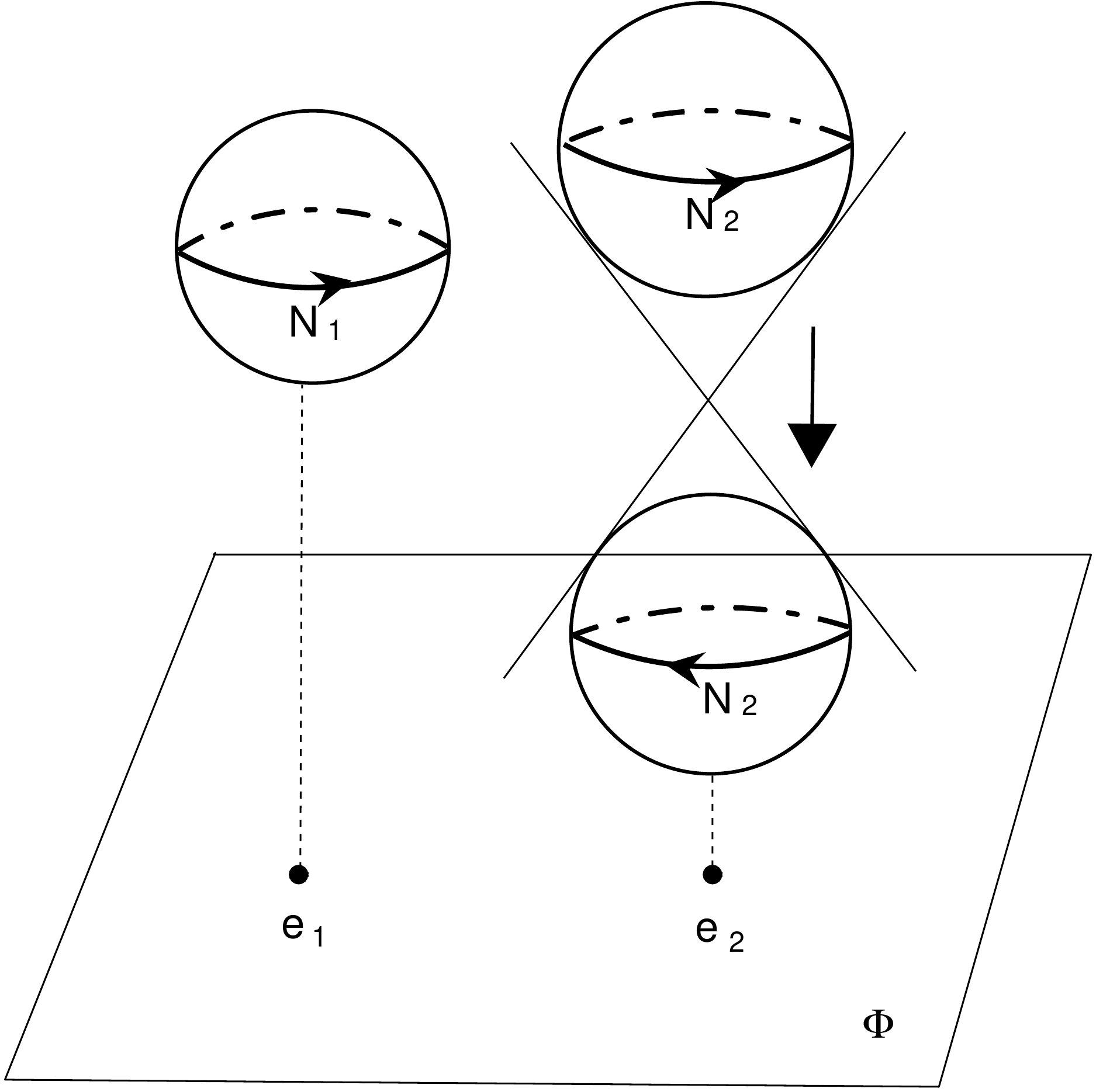}}
\caption[Flop of a $S^2$ turns branes wrapped on it into anti-branes.]{Start by wrapping $N_1, N_2 >0$ branes on $S^2$'s located at resolved singular points $x=e^1, e^2$. As we flop the second $S^2$, the second stack of branes will turn into a stack of anti-branes, in order to hold the same Gauss's law holds true for RR-charge.}
\label{fig:flop}
\end{figure}

\subsection{Supersymmetry breaking by background Neveu-Schwarz fluxes \label{subsec:susyNSbreak}}

Now consider the same geometry as in the previous subsection, but with a
holomorphically varying NS $B$-field introduced. Wrapping branes on the
conifold gives rise to supersymmetric theories. However, in the case of anti-branes,
supersymmetry is in fact broken. This arises from the fact that, while branes
preserve the same half of the background ${\mathcal{N}=2}$ supersymmetry as
the $B$-field, anti-branes preserve an opposite half.

As in the previous section, this section considers branes and anti-branes on the
conifold geometry, but
now with the holomorphically varying $B$-field given by:
\begin{equation}
{\ B(x )=t_{0}+t_{2} x^{2}.}  \label{BOpen}
\end{equation}%
This can be studied from the perspective of the IR effective field theory of
the glueball superfield $S$. Because of the underlying ${\mathcal{N}=2}$
structure of this theory, we will have a valid IR description regardless of
whether it is branes or anti-branes which are present.

Recall from \eqref{delW} that the superpotential in the dual geometry is
\begin{equation}
{{\mathcal{W}}(S)=-\oint_{A}B(x)ydx+N{\frac{\partial {\mathcal{F}}_{0}}{%
\partial S}}.}  \label{supn}
\end{equation}%
 The first term may be explicitly calculated as:
\begin{equation*}
\oint_{A}B(x)ydx=t_{0}S+t_{2}\frac{{S^{2}}}{m}.
\end{equation*}%
The scalar potential is again given by \eqref{Veff} with the same metric and
prepotential $\mathcal{F}_{0}$, but now with superpotential \eqref{supn}.
The vacua which extremize the potential $\partial _{S} V_{\mathrm{%
eff}}=0$ satisfy one of following:
\begin{eqnarray}
-\left( t_{0}+2t_{2}\frac{{S}}{m}\right) +N\tau &=&0, \label{vacua1}\\
-\left( t_{0}+2t_{2}\frac{{S}}{m}\right) +N{\bar{\tau}}+4\pi i(\tau -\bar{%
\tau})t_{2}\frac{{S}}{m}&=&0.  \label{vacua2}
\end{eqnarray}%
The solution to \eqref{vacua1} also satisfies $\partial \mathcal{W}=0$, and corresponds to
the case where branes are present, with
\begin{equation*}
\mathrm{Im}[\alpha ]\gg 0,
\end{equation*}%
where
\begin{equation}\alpha =t_{0}+2t_{2}\frac{S}{m}.\end{equation}
Large
positive values of $\mathrm{Im}[\alpha ]$ give $|S/m|\ll |\Lambda _{0}^{2}|$
within the allowed region. This vacuum is manifestly supersymmetric.

Instead, study anti-branes by allowing the geometry to undergo a flop,
so that
\begin{equation}
\mathrm{Im}[\alpha]\ll0.
\end{equation}
Then the supersymmetric solution is unphysical, and we instead study
solutions to (\ref{vacua2}).  One
can directly observe the fact that supersymmetry is broken in this vacuum by
computing the tree-level masses of the bosons and fermions in the theory,
and showing that there is a nonzero mass splitting.

The fermion masses may be read off from the $\mathcal{N}=1$ Lagrangian as
\begin{eqnarray*}
{\Lambda ^{-4}m_{\psi }} &{=}&\frac{{1}}{{{2i\left( \mathrm{Im}\tau \right)
^{2}}}}\frac{{1}}{{2\pi iS}}{\left( t_{0}+N\bar{\tau}+2t_{2}\frac{{S}}{m}%
\right) +\ }\frac{{1}}{{{\mathrm{Im}\tau }}}\frac{{{2t_{2}}}}{m} \\
{\Lambda ^{-4}m_{\lambda }} &{=}& \frac{{1}}{{{2i\left( \mathrm{Im}\tau
\right) ^{2}}}}\frac{{1}}{{2\pi iS}}{\overline{\left( t_{0}+N\tau +2t_{2}%
\frac{{S}}{m}\right) }},
\end{eqnarray*}%
while the bosonic masses are computed to be
\begin{equation*}
\Lambda ^{-4}m_{b,\pm }^{2}=\frac{{1}}{{\mathrm{Im}\tau }}\left( \partial
\bar{\partial}{V_{\mathrm{eff}}}\pm \left\vert \partial \partial {V_{\mathrm{%
eff}}}\right\vert \right) .
\end{equation*}%

By evaluation of the masses in the brane vacuum, it follows that $\lambda $ is a
massless fermion which acts as a partner of the massless gauge field $A$, while $\psi $ is a
superpartner to $S$. In other words, supersymmetry pairs up the bosons with
fermions of equal mass.

Evaluating the masses in the anti-brane vacuum, $\psi $ becomes the massless
goldstino. However, there is no longer a bose/fermi degeneracy like where
the background $B$-field was constant. Instead,
\begin{equation}
{m_{b,\pm }^{2}=|m_{\lambda }|^{2}\pm 4\pi \Lambda ^{4}|m_{\lambda }\partial
\alpha |.}  \label{onecutmass}
\end{equation}%
This mass splitting shows quite explicitly that all supersymmetries are
broken in this vacuum. Since this supersymmetry breaking can occur within a conifold, we call this {\it domestic} supersymmetry breaking.

\subsection{Multi-cut geometry and supersymmetry breaking \label{subsec:metaNS}}

Previously we have focused on the case where all gauge
couplings have the same sign, positive or negative. More generally, expand consideration to the more general case in which both signs are present. This gives {\it inter-conifold} supersymmetry breaking, involving D-brane stacks on multiple conifolds. The mass splittings of bosons and fermions are explicitly computed in \cite{absv2}. Interestingly, the
vacuum energy density formula is now given by:
\begin{equation}
\frac{{{{V_{\mathrm{eff}}}_{\ast }}}}{{{\Lambda ^{4}}}}{={2\sum_{i}N_{i}{%
\left( |\mathrm{Im}\,\alpha _{i}|-\mathrm{Im}\,\alpha _{i}\right) }}+\left(
\sum_{i,j}^{\delta _{i}>0,\delta _{j}<0}{\frac{2}{\pi }}N_{i}N_{j}\log
\left\vert {\frac{\Lambda _{0}}{\Delta _{ij}}}\right\vert \right) .}
\label{multiV}
\end{equation}%
Here, the first term is the brane tension contribution from each flopped $%
{S^2}$ with negative $g_{\mathrm{YM}}^{2}$ and the second term suggests that opposite brane types interact
to contribute a \textit{repulsive} Coulomb potential energy,as in the cases
considered in \cite{abf}.

\subsection{Decay mechanism for non-supersymmetric systems \label{subsec:metaNSdecay}}

It is straightforward to see how the non-supersymmetric systems studied in
this section can decay. This is particularly clear in the UV picture. If the
gauge coupling constants are all negative, the branes want to sit at the
critical point $x=e_i$ with the smallest $|\mathrm{Im} B(e_i)|$, minimizing vacuum energy according to (\ref{multiV}). Thus we expect that in
this case the system will decay to the $U(N)$ theory of
antibranes in a holomorphic $B$-field background. Although this breaks
supersymmetry, it is completely stable. Considering that RR charge has
to be conserved, no further decay is possible.

If there are some critical points at which $\mathrm{Im}B(e_i)$ is positive,
there is no unique stable vacuum. Instead, there are precisely as many vacua as number of
ways distribute $N$ branes amongst the critical points $x=e_i$ where $%
\mathrm{Im} B(e_i)>0$. Any of these numerous supersymmetric vacua
could be the end point of the decay process.

\chapter{A Dirac neutrino model in an F-theory $SU(5)$ Grand Unified Theory Model\label{Ch:DiracF}}

The discovery of neutrino oscillations \cite{homestake98,superkamiokande98}
has revealed that neutrinos have small but non-zero mass. However, massive
neutrinos cannot be explained in the Standard Model of particle physics without
introducing extra ingredients. As such, neutrino physics offers a concrete
and exciting window into physics beyond the Standard Model.

Aspects of flavor physics in
F-theory Grand Unified Theories have been studied in \cite{HVCKM}, where it was shown that
with the minimal number of geometric ingredients necessary for achieving one
heavy generation, the resulting flavor hierarchies in the quark and charged
lepton sectors are in accord with observation. The aim of this chapter is to
extend this minimal framework to include a neutrino sector with viable
flavor physics using the supersymmetry breaking and mediation mechanisms of \cite{HVweak}.

We present a Dirac neutrino model in a minimal $SU(5)$ F-theory GUT, which leads
to a phenomenologically consistent model of neutrino flavor. (This work was previously published in \cite{fgutnus}, which also studies a Majorana neutrino scenario.) Integrating out massive Kaluza-Klein modes
generates higher dimension operators which generate viable neutrino masses.
The neutrino mass scale $m_{\nu }$ is roughly related to the weak scale $M_{\mbox{\tiny weak}}$ and
a scale close to $M_{\rm GUT}$ through the numerology of the seesaw mechanism:
\begin{equation}
m_{\nu }\sim \frac{M_{\mbox{\tiny weak}}^{2}}{\Lambda _{\mbox{\tiny UV}}}.
\end{equation}%
In the Dirac scenario, the D-term
\begin{equation}
\lambda _{ij}^{\mbox{\tiny Dirac}}\int \mathrm{d}^{4}\theta \frac{%
H_{d}^{\dag }L^{i}N_{R}^{j}}{\Lambda _{\mbox{\tiny UV}}}  \label{e:DIR}
\end{equation}%
is generated by integrating out massive modes on the Higgs down curve.
Supersymmetry breaking leads to an F-term for $H_{d}^{\dag }$ of order ${%
F_{H_{d}}}\sim \mu H_{u}\sim M_{\mbox{\tiny weak}}^{2}$, inducing a
Dirac mass.

The supersymmetry breaking sector of \cite{HVweak} naturally
enters the discussion of neutrino physics. In \cite{HVweak}, the absence of
a bare $\mu $ term in the low energy theory was ascribed to the presence of
a $U(1)$ Peccei-Quinn symmetry, derived from an underlying $E_{6}$ GUT
structure.

Estimating the form of the Yukawa matrices for the operator
 \eqref{e:DIR}, we find that in both scenarios the
neutrinos exhibit a ``normal" hierarchy, where the two lightest neutrinos
are close in mass. The participation of Kaluza-Klein modes dilutes the mass
hierarchy in comparison to the quark and charged lepton sectors. More
precisely, the resulting neutrino mass hierarchy is roughly:
\begin{equation}
m_{1}:m_{2}:m_{3} \sim \alpha_{\rm GUT}:\alpha^{1/2}_{\rm GUT}:1
\end{equation}
which is in reasonable accord with the observed neutrino mass splittings.

The structure of the neutrino mixing matrix displays a mild hierarchical
structure. The two mixing angles $\theta _{12}$ and $\theta _{23}$ are found
to be comparable, and in rough agreement with experiments. The mixing angle $%
\theta _{13}$, which measures mixing between the heaviest and lightest
neutrino (in our normal hierarchy), is predicted to be roughly given (in
radians) by:%
\begin{equation}
\theta _{13}\sim \theta _{C}\sim \alpha _{\rm GUT}^{1/2}\sim 0.2\mbox{\tiny ,}
\end{equation}%
where $\theta _{C}$ denotes the Cabibbo angle.

The organization of the rest of the chapter is as follows. Section \ref{FREV}
provides a short review of those aspects of F-theory GUTs which are of
relevance to neutrino physics. A Dirac neutrino model is presented in section \ref{sec:DIRAC}. Our results for the neutrino masses and
mixing angles are compared with experiments in section \ref{sec:OBS}.

\section{Minimal F-theory Grand Unified Theories\label{FREV}}

In this section we briefly review the main features of minimal F-theory
GUTs, focusing on those aspects of particular relevance to neutrino
physics.\ For further background and review, see \cite{HeckmanReview}. We also discuss in greater detail the role of the anomalous global $U(1)$
Peccei-Quinn symmetry in the supersymmetry breaking sector of the low energy
theory, and its interplay with the neutrino sector.

\begin{figure}[t]
\centerline{\includegraphics[width=0.4\textwidth]{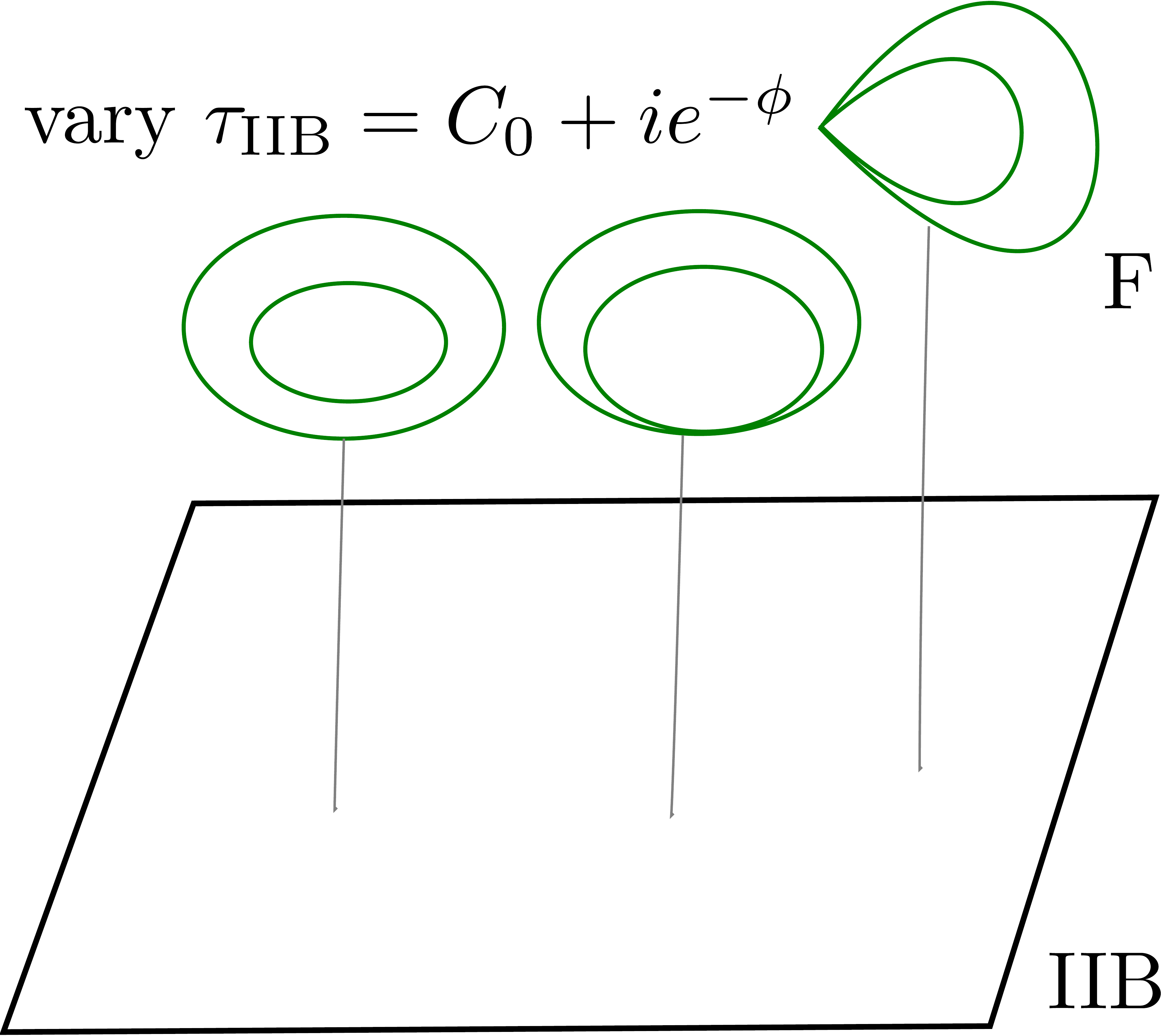}}
\caption[F-theory as a non-perturbative extension of type IIB string theory]{F-theory is a non-perturbative extension of type IIB string theory with coupling constant $\tau_{\rm IIB}$ allowed to vary over a complex number with $SL(2,Z)$ symmetry of a \textcolor{green}{2-torus}}
\label{fig:IIBF}
\end{figure}

F-theory is defined as a strongly coupled formulation of IIB string theory
in which the profile of the axio-dilaton $\tau _{IIB}$ is allowed to vary
over the ten-dimensional spacetime (see Figure \ref{fig:IIBF}.).

In F-theory GUTs, the gauge degrees of freedom of the GUT\ group propagate
in the bulk of the seven-brane wrapping a complex surface $S$. The type of
singular fibers over $S$ decides the GUT group, which in the present, minimal case corresponds
to $SU(5)$.
The chiral matter and Higgs fields of the MSSM\ localize on 2d Riemann surfaces
(complex curves) in $S$. The Yukawa couplings of the model localize near
points where at least three such matter curves intersect.

Breaking the GUT group requires introducing a non-trivial flux in the $%
U(1)_{Y}$ hypercharge direction of the GUT group \cite%
{BHVII,DonagiWijnholtBreakGUT}. The resulting unbroken gauge group in four
dimensions is then given by $SU(3)_{C}\times SU(2)_{L}\times U(1)_{Y}$.
Doublet-triplet splitting in the Higgs sector can be achieved by requiring
that this flux pierces the Higgs up and Higgs down curves.

Generating an appropriate value for the $\mu $ term in F-theory GUTs
requires a specific scale of supersymmetry breaking $\sqrt{F}\sim
10^{8}-10^{9}$ GeV. F-theory GUTs appear to more naturally accommodate
minimal gauge-mediated supersymmetry breaking scenarios.

We shall focus our attention on vacua with a \textit{minimal} number of
additional geometric and field theoretic ingredients required to obtain
phenomenologically viable low energy physics.

\subsection{$U(1)_{\rm PQ}$ and neutrinos \label{U1PQNEUT}}

Selection rules in string based constructions sometimes forbid
interaction terms in the low energy theory. In the specific context of
F-theory GUTs, the $U(1)_{\rm PQ}$ symmetry plays an especially prominent role
in that it forbids a bare $\mu $ and $B\mu $ term in the low energy theory.
Indeed, $U(1)_{\rm PQ}$ symmetry breaking and supersymmetry breaking are tightly
correlated in the deformation away from gauge mediation \cite{HVweak}.

In order to explain why $\mu $ can be far smaller than the GUT scale, we
forbid the bare $\mu $-term
\begin{equation}
\mu H_{u}H_{d}\mbox{\tiny ,}
\end{equation}%
by imposing a global $U(1)_{\rm PQ}$ symmetry under which the Higgs up and Higgs
down have $U(1)_{\rm PQ}$ charges $q_{H_{u}}$ and $q_{H_{d}}$ with $%
q_{H_{u}}+q_{H_{d}}\neq 0$.

In the context of F-theory GUTs, correlating the value of the $\mu $ term
with supersymmetry breaking is achieved using the higher
dimensional operator:%
\begin{equation}
L_{\rm effective}\supset \int \mathrm{d}^{4}\theta \frac{X^{\dag }H_{u}H_{d}}{\Lambda
_{\mbox{\tiny UV}}}\mbox{\tiny .} \label{leffhigh}
\end{equation}%
Here, $X$ is a chiral superfield localized on a matter
curve normal to the GUT seven-brane. The curves $X$, $H_{u}$ and $H_{d}$
have a triple intersection and the operator \eqref{leffhigh} results from
integrating out Kaluza-Klein modes on the curve on which $X$ localizes. This
necessarily requires that $X$ be charged under $U(1)_{\rm PQ}$ with charge:%
\begin{equation}
q_{X}=q_{H_{u}}+q_{H_{d}}\mbox{\tiny .}
\end{equation}%
When $X$ develops a supersymmetry breaking vev:%
\begin{equation}
\left\langle X\right\rangle =x+\theta ^{2}F_{X}\mbox{\tiny ,}
\end{equation}%
inducing an effective $\mu $ term of order:%
\begin{equation}
\mu \sim \frac{\overline{F_{X}}}{\Lambda _{\mbox{\tiny UV}}}\mbox{\tiny .}
\end{equation}%
Using the fact that $\Lambda _{\mbox{\tiny UV}%
}\mathrel{\mathstrut\smash{\ooalign{\raise2.5pt\hbox{$<$}\cr\lower2.5pt%
\hbox{$\sim$}}}}M_{\rm GUT}$, generating a value for the $\mu $ term near the
scale of electroweak symmetry breaking requires $\sqrt{F_{X}}\sim
10^{8} - 10^{9}$ GeV \cite{HVweak}.

As explained in \cite{HVweak}, this type of structure is compatible with an
underlying $E_{6}$ symmetry. The $U(1)_{\rm PQ}$ charges of the various
fields are:

\begin{equation}
\begin{tabular}{l|llllllll}
& $X$ & $Y$ & $Y^{\prime }$ & $H_{u}$ & $H_{d}$ & $\mathbf{10}_{M}$ & $%
\mathbf{\overline{5}}_{M}$ & $N_R$\\ \hline
$U(1)_{\rm PQ}$ & $-4$ & $+2$ & $+2$ & $-2$ & $-2$ & $+1$ & $+1$ & $-3$
\end{tabular}%
\label{PQcharge}
\end{equation}
where $Y$ and $%
Y^{\prime }$ denote the messenger fields of the gauge mediation sector. In
addition to forbidding a bare $\mu $ term, a $\mathbb{Z}_{2}$ subgroup of $%
U(1)_{\rm PQ}$ can naturally be identified with matter parity of the MSSM.
Indeed, by inspection of the above charges, note that the charges of the
MSSM\ chiral matter are all odd, while the Higgs fields are even.

\section{A Dirac scenario\label{sec:DIRAC}}

In this section we study minimal F-theory GUT scenarios which incorporate
Dirac masses through higher dimension operators of the effective theory. The
right-handed neutrinos correspond to four-dimensional zero modes of the
compactification. Kaluza-Klein mode excitations of the higher
dimensional theory play a prominent role in setting the overall mass
scale of the neutrino sector.
When the Higgs down, lepton doublet and right-handed neutrino
curves intersect at a point, the required D-term is generated by integrating out
massive modes localized on the Higgs down curve.

A suggestive link between the neutrino, weak and GUT scales is present in
Dirac scenarios where the Dirac mass term is generated by the higher
dimension operator
\begin{equation}
\int \mathrm{d}^{4}\theta \frac{H_{d}^{\dag }LN_{R}}{\Lambda _{%
\mbox{\tiny
UV}}}.  \label{ourOP}
\end{equation}%
This operator is generated in an analogous fashion
to the Giudice-Masiero operator $X^{\dag }H_{u}H_{d}/\Lambda _{%
\mbox{\tiny
UV}}$ obtained in \cite{HVweak} where $\Lambda _{\mbox{\tiny UV}}$ is close
to $M_{\rm GUT}$. \textit{Moreover the resulting scale of the neutrino mass
is automatically right}: Indeed, the most important feature of the usual GUT
scale seesaw is that:
\begin{equation}
m_{\nu }\sim \frac{M_{\mbox{\tiny weak}}^{2}}{\Lambda _{\mbox{\tiny UV}}}%
\sim \frac{v_{u}^{2}}{\Lambda _{\mbox{\tiny UV}}}\sim \frac{\overline{%
F_{H_{d}}}}{\Lambda _{\mbox{\tiny UV}}},
\end{equation}%
where as usual, $v_{u}$ denotes the scale of the Higgs up vev, and $%
F_{H_{d}} $ denotes the F-term component of the $H_{d}$ superfield. Note
that $F_{H_{d}}$ converts the D-term to a Dirac mass term for the neutrinos:
\begin{equation}
\int \mathrm{d}^{4}\theta \frac{H_{d}^{\dag }LN_{R}}{\Lambda _{%
\mbox{\tiny
UV}}}\rightarrow \int \mathrm{d}^{2}\theta \frac{\mu \langle H_{u}\rangle
LN_{R}}{\Lambda _{\mbox{\tiny UV}}}.  \label{ourfinalOP}
\end{equation}%
This last equality follows from the fact that the MSSM superpotential
contains the $\mu $-term
\begin{equation}
W_{MSSM}\supset \mu H_{u}H_{d},
\end{equation}%
so that the F-term equation of motion yields:
\begin{equation}
\overline{F_{H_{d}}}\sim \frac{\partial W_{MSSM}}{\partial H_{d}}\sim \mu
\left\langle H_{u}\right\rangle \sim 10^{5} \mbox{GeV}^{2}%
\mbox{\tiny
 }.
\end{equation}%
Here we have used the fact that the $\mu $ parameter is typically between $%
500-1000$ GeV in F-theory GUTs\cite{HVweak}.

\begin{figure} [!b]
\centerline{\includegraphics[width=1.0\textwidth]{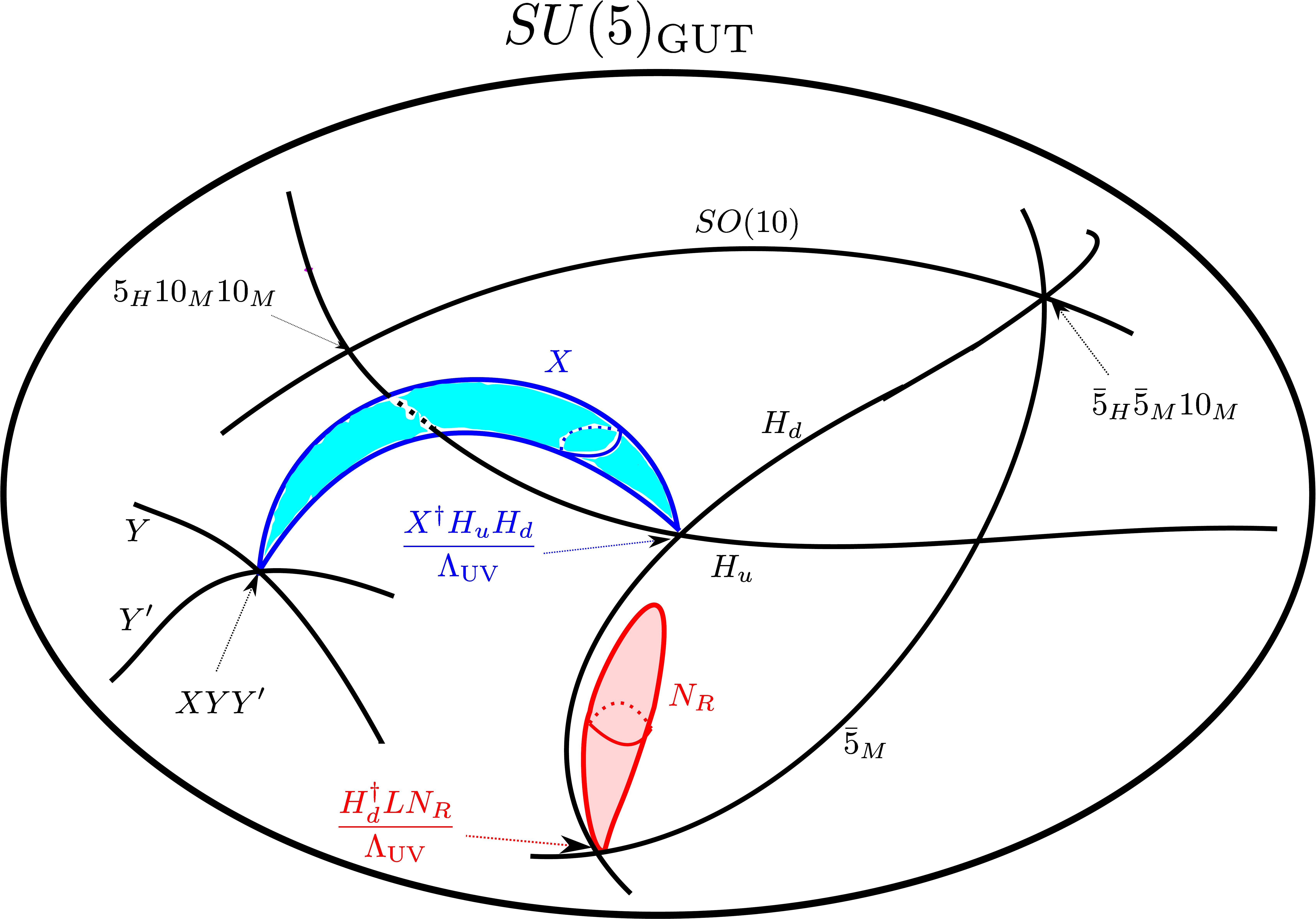}}
\caption[Matter fields and Yukawa interaction terms in an F-theory $SU(5)$ Grand Unified Theory model with Dirac neutrinos.]{Matter fields and Yukawa interaction terms in an F-theory $SU(5)$ Grand Unified Theory model with Dirac neutrinos. \colorbox{pink}{\color{red}Right-handed neutrino $N_R$} and \colorbox{lightblue}{\color{blue}messenger $X$} do not transform under $SU(5)$. Corresponding matter curves for these $SU(5)$ singlets do not live on the $SU(5)$ GUT brane and stick out of the page.}
\label{fig:Dirac}
\end{figure}

 \subsection{Generating higher dimensional operators}

We now demonstrate that the higher dimension operator:
\begin{equation}
L \supset \frac{\lambda_{ij}^{\mbox{\tiny Dirac}}}{\Lambda_{\mbox{\tiny UV}}}\int%
\mathrm{d}^{4}\theta H_{d}^{\dag}L^{i}N_{R}^{j}  \label{cubic2}
\end{equation}
is generated by integrating out massive modes localized on the Higgs down
curve. Here the right-handed neutrinos localize on a curve which is normal
to the GUT seven-brane. See Figure \ref{fig:Dirac} for a depiction of a
minimal F-theory GUT which contains a Dirac neutrino sector.
This operator can originate from a point of triple-intersection of the Higgs down, lepton doublet and right-handed neutrino
curves.

The higher-dimensional action is expressed in
terms of an infinite collection of $\mathcal{N}=1$ four-dimensional chiral
superfields, each labeled by points of the internal directions of the
compactification. The operator of (\ref{cubic2}) is obtained by
integrating out massive modes localized on the Higgs down curve. Labeling
the higher-dimensional fields by points of the threefold base,
the relevant interaction terms are given by:%
\begin{equation}
L\supset \int_{B_{3}}\mathrm{d}^{4}\theta \mathcal{H}_{d}^{\dag }\mathcal{H}%
_{d}+\int_{B_{3}}\mathrm{d}^{2}\theta \mathcal{H}_{d}^{c}\mathcal{LN}%
+\int_{B_{3}}\mathrm{d}^{2}\theta \mathcal{H}_{d}^{c}\overline{\partial }_{%
\mathcal{A}}\mathcal{H}_{d}\mbox{\tiny .}  \label{tendeff}
\end{equation}%
The F-term equation of motion for $\mathcal{H}_{d}^{c}$ yields
\begin{equation}
\overline{\partial }_{\mathcal{A}}\mathcal{H}_{d}+\mathcal{LN}=0%
\mbox{\tiny
,}
\end{equation}%
so that
\begin{equation}
\mathcal{H}_{d}=H_{d}-\frac{1}{\overline{\partial }_{\mathcal{A}}^{\prime }}%
\mathcal{LN}\mbox{\tiny ,}  \label{shifter}
\end{equation}%
where $H_{d}$ denotes the four-dimensional massless mode solution. Plugging $%
\mathcal{H}_{d}$ back into the effective action of (\ref{tendeff}), one
obtains the effective operator
\begin{equation}
\frac{\lambda _{ij}^{\mbox{\tiny Dirac}}}{\Lambda _{\mbox{\tiny UV}}}\int
\mathrm{d}^{4}\theta H_{d}^{\dag }L^{i}N_{R}^{j}=\int_{B_{3}}\mathrm{d}%
^{4}\theta H_{d}^{\dag }\frac{1}{\overline{\partial }_{\mathcal{A}}^{\prime }%
}L^{i}N_{R}^{j}\mbox{\tiny .}
\end{equation}
In other words, the relevant Yukawa matrix is given by the overlap integral
\begin{equation}
\frac{\lambda _{ij}^{\mbox{\tiny Dirac}}}{\Lambda _{\mbox{\tiny UV}}}%
=\int_{B_{3}}\overline{\Psi }_{H_{d}}\frac{1}{\overline{\partial }_{\mathcal{%
A}}^{\prime }}\Psi _{L}^{i}\Psi _{N}^{j}\mbox{\tiny ,}  \label{OURDIRAC}
\end{equation}%
where the $\Psi $'s denote the zero mode wave functions. This can be
rewritten in bra-ket notation by inserting a complete basis of states, so
that the Dirac Yukawa reduces to a sum over massive states $\left\vert \Psi
_{\mathcal{H}}\right\rangle $
\begin{equation}
\frac{\lambda _{ij}^{\mbox{\tiny Dirac}}}{\Lambda _{\mbox{\tiny UV}}}=%
\underset{\Psi _{\mathcal{H}}}{\sum }\left\langle \Psi _{H_{d}}|\Psi _{%
\mathcal{H}}\right\rangle \frac{1}{M_{\Psi _{\mathcal{H}}}}\left\langle \Psi
_{\mathcal{H}}|\Psi _{L}^{i}\Psi _{N}^{j}\right\rangle \mbox{\tiny .}
\label{MASSIVESUM}
\end{equation}%
It follows that to estimate the structure of $\lambda _{ij}^{%
\mbox{\tiny
Dirac}}/\Lambda _{\mbox{\tiny UV}}$, it is enough to compute the overlap of
the massive mode wave functions localized on the Higgs down curve with the
lepton doublet and neutrino zero mode wave functions  %
\begin{equation}
\left\langle \Psi _{\mathcal{H}}|\Psi _{L}^{i}\Psi _{N}^{j}\right\rangle
=\int_{\mathcal{U}_{B}}\overline{\Psi }_{\mathcal{H}}\Psi _{L}^{i}\Psi
_{N}^{j}\mbox{\tiny .}
\end{equation}%
Here in the above, $\mathcal{U}_{B}$ denotes a patch in $B_{3}$ containing
the neutrino interaction point.

\subsection{Neutrino Yukawa matrix}

The zero mode wave functions $\Psi _{L}^{i}$ and $\Psi _{N}^{j}$ can be
organized according to order in $z_{L,N}$ so that%
\begin{equation}
\Psi _{L}^{i}\sim \left( \frac{z_{L}}{R_{L}}\right) ^{3-i}\mbox{\tiny , }%
\qquad \Psi _{N}^{j}\sim \left( \frac{z_{N}}{R_{N}}\right) ^{3-j}%
\mbox{\tiny
 }.
\end{equation}%
Here $z_{L}$ (resp. $z_{N}$) denotes a local coordinate for the lepton
doublet (resp. neutrino) curve, and $R_{L}$ (resp. $R_{N}$) denotes the
characteristic length scale of this curve. The crucial point is that the
massive modes will overlap with the zero mode wave functions, inducing
maximal violation of the corresponding $U(1)$ coordinate rephasing
symmetries in the directions transverse to the Higgs down curve.\ Indeed,
the massive mode wave function $\Psi _{\mathcal{H}}^{I_{L},I_{N}}$ will
contain contributions of the form %
\begin{equation}
\Psi _{\mathcal{H}}^{I_{L},I_{N}}\supset \left( \frac{\overline{z_{L}}}{%
R_{\ast }}\right) ^{i}\left( \frac{\overline{z_{N}}}{R_{\ast }}\right)
^{j}\exp \left( -\frac{z_{L}\overline{z_{L}}}{R_{\ast }^{2}}-\frac{z_{N}%
\overline{z_{N}}}{R_{\ast }^{2}}\right)
\end{equation}%
for all $i\leq I_{L}$, $j\leq I_{N}$. It now follows that the overlap is
given by
\begin{equation}
\langle \Psi _{\mathcal{H}}^{I_{L},I_{N}}|\Psi _{L}^{i}\Psi _{N}^{j}\rangle
\sim \sqrt{\varepsilon _{L}^{3-i}\varepsilon _{N}^{3-j}}\theta
_{3-i}(I_{L})\theta _{3-j}(I_{N})\mbox{\tiny }.
\end{equation}%
Here $\theta _{3-i}(I)$ denotes a step function which is $1$ for $%
I\geq 3-i$, and $0$ for $I<3-i$, and we have introduced the small parameters %
\begin{equation}
\varepsilon _{L}\equiv \left( \frac{R_{\ast }}{R_{L}}\right) ^{2}%
\mbox{\tiny
, }\qquad \varepsilon _{N}\equiv \left( \frac{R_{\ast }}{R_{N}}\right) ^{2}%
\mbox{\tiny .}
\end{equation}%
Summing over all of the massive mode contributions in equation (\ref%
{MASSIVESUM}), it now follows that the Dirac matrix is given by
\begin{equation}
\frac{\lambda _{(\nu )}^{\mbox{\tiny Dirac}}}{\Lambda _{\mbox{\tiny UV}}}%
\sim \frac{\Sigma }{M_{\ast }}%
\begin{pmatrix}
\varepsilon _{L}\varepsilon _{N} & \varepsilon _{L}^{1/2}\varepsilon _{N} &
\varepsilon _{N} \\
\varepsilon _{L}\varepsilon _{N}^{1/2} & \varepsilon _{L}^{1/2}\varepsilon
_{N}^{1/2} & \varepsilon _{N}^{1/2} \\
\varepsilon _{L} & \varepsilon _{L}^{1/2} & 1%
\end{pmatrix}%
\sim \frac{\Sigma }{M_{\ast }}%
\begin{pmatrix}
\varepsilon ^{2} & \varepsilon ^{3/2} & \varepsilon \\
\varepsilon ^{3/2} & \varepsilon & \varepsilon ^{1/2} \\
\varepsilon & \varepsilon ^{1/2} & 1%
\end{pmatrix}%
\mbox{\tiny ,}  \label{ourLAMBDADIR}
\end{equation}%
where $\Sigma $ denotes the contribution from the convolution of the wave
functions by the Green's function, and where the final relation uses
the approximation $\varepsilon _{L}\sim \varepsilon _{N}\sim \varepsilon
\sim \sqrt{\alpha _{\rm GUT}}$.

\section{Comparison with experiments \label{sec:OBS}}

Let us first examine the PMNS neutrino mixing matrix \cite{Pontecorvo,MNS} :
\begin{equation}
U_{\mathrm{PMNS}}=U_{L}^{(l)}\left( U_{L}^{(\nu)}\right) ^{\dag} = \left(
\begin{array}{ccc}
c_{12}c_{13} & s_{12}c_{13} & s_{13}e^{-i\delta} \\
-s_{12}c_{23}-c_{12}s_{23}s_{13}e^{i\delta} &
c_{12}c_{23}-s_{12}s_{23}s_{13}e^{i\delta} & s_{23}c_{13} \\
s_{12}s_{23}-c_{12}c_{23}s_{13}e^{i\delta} &
-c_{12}s_{23}-s_{12}c_{23}s_{13}e^{i\delta} & c_{23}c_{13}%
\end{array}
\right) \cdot D_{\alpha}  \notag
\end{equation}
where $D_{\alpha}=$ diag$(e^{i\alpha_{1}/2},e^{i\alpha_{2}/2},1)$, $%
c_{ij}=\cos\theta_{ij}$ and $s_{ij}=\sin\theta_{ij}$. $\delta$, $\alpha_1$
and $\alpha_2$ are CP violating phases.

By diagonalizing the Yukawa interaction matrix of (\ref{ourLAMBDADIR}), we get
\begin{equation}
U_{\mathrm{PMNS}}^{\mathrm{F-th}} \sim\left(
\begin{array}{ccc}
U_{e1} & \varepsilon^{1/2} & \varepsilon \\
\varepsilon^{1/2} & U_{\mu 2} & \varepsilon^{1/2} \\
\varepsilon & \varepsilon^{1/2} & U_{\tau 3}%
\end{array}
\right)
\end{equation}
with $\varepsilon \sim \sqrt{\alpha_{\rm GUT} }\sim 0.2$ for the Dirac scenario of the previous section.
Ignoring CP violating phases, each entry has following magnitude
\begin{equation}
\left\vert U_{\mathrm{PMNS}}^{\mathrm{F-th}}\right\vert \sim%
\begin{pmatrix}
0.87 & 0.45 & 0.2 \\
0.45 & 0.77 & 0.45 \\
0.2 & 0.45 & 0.87%
\end{pmatrix}%
,
\end{equation}
in remarkable agreement with experimental results in \cite{GonMal,Gon}.
\begin{equation}
\left\vert U_{\mathrm{PMNS}}^{3\sigma}\right\vert \sim\left(
\begin{array}{ccc}
0.77-0.86 & 0.50-0.63 & 0.00-0.22 \\
0.22-0.56 & 0.44-0.73 & 0.57-0.80 \\
0.21-0.55 & 0.40-0.71 & 0.59-0.82%
\end{array}
\right),
\end{equation}
displaying a remarkable agreement. Note that we predict a rather large value
of upper right corner, $\sin \theta_{13} \sim 0.2$, and we are delighted
that there is new experimental result confirming the largeness of $%
\theta_{13}$ angle \cite{recentMINOS}.

Next, consider the ratio of neutrino mass eigenvalues. Our model
predicts mild ``normal hierarchy'' $m_{1}:m_{2}:m_{3}\sim\varepsilon^{2}:%
\varepsilon:1 $ with $\varepsilon \sim \sqrt{\alpha_{\rm GUT} } \sim 0.2$. If one
assumes normal hierarchy, the experimental data automatically gives $m_{3}^{%
\mathrm{observe}} \sim\sqrt{\Delta m_{31}^{2}}\sim50\pm4\ {\rm meV}$, $m_{2}^{\mathrm{observe}} \sim\sqrt{\Delta m_{21}^{2}}\sim8.7\pm0.4%
\ {\rm meV}$, and our F-theory model predicts the smallest mass eigenvalue to be $%
m_{1}^{\mathrm{F-th}}\sim1-3 \ {\rm meV}.$

The Dirac neutrino scenario prohibits double beta decay, but the alternative Majorana scenario in \cite{fgutnus} predicts the double beta decay mass $\left\vert
m_{\beta\beta}\right\vert ^{2}=\left\vert \underset{i=1}{\overset{3}{\sum}}%
m_{i}\left( U_{ei}^{\mathrm{PMNS}}\right) ^{2}\right\vert ^{2}$ to be $%
m_{\beta\beta}^{\max}\sim6\ {\rm meV}$. This may be observed
within ten years. The EXO experiment is expected to be sensitive down to
4-40 meV \cite{EXO}.
For single beta decay $\left\vert m_{\beta}\right\vert^{2}= \underset{i=1}{%
\overset{3}{\sum}}m_{i}^2 \left\vert U_{ei}^{\mathrm{PMNS}}\right\vert^{2}$,
we predict $\left\vert m_{\beta}^{\mathrm{F-th}}\right\vert \sim5-10%
\ {\rm meV}$, which is too small to be observed soon. The KATRIN
experiment is expected to be sensitive down to 0.2 eV \cite{KATRIN}.

\chapter{Conclusion and open problems\label{Conclusion}}

We discussed the importance and relevance of string theory in the realm of theoretical physics. If one wants to connect string theory to the real world, the first job
is to get rid of extra dimensions and to break supersymmetry while maintaining
stability of the vacuum. This thesis discussed various supersymmetry breaking mechanisms in heterotic and type IIB string theories, and generation of neutrino mass scale due to supersymmetry breaking in an F-theory $SU(5)$ Grand Unified Theory model.

We considered BPS and non-BPS states in heterotic string theory
compactified on $T^4$. A non-BPS state can decay into a set of BPS
states with the same total charge and smaller total mass. We organized
conservation of the charges using a set
of eight $16 \times 16$ transformation matrices and constrained possible decay modes allowed by charge
conservation. We constructed a non-BPS state which is rather robust against decays into BPS states. We identified its huge
stability region in moduli space of $T^4$, proving that no other
decays into BPS states are possible. We constructed a non-BPS state which does not require fine tuning of moduli for its stability, and it will be interesting to use this in a
string-inspired model building.  The study of non-BPS objects and
their stability may also provide non-trivial tests of weak-strong
duality between heterotic string theory compactified on $T^4$ and type
IIA string theory on an orbifold limit of a K3 surface.

With the similar objective of breaking supersymmetry without sacrificing
stability, we proposed various mechanisms for metastable and exactly stable vacua in type IIB string theory. These goals were achieved by
 wrapping D-branes and
anti-D-branes on cycles in a non-compact Calabi-Yau three-fold. A geometric transition was employed to take the large $N$ holographic dual to a
flux-only picture with no branes in a new geometry, corresponding to a low energy effective theory in IR.
We further investigated its phase structure and loss of classical stability.

In order to determine the global phase structure, it remains to compute explicit expressions for the periods in
regions of moduli space other than in the locus where the branch cuts are small and far apart. We studied the properties of the moduli space and performed various integrals in the real locus. It would be interesting to extend these results by determining whether large $N$ duality still holds in other regions of moduli space. It remains an open problem to determine whether the (non-) supersymmetric solutions exist and whether they are stable.

We also considered meta- and exactly stable non-supersymmetric systems of type IIB string theory by turning on a
 Neveu-Schwarz field. Variation of this field can drive the gauge coupling squared into negative, which
corresponds to a flop of the complexified K\"ahler class.

Our metastable vacua in the type IIB string theory setup has dual descriptions in type IIA string theory and M-theory, expressed in terms of
D4-branes and anti-D4-branes hanging between 2 NS5-branes \cite{IIAMextended,IIAM}, and in terms of M5-branes, respectively. It would be interesting to determine the
brane configuration of the exactly stable supersymmetry-breaking vacua in type IIA string theory and M-theory.

Finally, we showed that an F-theory Grand Unified Theory model yields neutrino masses with mass ratios
$m_{1}:m_{2}:m_{3}\sim \alpha _{\rm GUT}:\alpha _{\rm GUT}^{1/2}:1$ and a mixing matrix with large $\theta _{13}$. The supersymmetry breaking mechanism provided in \cite{HVweak} solves Higgs mass term $\mu$ problem and provides the correct neutrino mass scale.

We plan to continue our investigations by building string phenomenology models using the above meta- and exactly stable supersymmetry breaking configurations as a hidden sector. Also, we would like to determine whether our results carry over to a compact geometry, in order to have realistic gravity in 4D.
F-theory {\it local} GUT\ models still need to pass the {\it global} consistency test.
It is therefore crucial to understand what constraints global geometry imposes on
F-theory model-building and the
phenomenological implications of the decoupling limit.

\nocite{} {
 \ssp
\bibliography{SeoThesisBib}
}

\end{document}